\documentclass[a4paper,11pt]{article}
\usepackage{jheppub} 

\pdfoutput=1

\usepackage{mathrsfs,amsmath,amsthm,amssymb,xcolor,epstopdf,verbatim,hyperref,enumerate,graphicx,hyperref,subfigure,bm,bbm,amsfonts,subdepth,cleveref,setspace,booktabs,tabularx,placeins,multirow,mathtools,units}
\usepackage{cleveref}
\usepackage[english]{babel}
\usepackage[sort&compress]{natbib}
\usepackage[normalem]{ulem}
\usepackage{dcolumn}
\graphicspath{{Figs/}}
\usepackage{float}

\usepackage{tabularx}

\newcommand{\drawsquare}[2]{\hbox{%
		\rule{#2pt}{#1pt}\hskip-#2pt
		\rule{#1pt}{#2pt}\hskip-#1pt
		\rule[#1pt]{#1pt}{#2pt}}\rule[#1pt]{#2pt}{#2pt}\hskip-#2pt
	\rule{#2pt}{#1pt}}

\newcommand{\fund}{\raisebox{-.5pt}{\drawsquare{6.5}{0.4}}}
\newcommand{\Ysymm}{\raisebox{-.5pt}{\drawsquare{6.5}{0.4}}\hskip-0.4pt%
	\raisebox{-.5pt}{\drawsquare{6.5}{0.4}}}
\newcommand{\Yasymm}{\raisebox{-3.5pt}{\drawsquare{6.5}{0.4}}\hskip-6.9pt%
	\raisebox{3pt}{\drawsquare{6.5}{0.4}}}
\newcommand{\antifund}{\overline{\fund}}

 \renewcommand{\thetable}{\arabic{table}} 
 \def\ov{\overline}
\usepackage{jheppub} 
\def\yzero{\smash{\hbox{$y\kern-4pt\raise1pt\hbox{${}^\circ$}$}}}

\def\ov{\overline}
\def\s2{\frac{1}{\sqrt2}}

\def\beq{\begin{equation}}
	\def\eeq{\end{equation}}
\def\beqa{\begin{eqnarray}}
	\def\eeqa{\end{eqnarray}}

\newcommand{\Zbb}{\mathbb{Z}}
\newcommand{\Tbb}{\mathbb{T}}

\begin{document}
	
\title{The ${\cal N}=1$ supersymmetric Pati-Salam models with extra $SU(2)_{L_2/R_2}$ gauge symmetry from intersecting D6-branes}

\author[a,b,1]{Haotian Huangfu}
\author[c,2]{Tianjun Li}
\author[a,b,3]{Qi Sun}
\author[d,4]{Rui Sun}

\affiliation[a]{CAS Key Laboratory of Theoretical Physics, Institute of Theoretical Physics,\\
		Chinese Academy of Sciences, Beijing 100190, P. R. China}
\affiliation[b]{School of Physical Sciences, University of Chinese Academy of Sciences,\\
		No.19A Yuquan Road, Beijing 100049, P. R. China}
\affiliation[c]{School of Physics, Henan Normal University, Xinxiang 453007, P. R. China} 
\affiliation[d]{School of Mathematical Sciences, University of Chinese Academy of Sciences,\\
		No.19A Yuquan Road, Beijing 100049, P. R. China}

\emailAdd{huangfuhaotian@itp.ac.cn}
\emailAdd{tli@itp.ac.cn}
\emailAdd{sunqi@itp.ac.cn}
\emailAdd{sunrui24@ucas.ac.cn}

\abstract{

By introducing an extra stack of D6-branes to standard ${\cal N}=1$ supersymmetric Pati-Salam models, we extend the landscape of its complete search.
In this construction, the $d$-stack of D6-branes is introduced besides the standard $a,~b,~c$-stacks.
More intersections from the extra stacks of D6-branes appear, and thus Higgs/Higgs-like particles arise from more origins. 
Among these models, we complete with 28 classes of ${\cal N}=1$ supersymmetric Pati-Salam models under gauge symmetries $SU(4)_C\times SU(2)_L\times SU(2)_{R_1}\times SU(2)_{R_2}$ and $SU(4)_C\times SU(2)_{L_1}\times SU(2)_{R}\times SU(2)_{L_2}$, where $d$-stack of D6-branes carries the gauge symmetries $SU(2)_{R_2}$ and $SU(2)_{L_2}$, respectively.
The $SU(2)_{L_1/R_1} \times SU(2)_{L_2/R_2}$ can be broken down to the diagonal $SU(2)_{L/R}$ gauge symmetry via bifundamental Higgs fields.
In such a way, we for the first time successfully constructed three-family supersymmetric Pati-Salam models from non-rigid D6-branes with extra $d$-stacks of D6-branes as visible sectors.
Interestingly, by introducing extra stack of D6-branes to the standard supersymmetric Pati-Salam models, the number of filler brane reduces in general, and eventually the models without any $USp(N)$ gauge symmetry present.
This reduces the exotic particles from filler brane intersection yet provides more vector-like particles from ${\cal N}=2$ subsector that are useful in renormalization group equation evolution as an advantage.
Moreover, interesting degeneracy behavior with the same gauge coupling ratio exists in certain class of models.

}

\maketitle

\section{Introduction}

Brane intersection theory has been extensively explored, bridging the theoretical work of model building and phenomenological analysis, as one of the most crucial topics in string phenomenology~\cite{Berkooz:1996km, Blumenhagen:2000wh,  Klein:2000qw, Aldazabal:2000cn,Aldazabal:2000dg,Ibanez:2001nd,Cvetic:2001tj, Blumenhagen:2002gw}.
One of the main goals of string phenomenology is to realize ${\cal N}=1$ Minimal Supersymmetric Standard Models (MSSM) and Standard Models (SM), from Type I, Type IIA and Type IIB string theories, such as in~\cite{Lykken:1998ec,Cvetic:1999hb,Aldazabal:2000sa}.
In particular, in order to acquire these quasi-realistic and realistic models, chiral fermions are required to arise from either singularities of worldvolumes on D-branes or intersections between distinct D-brane stacks in the internal space.

Moreover, Grand Unified Theories (GUTs) propose that all gauge couplings from MSSM are unified at GUT scale $M_{GUT}\sim 2\times 10^{16}$ GeV, as given in~\cite{Ellis:1990wk,Langacker:1991an,Amaldi:1991cn},
while in weakly coupled heterotic string theories, the typical string scale is $M_{string}\approx g_{string}\times 5.27\times 10^{17}$ GeV~\cite{Dienes:1996du,Kaplunovsky:1992vs}.
Considering the fact that $g_{string}\sim {\cal O}(1)$, we apply $M_{string}\sim 5\times10^{17}$ GeV.
In the internal space, the intersecting D6-branes are not transversal, and thus the typical string scale is close to the reduced Planck scale $M_{\rm Pl}\simeq 2.43 \times 10^{18}$ GeV~\cite{Cvetic:2004ui}.
It brings the gauge hierarchy problem considering large reduced Planck scale corrections at the loop level and was later solved with numerous MSSM models and GUTs constructed in ~\cite{Cvetic:2001nr,Cvetic:2002pj,Cvetic:2003xs,Cvetic:2002qa,Honecker:2003vq,Li:2003xb,Cvetic:2004ui,Cvetic:2004nk,Blumenhagen:2005mu,Chen:2005mm,Chen:2007zu,Chen:2007px,Chen:2005mj,Chen:2005aba,Chen:2007ms,Chen:2008rx}.
Among these models, ${\cal N}=1$ supersymmetric three-family $SU(5)$ models from Type IIA orientifolds on $\Tbb^6 /(\Zbb_2 \times \Zbb_2)$ with intersecting D6-branes were systematically investigated~\cite{Cvetic:2002pj}.
Later in~\cite{Cvetic:2004ui}, three-family supersymmetric Pati-Salam models were systematically studied.
The Pati-Salam gauge symmetry $SU(4)_C\times SU(2)_L\times SU(2)_R$ breaks down to $SU(3)_C\times SU(2)_L\times U(1)_{B-L}\times U(1)_{I_{3R}}$ via branes splitting, and continues to break down to the SM gauge symmetry $SU(3)_C\times SU(2)_L\times U(1)_Y$ via Higgs mechanism.
Various ways of supersymmetry breaking were also discussed in~\cite{Bachas:1995ik,Taylor:1990wr,Angelantonj:2000hi}.


These study lead to the continuous investigations of string landscape as one of the most important topics in searching for the complete list of realistic particle physics vacua.
The statistics of such String vacua are investigated through the study of ensembles of eﬀective Lagrangians, with String vacua classified in~\cite{Douglas:2003um}. 
The statistics of four-dimensional, supersymmetric intersecting D-brane models constructed in a toroidal orientifold background were also investigated in \cite{Gmeiner:2005vz}. In particular, the tadpole and K-theory constraints were utilized to reduce the moduli space of string landscape with searching algorithms constructed. It was concluded that the 
frequency of MSSM-like models with three generations are drawn to be one in a billion.

For the model building of  Type IIA orientifolds on $\Tbb^6 /(\Zbb_2 \times \Zbb_2)$ with intersecting D6-branes, an exact count of the
number of distinct (i.e. gauge invariant) intersecting brane configurations
was recently obtained in \cite{Loges:2022mao}.
In which, dynamic programming is utilized to efficiently count the number of 
solutions of the tadpole, K-theory and supersymmetry consistency conditions with 134 billion intersecting brane models obtained.  
The distributions of four-dimensional gauge group and complex structure moduli in this intersecting brane vacua are also presented. 
As discussed in \cite{Gmeiner:2005vz}, the careful analysis of the constraints from tadpole, K-theory and supersymmetry consistency will 
efficiently sharp the upper and lower bounds of the string landscape.


Along this direction, the search for concrete supersymmetric Pati-Salam models was carried out in many following works such as in~\cite{Li:2019nvi}, and later generalized in~\cite{Li:2021pxo} that ${\cal N}=1$ supersymmetric $SU(12)_C\times SU(2)_L\times SU(2)_R$ models, $SU(4)_C\times SU(6)_L\times SU(2)_R$ models and $SU(4)_C\times SU(2)_L\times SU(6)_R$ models were constructed from Type IIA orientifolds on $\Tbb^6 /(\Zbb_2 \times \Zbb_2)$ with intersecting D6-branes.
In which these generalized gauge symmetries can be further broken into the standard Pati-Salam gauge symmetry via Higgs mechanism.
Along this way, powerful machine learning methods were also utilized with realistic intersecting D-brane models constructed in~\cite{Halverson:2019tkf} and \cite{Loges:2021hvn}.
The Yukawa couplings are discussed in~\cite{Honecker:2012jd,Anastasopoulos:2009mr,Anastasopoulos:2010ca,Anastasopoulos:2010hu,Anastasopoulos:2011zz,Ecker:2015vea,Marchesano:2022qbx,Antoniadis:2004dt,Antoniadis:2010zze,Kiritsis:2007zz,Kiritsis:2003mc}, and the masses and mixings of SM fermions from intersecting D6-branes were also studied in ~\cite{Sabir:2022hko,Sabir:2024cgt}.
As a milestone in the study of supersymmetric Pati-Salam models, a systematic method was developed in~\cite{He:2021gug} with the landscape of supersymmetric Pati-Salam models completed. This is investigated under the three-generation condition $I_{ab}+I_{ab'}=3$, and $I_{ac}=-3, I_{ac'}=0$ or $I_{ac}=0~,~I_{ac'}=-3$.
In this complete search of three-generation supersymmetric Pati-Salam models, a total of 202752 models were discovered, classified into 33 types from the perspective of gauge coupling relations.
Besides, string-scale gauge coupling unification for the 33 classes of models can be achieved, and one can refer to~\cite{He:2021kbj,Li:2022cqk} for details.
In recent work, by generalizing the three-generation condition to $I_{ac}=-(3+h)$ and $I_{ac'}=h$ where $h$ is a positive integer, another 4 types of supersymmetric Pati-Salam models were found in~\cite{Li:2025xpn}. This extended the landscape of supersymmetric Pati-Salam models to the general framework proposed in~\cite{Cvetic:2004ui}, which was expected to be difficult to construct.

An intriguing question was raised in~\cite{Cvetic:2002pj}:
is it possible to exclude the situation where there are more than three stacks (noted as $a$-, $b$- and $c$-stacks) of branes not parallel to the orientifold planes, with each moduli equation satisfied.
And the answer to this question appears to be negative. 
In~\cite{Mansha:2023kwq}, such $d$-stacks of D6-branes were first introduced in hidden sectors in three-family supersymmetric Pati-Salam models.
More details are shown in Section III of~\cite{Mansha:2023kwq}.
The gauge couplings remain the same as those in~\cite{He:2021gug} since the visible sectors are fixed the same.
However, a class of consistent three-family supersymmetric Pati-Salam models was constructed from rigid intersecting D6-branes on the factorizable $\Tbb^6 /(\Zbb_2 \times \Zbb'_2)$ orientifold with discrete torsion in~\cite{Mansha:2025yxm}.
Among all these models, additional $d$-stacks of rigid branes were introduced in visible sectors to yield right-handed chiral fermions by intersecting with $a$-stacks of branes as well as the intersection of $a$- and $c$-stacks of branes.
Therefore, the three-generation condition was modified to $I_{ab}+I_{ab'}=-(I_{ac}+I_{ac'}+I_{ad}+I_{ad'})=\pm 3$, and the extended Pati-Salam gauge symmetry was written as $SU(4)_C\times SU(2)_L\times SU(2)_{R_1}\times SU(2)_{R_2}$ accordingly.
In which, the $c$- and $d$-stacks of the branes represent the $SU(2)_{R_1}$ and $SU(2)_{R_2}$ gauge symmetries, respectively. 

Inspired by~\cite{Mansha:2025yxm}, here we for the first time successfully construct three-family supersymmetric Pati-Salam models from non-rigid D6-branes with extra $d$-stacks of D6-branes as visible sectors. 
There are classes of extended Pati-Salam models with $d$-stacks of branes that yield right-handed chiral fermions by intersecting with $a$-stacks of branes, as well as classes of extended Pati-Salam models with $d$-stacks of branes that yield left-handed chiral fermions by intersecting with $a$-stacks of branes in our investigation.
These new classes of models are distinct from the 37 models aforementioned from~\cite{He:2021gug} and~\cite{Li:2025xpn}.
Considering the equal positions of certain stacks of branes, D6-brane Sign Equivalence Principle (DSEP), T-duality and other symmetries, there are 98304 equivalent models with the same gauge coupling relation for each class of models, greatly extending the landscape of Pati-Salam models.

From such construction, both the extended Pati-Salam gauge symmetries $SU(4)_C\times SU(2)_L\times SU(2)_{R_1}\times SU(2)_{R_2}$ and $SU(4)_C\times SU(2)_{L_1}\times SU(2)_{R}\times SU(2)_{L_2}$ can be broken down to the standard Pati-Salam gauge symmetry $SU(4)_C\times SU(2)_L\times SU(2)_R$ via $SU(2)_{R_1}\times SU(2)_{R_2}\to SU(2)_R$ and $SU(2)_{L_1}\times SU(2)_{L_2}\to SU(2)_L$, respectively, at first, and then yield to the SM gauge symmetry $SU(3)_C\times SU(2)_L\times U(1)_Y$ via D6-branes splitting and Higgs mechanism, while the massless vector-like Higgs fields are allowed.
Moreover, their gauge coupling relation can be realized at string and GUT scales via two-loop Renormalization Group Equation (RGE) running by introducing additional vector-like particles from the ${\cal N}=2$ subsector and/or adjoint chiral multiplets from $SU(4)_C$ and $SU(2)_L$ gauge symmetries as discussed in~\cite{He:2021kbj,Li:2022cqk}.

This paper is organized as follows.
In Section 2, we first review the basics of model building on Type IIA $\Tbb^6 /(\Zbb_2 \times \Zbb_2)$ orientifolds with intersecting D6-branes.
In Section 3, the basic tadpole cancellation conditions, supersymmetry constraints and three-generation conditions for three-family supersymmetric Pati-Salam models in our construction are given.
In Section 4, we present the representative models under gauge symmetry $SU(4)_C\times SU(2)_L\times SU(2)_{R_1}\times SU(2)_{R_2}$ and $SU(4)_C\times SU(2)_{L_1}\times SU(2)_{R}\times SU(2)_{L_2}$ with their spectra explicitly.
In Section 5, we present the landscape of the extended Pati-Salam models by generalizing the Pati-Salam gauge symmetry with extra $SU(2)_{L_2/R_2}$ gauge symmetry, with the full list of detailed independent models given in the Appendix.
In Section 6, we present that the two-loop RGEs can be performed and the gauge coupling relations at string scale for representative models can be realized more efficiently with extra $d$-stack of D6-branes.
In Section 7, we eventually arrive at the conclusion of our study with the outlook given and other String landscape discussed.

\section{Basics on $\Tbb^6 /(\Zbb_2 \times \Zbb_2)$ orientifolds with intersecting D6-branes}

In brane intersection theory, D6-branes are set to intersect at generic angles on Type IIA $\Tbb^6 /(\Zbb_2 \times \Zbb_2)$ orientifolds, to obtain supersymmetric models.
The six-torus $\Tbb^{6}$ can be considered as the direct product of three two-tori, namely $\Tbb^{6} = \Tbb^{2} \times \Tbb^{2} \times \Tbb^{2}$.
Each two-torus can be described with the real coordinates $(x_i, y_i)$ or the complex coordinate $z_i$, $i=1,\; 2,\; 3$, respectively.
The two generators $\theta$ and $\omega$ of the orbifold group $\Zbb_{2} \times \Zbb_{2}$, corresponding to the twist vectors $(1/2,-1/2,0)$ and $(0,1/2,-1/2)$ respectively, acting on the complex coordinates $z_i$, such that
\beqa
& \theta: & (z_1,z_2,z_3) \mapsto (-z_1,-z_2,z_3)~,~ \nonumber \\
& \omega: & (z_1,z_2,z_3) \mapsto (z_1,-z_2,-z_3)~.~\,
\label{orbifold} \eeqa 
The orientifold projection is introduced by gauging the $\Omega R$ symmetry.
Here, $\Omega$ is the world-sheet parity and $R$ is the inversion on each real axis, which acts as
\beqa
R: (z_1,z_2,z_3) \mapsto ({\ov z}_1,{\ov z}_2,{\ov
	z}_3)~,~\, 
\eeqa
or $(x_i,\; y_i)\mapsto(x_i,-y_i)$ on each two-torus. 
Hence, we acquire four types of orientifold 6-planes (O6-planes) and result in four types of actions $\Omega R$, $\Omega R\theta$, $\Omega R\omega$, and $\Omega R\theta\omega$, respectively. 
Here, the O6-planes carry negative Ramond-Ramond (RR) charges.
In order to cancel them and conserve the RR field flux in the compact $\Tbb^6 /(\Zbb_2 \times \Zbb_2)$ without boundary, stacks of D6-branes wrapping on factorized three-cycles are introduced.
There are two possible complex structures consistent on each two-torus under orientifold projection:
rectangular or tilted~\cite{Chen:2007zu,Blumenhagen:2000ea,Cvetic:2001nr,Cvetic:2002pj}. 
The homology classes of the three cycles wrapped by D6-brane
stacks can be denoted by $n_a^i[a_i]+m_a^i[b_i]$ for rectangular tori and $n_a^i[a'_i]+m_a^i[b_i]$ for tilted tori, which are related by the fact $[a_i']=[a_i]+\frac{1}{2}[b_i]$.
Two wrapping numbers $(n_a^i,l_a^i)$ regarding the different wrapping directions on the tori are used to label a generic one cycle on both tilted and untitled tori. 
Note that in the rectangular case $l_{a}^{i}\equiv m_{a}^{i}$, while one has $l_{a}^{i}\equiv 2\tilde{m}_{a}^{i}=2m_{a}^{i}+n_{a}^{i}$ for the tilted tori. Thus, it is important to remind that $l_a^i-n_a^i$ must be even for the tilted two-torus.
It is also often required that ${m}_{a}^{i}$ and ${n}_{a}^{i}$ be relatively co-prime to avoid multiply wrapped branes~\cite{Cvetic:2002pj}, which can be represented by
\beq
l_{a}^{i}\equiv2^{\beta^i}(m_{a}^{i}+\frac{\beta^i}{2}n_{a}^{i}),\; \beta^i=
    \begin{cases}
    0,\; \mathrm{rectangular}, \\
    1,\; \mathrm{tilted}. 
    \end{cases}
\eeq
The $a$-stack of $N_a$ D6-branes along the cycle $(n_a^i,l_a^i)$ will be mapped to $a'$-stack of $N_a$ D6-branes as images with wrapping numbers $(n_a^i,-l_a^i)$, under the action of $\Omega R$. 
The homology three-cycles can be denoted as
\beq
[\Pi_a]=\prod_{i=1}^{3}\left(n_{a}^{i}[a_i]+2^{-\beta_i}l_{a}^{i}[b_i]\right)\quad \text{and} \quad
\left[\Pi_{a'}\right]=\prod_{i=1}^{3}
\left(n_{a}^{i}[a_i]-2^{-\beta_i}l_{a}^{i}[b_i]\right)~,~\, \eeq
where $\beta_i=0$ while the $i$-th torus is rectangular and while the $i$-th torus is tilted $\beta_i=1$, as mentioned above.
The homology three-cycles wrapped by the four orientifold planes are  denoted as follows
\beq \Omega R: [\Pi_{1}]= 2^3
[a_1]\times[a_2]\times[a_3]~,~\, \label{Pi1} \eeq \beq
\Omega R\omega:
[\Pi_{2}]=-2^{3-\beta_2-\beta_3}[a_1]\times[b_2]\times[b_3]~,~\,
\label{Pi2} \eeq \beq \Omega R\theta\omega: [\Pi_{3}]=-2^{3-\beta_1-\beta_3}[b_1]\times[a_2]\times[b_3]~,~\,
\label{Pi3} \eeq \beq
\Omega R\theta:  [\Pi_{4}]=-2^{3-\beta_1-\beta_2}[b_1]\times[b_2]\times[a_3]~.~\,
\label{Pi4} \eeq 
In addition, $[\Pi_{O6}]=[\Pi_{\Omega R}]+[\Pi_{\Omega R\omega}]+[\Pi_{\Omega
	R\theta\omega}]+[\Pi_{\Omega R\theta}]$ represents the sum of the four O6-plane homology three-cycles.
And the intersection numbers between different stacks can be introduced with  $k=\beta_1+\beta_2+\beta_3$ denoting  the total number of tilted two-tori. Recall the rules of Grassmann algebra, we have 
\beq
[a_i][b_j]=-[b_j][a_i]=\delta_{ij},\; [a_i][a_j]=[b_i][b_j]=0~.~\,
\eeq
Therefore, the intersection numbers between different stacks of D6-branes  can be written  in terms of wrapping numbers, such that 
\beq
I_{ab}=[\Pi_a][\Pi_b]=\prod_{i=1}^3[2^{-\beta_i}(n_a^il_b^i-n_b^il_a^i)]
=2^{-k}\prod_{i=1}^3(n_a^il_b^i-n_b^il_a^i)~,~\,
\label{Iab}
\eeq \beq
I_{ab'}=[\Pi_a]\left[\Pi_{b'}\right]=
\prod_{i=1}^3[-2^{-\beta_i}(n_a^il_b^i+n_b^il_a^i)]
=-2^{-k}\prod_{i=1}^3(n_{a}^il_b^i+n_b^il_a^i)~,~\,
\label{Iabp}
\eeq 
\beq
I_{aa'}=[\Pi_a]\left[\Pi_{a'}\right]
=\prod_{i=1}^3(-2^{1-\beta_i} n_a^il_a^i)
=-2^{3-k}\prod_{i=1}^3(n_a^il_a^i)~,~\,
\eeq 
\beq {I_{aO6}=[\Pi_a][\Pi_{O6}]=2^{3-k}(-l_a^1l_a^2l_a^3
	+l_a^1n_a^2n_a^3+n_a^1l_a^2n_a^3+n_a^1n_a^2l_a^3)}~,~\,
\label{intersections} \eeq

In terms of intersection numbers, we acquire the generic massless particle spectrum for the intersecting D6-branes at generic angles, as shown in Table \ref{spectrum}. It is suitable for both the case of all rectangular tori and the case when a tilted tori is present. 

\begin{table}[h] 
	\renewcommand{\arraystretch}{1.25}
	\begin{center}
		\begin{tabular}{|c|c|}
			\hline {\bf Sector} & \phantom{more space inside this box}{\bf
				Representation}
			\phantom{more space inside this box} \\
			\hline\hline
			$aa$   & $U(N_a/2)$ vector multiplet  \\
			& 3 adjoint chiral multiplets  \\
			\hline
			$ab+ba$   & $I_{ab}$ $(\fund_a,\antifund_b)$ fermions   \\
			\hline
			$ab'+b'a$ & $I_{ab'}$ $(\fund_a,\fund_b)$ fermions \\
			\hline $aa'+a'a$ &$\frac 12 (I_{aa'} - \frac 12 I_{a,O6})\;\;
			\Ysymm\;\;$ fermions \\
			& $\frac 12 (I_{aa'} + \frac 12 I_{a,O6}) \;\;
			\Yasymm\;\;$ fermions \\
			\hline
		\end{tabular}
	\end{center}
    \label{tab2}
	\caption{ 
	The general massless particle spectrum arising from intersecting D6-branes at generic angles. The second column specifies the representations of the gauge group $\mathrm{U}(N_a/2)$ under the $\mathbb{Z}_2\times \mathbb{Z}_2$ orbifold projections.  
}
	\label{spectrum}
\end{table}

\section{Three-family supersymmetric Pati-Salam configuration} 

In this section, we will focus on the principle conditions in the model building of three-family supersymmetric Pati-Salam models: three generation condition, 
RR tadpole cancellation conditions, and supersymmetry constraints.

\subsection{Generalized configurations with an extra stack of D6-branes}\label{3cond}
\subsubsection*{Extended three-generation conditions}

Firstly, supersymmetric Pati-Salam models with three families of chiral fermions will be obtained through the D6-branes intersection.
Instead of the previously studied three-generation condition $I_{ab}+I_{ab'}=3$ and $I_{ac}+I_{ac'}=-3$ in~\cite{Li:2025xpn}, 
here we introduce the $d$-stack to the supersymmetric Pati-Salam models. 
The three generations of chiral fermions arise from the intersection of $a$ and $c/d$-stacks, so that the three-generation conditions can be instead represented by~\cite{Mansha:2025yxm}
\beq
I_{ab}+I_{ab'}=3 ~,~
I_{ac}+I_{ac'}+I_{ad}+I_{ad'}=-3~.~
\label{3g}
\eeq
In our convention, the first and second equations in (\ref{3g}) refer to the left-handed and right-handed chiral fermions, respectively, which are expressed in terms of the intersection numbers.

\subsubsection*{Generalized RR tadpole cancellation conditions}

For convenience of subsequent discussion, the products of wrapping numbers are introduced as follows. 
Consider the $a$-stack as an example 
\beq
\begin{array}{rrrr}
	A_a \equiv -n_a^1n_a^2n_a^3, & B_a \equiv n_a^1l_a^2l_a^3,
	& C_a \equiv l_a^1n_a^2l_a^3, & D_a \equiv l_a^1l_a^2n_a^3, \\
	\tilde{A}_a \equiv -l_a^1l_a^2l_a^3, & \tilde{B}_a \equiv
	l_a^1n_a^2n_a^3, & \tilde{C}_a \equiv n_a^1l_a^2n_a^3, &
	\tilde{D}_a \equiv n_a^1n_a^2l_a^3.\,
\end{array}
\label{variables}\eeq 
The same expression applies to the other stack's products of wrapping numbers as well, denoted by subscripts $b$, $c$, and $d$.

As we discussed above in the last section, negatively charged O6-planes are introduced in the spacetime, to be precise, carrying -4 RR charges each. Such that the total amount of RR charges results in zero in a compact space to cancel the anomalies.
In total, this contributes with
\beq
-4\left[\Pi_{O6}\right]=-4(\left[\Pi_1\right]+\left[\Pi_2\right]+\left[\Pi_3\right]+\left[\Pi_4\right]) ~.~\,
\eeq
Taking the $a$-stack of D6-branes as an example, the $a$-stack and its $\omega R$ images together provide charges of $N_a [\Pi_a]+N_a\left[\Pi_{a'}\right]$ totally.
Therefore, the RR tadpole cancellation conditions read
\beq
\sum_a N_a [\Pi_a]+\sum_a N_a\left[\Pi_{a'}\right]-4\left[\Pi_{O6}\right]=0,
\eeq
where $a$ stands for the contributions from all stacks of D6-branes.
This can be reformulated with the product of wrapping number (\ref{variables}) in 
\beq
16 + \sum_a N_a A_a=16 + \sum_a N_a B_a=16 + \sum_a N_a C_a=16 + \sum_a N_a D_a=0
\eeq
Moreover, considering the $2N^{(i)}$ filler branes wrapping along the $i$-th O6-planes especially, the tadpole cancellation conditions read
\beq
\begin{array}{c}
    2^k N^{(1)}=16 + \sum_a N_a A_a ~,~\,\\
    2^k N^{(2)}=16 + \sum_a N_a B_a ~,~\,\\
    2^k N^{(3)}=16 + \sum_a N_a C_a ~,~\,\\
    2^k N^{(4)}=16 + \sum_a N_a D_a ~,~\,
\label{tadpole}\end{array}
\eeq
while the possible configurations of the filler branes are shown in Table~\ref{orientifold1}. 

\begin{table}[h] 
	\begin{center}
		\begin{tabular}{|c|c|c|}
			\hline
			Orientifold Action & O6-Plane & $(n^1,l^1)\times (n^2,l^2)\times
			(n^3,l^3)$\\
			\hline
			$\Omega R$& 1 & $(2^{\beta_1},0)\times (2^{\beta_2},0)\times
			(2^{\beta_3},0)$ \\
			\hline
			$\Omega R\omega$& 2& $(2^{\beta_1},0)\times (0,-2^{\beta_2})\times
			(0,2^{\beta_3})$ \\
			\hline
			$\Omega R\theta\omega$& 3 & $(0,-2^{\beta_1})\times
			(2^{\beta_2},0)\times
			(0,2^{\beta_3})$ \\
			\hline
			$\Omega R\theta$& 4 & $(0,-2^{\beta_1})\times (0,2^{\beta_2})\times
			(2^{\beta_3},0)$ \\
			\hline
		\end{tabular}
	\end{center}
	\caption{Possible wrapping numbers of the four O6-planes.} \vspace{0.4cm}
	\label{orientifold1}
\end{table}

\subsubsection*{Supersymmetry condition with an extra stack of D6-brane}
\label{sec:susy}

In order to preserve ${\cal N }=1$ supersymmetry, the relative angles to the orientifold plane of any D6-branes on the three two-tori, $\theta_1$, $\theta_2$ and $\theta_3$, should be the group elements of $SU(3)$.
This requires that the rotation matrix acting on the complexified compact coordinates must be unimodular~\cite{Cvetic:2002pj}, leading to
\beq
\theta_1 +\theta_2+\theta_3=0\mod 2\pi ~,~\,
\eeq
which is equivalent to $\sin(\theta_1+\theta_2+\theta_3)=0$ and $\cos(\theta_1+\theta_2+\theta_3)>0$.
Denoting $L^i(n^i,l^i)=\sqrt{(2^{-\beta_i}l^i R^i)^2+(n^i R^i)^2}$ as the length of the $i$-th torus, we obtain
\beq
\sin{\theta_i}=\frac{2^{-\beta_i}l^i R^i_2}{L^i(n^i,l^i)}~,~\,
\cos{\theta_i}=\frac{n^i R^i_2}{L^i(n^i,l^i)}~,~\,
\tan{\theta_i}=\frac{l^i R^i_2}{2^{\beta_i}n^i R^i_2}~.~\,
\eeq
Applying the products defined in (\ref{variables}), for all stacks of D6-branes the supersymmetry condition can be represented by
\beq
\begin{array}{c}
	x_A\tilde{A}_a+x_B\tilde{B}_a+x_C\tilde{C}_a+x_D\tilde{D}_a=0~,~\, \\ \frac{A_a}{x_A}+\frac{B_a}{x_B}+\frac{C_a}{x_C}+\frac{D_a}{x_D}<0~,~\,
	\label{susyconditions}
\end{array} 
\eeq
where $x_A=\lambda$ is  constant, $x_B=\lambda\, 2^{\beta_2+\beta_3}/\chi_2 \chi_3$, $x_C=\lambda\, 2^{\beta_1+\beta_3}/\chi_1 \chi_3$, and $x_D=\lambda \,2^{\beta_1+\beta_2}/\chi_1 \chi_2$ denoting a set of non-trivial solutions to the equation in (\ref{susyconditions}), and $\chi_i=R_i^2/R_i^1$ represents the complex structure modulus for the $i$-th torus.
It is obvious that $\chi_i$ should be real and positive, which directly leads  $x_A$, $x_B$, $x_C$ and $x_D$ to share the same sign.

\subsection{Gauge coupling relation with an extra $d$-stack of branes}

Recall that the Kähler potential of the Pati-Salam models is represented by~\cite{Cvetic:2003yd}
\beq
K=-\ln{(S+\bar{S})}-\sum_{i=1}^3 \ln{(U^i+\bar{U}^i)}~,~
\eeq
where $S$ and $U^i$ stand for dilaton and complex structure moduli, and the relations of their real parts are as follows
\beq
\mathrm{Re}(S)=\frac{M_s^3 R_1^1 R_2^1 R_3^1}{2\pi g_s}~,~ 
\mathrm{Re}(U^i)=\mathrm{Re}(S)\chi_j\chi_k~.~
\label{SU}
\eeq
We have already mentioned in Section \ref{sec:susy} that $R_i^1$ and $R_i^2$ are geometric parameters, together with the ratio $\chi_i={R_i^2}/{R_i^1}$, describing the $i$-th two-torus, where $g_s$ is the string coupling. 

For the $x$-stacks of D6-branes ($x$ is $a,\ b,\ c$ or $d$), carrying a gauge symmetry of $U(n)$, we have the following holomorphic gauge kinetic function represented by 
\beq
f_x=\frac{1}{4\kappa_x}(n_x^1 n_x^2 n_x^3 S-\sum_{i=1}^3 2^{-\beta_j-\beta_k}n_x^i l_x^j l_x^k U^i)\quad (i\neq j\neq k)~.~
\eeq
Here, we take $\kappa_x=1$ with respect to $SU(n)$ symmetry. Meanwhile, we have the relation between the gauge coupling constant and the kinetic function of $x$-stack of D6-branes as
\beq
g_{D6_x}^{-2}=|\mathrm{Re}(f_x)|~.~
\eeq
In the case of model building from symmetry breaking to the diagonal subgroup with
$SU(2)_{R_1}\times SU(2)_{R_2}\xrightarrow[]{}SU(2)_R$, then $g_R^{-2}$ can be written in terms of $g_c^{-2}$ and $g_d^{-2}$ as~\cite{Mansha:2025yxm}
\beq
g_R^{-2}= g_c^{-2}+g_d^{-2}=|\mathrm{Re}(f_c)|+|\mathrm{Re}(f_d)|~,~
\eeq
Alternatively, this can be represented by the gauge kinetic function:
\beq
f_R=f_c+f_d~.~
\eeq
The holomorphic gauge kinetic function of the $U(1)_Y$ gauge symmetry can be expressed as a linear combination of the kinetic function related to $SU(4)_C$ and $SU(2)_R$ as~\cite{Chen:2007zu,Blumenhagen:2000ea}
\beq
f_Y=\frac{3}{5}(\frac{2}{3}f_a+f_R)=\frac{3}{5}(\frac{2}{3}f_a+f_c+f_d)~.~
\eeq
Then the tree-level MSSM gauge couplings of the Pati-Salam models can be expressed as
\beq
g_a^2=\alpha g_b^2=\beta \frac{5}{3}g_Y^2=\gamma[\pi e^{\phi_4}]~,~
\label{gabY}
\eeq
where $g_a^2$, $g_b^2$ and $\frac{5}{3}g_Y^2$ are strong, weak and hypercharge gauge couplings, respectively, while the ratios $\alpha$, $\beta$ and $\gamma$ are different for each independent class of models. 

In addition, according to the supersymmetry conditions we discussed above, the complex structure moduli $\chi_1$, $\chi_2$ and $\chi_3$ are completely determined by the gauge symmetry, while there remains the dilation degree of freedom to be stabilized.
In order to complete the stabilization, the beta functions of $2N^{(i)}$ filler branes shall be calculated such that 
\beq
\begin{aligned}
   \beta_i^g
   & =-3\left( \frac{N^{(i)}}{2}+1\right) +2|I_{ai}|+|I_{bi}|+|I_{ci}|+|I_{di}|+3\left( \frac{N^{(i)}}{2}-1\right) \\
& =-6+2|I_{ai}|+|I_{bi}|+|I_{ci}|+|I_{di}|\quad (i=1,2,3,4)~.~ 
\end{aligned}
\eeq
We demand that the beta functions for the $USp(n)_i$ groups at the one-loop level be negative, which provides an approach to achieve gaugino condensations as shown in~\cite{Taylor:1990wr,Brustein:1992nk,deCarlos:1992kox}.

\section{Extended supersymmetric Pati-Salam models with $d$-stack of branes}\label{models}

To build a standard model or standard model-like theories from the D6-brane intersection, besides the $U(3)_C$ and $U(2)_L$ gauge symmetries, there are at least two additional $U(1)$ gauge groups present.
One $U(1)_{I_{3R}}$ is analogous to the third component of the right-handed weak isospin, while the other $U(1)_{L}$ is the lepton symmetry.
$Q_B$ is introduced as the charge of $U(1)_B$ in $U(3)_C\cong SU(3)_C\times U(1)_B$, and then the hypercharge should be
\beq 
Q_Y=Q_{I_{3R}}+\frac{Q_B-Q_L}{2} ~.~
\eeq
From the splitting of branes, $U(1)_{B-L}$ comes from $SU(4)_C$, and $U(1)_{I_{3R}}$ originates from $SU(2)_R$.
And no anomalies arise during the symmetry breaking from the Pati-Salam model to SM.

In the right-handed extension, the $a$-, $b$-, $c$- and $d$-stack of D6-branes corresponds to the $U(4)_C$, $U(2)_L$, $U(2)_{R_1}$ and $U(2)_{R_2}$ gauge symmetries, respectively. This leads to the total gauge symmetries $SU(4)_C\times SU(2)_L\times SU(2)_{R_1}\times SU(2)_{R_2}$.
In the right-handed $SU(2)_R$ extension, the $SU(2)_R$ gauge group is  obtained from the diagonal subgroup of $SU(2)_{R_1}\times SU(2)_{R_2}$ via the bifundamental Higgs fields.
Finally, a conventional gauge symmetry $SU(4)_C\times SU(2)_L\times SU(2)_R$ of the Pati-Salam model is achieved.
And the SM gauge symmetry can then be obtained by branes splitting and Higgs mechanism as before~\cite{Cvetic:2004ui}. 
The complete chain of symmetry breaking can be written as

{\small
\beq
\begin{aligned}
SU(4)\times SU(2)_L\times SU(2)_{R_1}\times SU(2)_{R_2}
& \xrightarrow{\text{bifundamental Higgs}}SU(4)\times SU(2)_L\times SU(2)_{R}\\
& \xrightarrow{a\to a_1+a_2}SU(3)_C\times SU(2)_L\times SU(2)_R\times U(1)_{B-L}\\
& \xrightarrow{c\to c_1+c_2}SU(3)_C\times SU(2)_L\times U(1)_{I_{3R}}\times U(1)_{B-L}\\
& \xrightarrow{\text{Higgs mechanism}}SU(3)_C\times SU(2)_L\times U(1)_Y ~.~
\end{aligned}
\label{chain}
\eeq
}

While in the left-handed extension, the $a$-, $b$-, $c$- and $d$-stack of D6-branes corresponds to the $U(4)_C$, $U(2)_R$, $U(2)_{L_1}$ and $U(2)_{L_2}$ gauge symmetries, respectively. This leads to the total gauge symmetries $SU(4)_C\times SU(2)_{L_1}\times SU(2)_{R}\times SU(2)_{L_2}$.
In the left-handed $SU(2)_L$ extension, the $SU(2)_L$ gauge group is  obtained from the diagonal subgroup of $SU(2)_{L_1}\times SU(2)_{L_2}$ via the bifundamental Higgs fields.
Finally, a conventional gauge symmetry $SU(4)_C\times SU(2)_L\times SU(2)_R$ of the Pati-Salam model is achieved.
And the SM gauge symmetry can then be obtained by branes splitting and Higgs mechanism as well. 
The complete chain of symmetry breaking is then written as

{\small
\beq
\begin{aligned}
SU(4)\times SU(2)_{L_1}\times SU(2)_{L_2}\times SU(2)_{R}
& \xrightarrow{\text{bifundamental Higgs}}SU(4)\times SU(2)_L\times SU(2)_{R}\\
& \xrightarrow{a\to a_1+a_2}SU(3)_C\times SU(2)_L\times SU(2)_R\times U(1)_{B-L}\\
& \xrightarrow{c\to c_1+c_2}SU(3)_C\times SU(2)_L\times U(1)_{I_{3R}}\times U(1)_{B-L}\\
& \xrightarrow{\text{Higgs mechanism}}SU(3)_C\times SU(2)_L\times U(1)_Y ~.~
\end{aligned}
\label{chain}
\eeq
}

\subsection{$SU(2)_R$ gauge from extra symmetry breaking}

In the D6-brane configuration with $d$-stack extension, 
we implement the deterministic algorithm given in~\cite{He:2021gug} and complet the full list of physical independent models(\emph{i.e., with} distinct gauge coupling relation) in $d$-stack extension and presented in the Appendix. 
In particular, we found there are 14 classes of independent models each in the  right- and left-handed gauge extension from the $d$-stack of D6-brane configurations, respectively.

\subsubsection{$SU(2)_R$ gauge from $c/d$-stack of D6-branes}
Firstly, in the right-handed $d$-stack extension of supersymmetric Pati-Salam models, the $SU(2)_R$ gauge is realized from symmetry breaking of 
$SU(2)_{R_1}\times SU(2)_{R_2}\to SU(2)_R$.
Here in this section, we discuss on  four representative Models 1-4 shown in Table~\ref{model1},~\ref{model2},~\ref{model3},~\ref{model4}.
They are independent with each other from the perspective of gauge coupling relation as well. 
In particular, the three-family of right-handed chiral fermions arises from the combined contribution of the intersections including $a$- and $c$-stacks and $a$- and $d$-stacks.
\begin{table}[h!]
\scriptsize
    \caption{D6-brane configurations and intersection numbers of Model 1, and its MSSM gauge coupling relation is
    $g^2_a=g^2_b=\frac{1}{2} g^2_R=\frac{5}{8} (\frac{5}{3} g^2_Y)=2\sqrt{2}\times3^{1/4} \pi  e^{\phi_4}$, for which 
    the second torus is tilted.}
	\label{model1}
	\begin{center}
		\begin{tabular}{|c||c|c||c|c|c|c|c|c|c|c|c|c|}
			\hline\rm{Model} 1 & \multicolumn{12}{c|}{$U(4)\times U(2)_L\times U(2)_{R_1}\times U(2)_{R_2}\times USp(2)^2 $}\\
			\hline \hline			\rm{stack} & $N$ & $(n^1,l^1)\times(n^2,l^2)\times(n^3,l^3)$ & $n_{\Ysymm}$& $n_{\Yasymm}$ & $b$ & $b'$ & $c$ & $c'$ &$d$ &$d'$ & 1 & 4\\
			\hline
			$a$ & 8 & $(1,-1)\times (1,1)\times (1,0)$ & 0 & 0  & 3 & 0 & -1 & 0 & -2 & 0 & 0 & 0\\
			$b$ & 4 & $(0,-1)\times (1,-1)\times (-1,3)$ & 2 & -2  & - & - & 0 & -2 & 2 & 1 & -3 & 0\\
			$c$ & 4 & $(-1,0)\times (-1,1)\times (1,1)$ & 0 & 0 & - & - & - & - & -2 & 1 & 0 & 1\\
            $d$ & 4 & $(-1,-1)\times (-1,-3)\times (0,-1)$ & -2 & 2  & - & - & - & - & - & - & 3 & -1\\
			\hline
			1 & 2 & $(1, 0)\times (2, 0)\times (1, 0)$& \multicolumn{10}{c|}{$x_A = \frac{1}{3} x_B = \frac{1}{3} x_C= \frac{1}{3}x_D$}\\
            4 & 2 & $(0, -1)\times (0, 2)\times (1, 0)$& \multicolumn{10}{c|}{$\beta^g_1=0$, \quad\quad$\beta^g_4=-4$}\\
			& & & \multicolumn{10}{c|}{$\chi_1=\frac{1}{\sqrt{3}}$,  $\chi_2=\frac{2}{\sqrt{3}}$, $\chi_3=\frac{1}{\sqrt{3}}$ }\\
			\hline
		\end{tabular}
	\end{center}
\end{table}
\begin{table}[h!]
\scriptsize
    \caption{D6-brane configurations and intersection numbers of Model 2, and its MSSM gauge coupling relation is
    $g^2_a=3g^2_b=\frac{3}{4} g^2_R=\frac{5}{6} (\frac{5}{3} g^2_Y)=2\sqrt{2}\times3^{3/4} \pi  e^{\phi_4}$, for which the third torus is tilted.}
	\label{model2}
	\begin{center}
		\begin{tabular}{|c||c|c||c|c|c|c|c|c|c|c|c|c|}
			\hline\rm{Model} 2 & \multicolumn{12}{c|}{$U(4)\times U(2)_L\times U(2)_{R_1}\times U(2)_{R_2}\times USp(2)^2 $}\\
			\hline \hline			\rm{stack} & $N$ & $(n^1,l^1)\times(n^2,l^2)\times(n^3,l^3)$ & $n_{\Ysymm}$& $n_{\Yasymm}$ & $b$ & $b'$ & $c$ & $c'$ &$d$ &$d'$ & 2 & 4\\
			\hline
			$a$ & 8 & $(-1,1)\times (0,-1)\times (1,-1)$ & 0 & 0  & 1 & 2 & -1 & 0 & 0 & -2 & 0 & 0\\
			$b$ & 4 & $(0,1)\times (-1,1)\times (-1,3)$ & 2 & -2  & - & - & 0 & -2 & -3 & -6 & 1 & 0\\
			$c$ & 4 & $(1,0)\times (1,-1)\times (1,1)$ & 0 & 0 & - & - & - & - & -1 & 0 & 0 & 1\\
            $d$ & 4 & $(-3,-1)\times (-1,0)\times (1,-1)$ & 2 & -2  & - & - & - & - & - & - & 1 & -3\\
			\hline
			2 & 2 & $(1, 0)\times (0, -1)\times (0, 2)$& \multicolumn{10}{c|}{$x_A = \frac{1}{3} x_B = x_C= x_D$}\\
            4 & 2 & $(0, -1)\times (0, 1)\times (2, 0)$& \multicolumn{10}{c|}{$\beta^g_2=-4$, \quad$\beta^g_4=-2$}\\
			& & & \multicolumn{10}{c|}{$\chi_1=\sqrt{3}$,  $\chi_2=\frac{1}{\sqrt{3}}$, $\chi_3=\frac{2}{\sqrt{3}}$ }\\
			\hline
		\end{tabular}
	\end{center}
\end{table}

\begin{table}[h!]
\scriptsize
    \caption{D6-brane configurations and intersection numbers of Model 3, and its MSSM gauge coupling relation is
    $g^2_a=6g^2_b=\frac{3}{2} g^2_R=\frac{5}{4} (\frac{5}{3} g^2_Y)=4\times3^{3/4} \pi  e^{\phi_4}$,
    for which the first torus is tilted.}
	\label{model3}
	\begin{center}
		\begin{tabular}{|c||c|c||c|c|c|c|c|c|c|c|}
			\hline\rm{Model} 3 & \multicolumn{10}{c|}{$U(4)\times U(2)_L\times U(2)_{R_1}\times U(2)_{R_2} $}\\
			\hline \hline			\rm{stack} & $N$ & $(n^1,l^1)\times(n^2,l^2)\times(n^3,l^3)$ & $n_{\Ysymm}$& $n_{\Yasymm}$ & $b$ & $b'$ & $c$ & $c'$ &$d$ &$d'$ \\
			\hline
			$a$ & 8 & $(1,-1)\times (0,-1)\times (-1,1)$ & 0 & 0  & 2 & 1 & -1 & 0 & -2 & 0\\
			$b$ & 4 & $(-1,-3)\times (-1,-2)\times (0,-1)$ & -5 & 5  & - & - & 4 & 0 & 6 & 12\\
			$c$ & 4 & $(-1,-1)\times (-1,2)\times (1,0)$ & 1 & -1 & - & - & - & - & 0 & -2\\
            $d$ & 4 & $(-1,-1)\times (-1,0)\times (3,-1)$ & -2 & 2  & - & - & - & - & - & -\\
			\hline
			   &  &  & \multicolumn{8}{c|}{$x_A = \frac{1}{2} x_B = x_C= \frac{1}{6}x_D$}\\
			& & & \multicolumn{8}{c|}{$\chi_1=\frac{2}{\sqrt{3}}$,  $\chi_2=\frac{1}{2\sqrt{3}}$, $\chi_3=\sqrt{3}$ }\\
			\hline
		\end{tabular}
	\end{center}
\end{table}

\begin{table}[h!]
\scriptsize
    \caption{D6-brane configurations and intersection numbers of Model 4, and its MSSM gauge coupling relation is
    $g^2_a=2 g^2_b=g^2_R=(\frac{5}{3} g^2_Y)=4\times3^{1/4} \pi  e^{\phi_4}$,
    for which the first torus is tilted.}
	\label{model4}
	\begin{center}
		\begin{tabular}{|c||c|c||c|c|c|c|c|c|c|c|}
			\hline\rm{Model} 4 & \multicolumn{10}{c|}{$U(4)\times U(2)_L\times U(2)_{R_1}\times U(2)_{R_2}  $}\\
			\hline \hline			\rm{stack} & $N$ & $(n^1,l^1)\times(n^2,l^2)\times(n^3,l^3)$ & $n_{\Ysymm}$& $n_{\Yasymm}$ & $b$ & $b'$ & $c$ & $c'$ &$d$ &$d'$ \\
			\hline
			$a$ & 8 & $(1,-1)\times (0,-1)\times (-1,1)$ & 0 & 0  & 3 & 0 & -1 & 0 & 0 & -2\\
			$b$ & 4 & $(-1,-1)\times (-3,-2)\times (0,-1)$ & 1 & -1  & - & - & 0 & -4 & 2 & 4\\
			$c$ & 4 & $(-1,-1)\times (-1,2)\times (1,0)$ & 1 & -1 & - & - & - & - & 2 & -4\\
            $d$ & 4 & $(3,1)\times (1,0)\times (1,-1)$ & 2 & -2  & - & - & - & - & - & -\\
			\hline
			   &  &  & \multicolumn{8}{c|}{$x_A = \frac{1}{2} x_B = x_C= \frac{3}{2}x_D$}\\
			& & & \multicolumn{8}{c|}{$\chi_1=2\sqrt{3}$,  $\chi_2=\frac{\sqrt{3}}{2}$, $\chi_3=\frac{1}{\sqrt{3}}$ }\\
			\hline
		\end{tabular}
	\end{center}
\end{table}
In Model 1 and 2, the supersymmetric Pati-Salam models are realized similar with the ones presented in~\cite{He:2021gug} while with extra gauge coupling relations appear.
In Model 3 and 4, the tadpole cancellation condition can be satisfied without filler brane from the $USp(N)$ gauge symmetry. This is distinct with the usual supersymmetric Pati-Salam models found in~\cite{He:2021gug}. It shows to us that by introducing extra $d$-stack of branes opens a new scenario in filler brane configuration. 
In particular, we observe that the ratio itself of gauge couplings $g_a^2:g_b^2:g_R^2:(\frac{5}{3}g_Y^2)$ of Model 4 is the same as that of Model 10  from~\cite{He:2021gug}, yet with different value of $g_a^2=4\times3^{3/4} \pi e^{\phi_4}$ with $d$-stack and $g_a^2=\frac{16\times2^{1/4}}{3\sqrt{3}} \pi e^{\phi_4}$ while without $d$-stack.
 
Note that while tree-level MSSM gauge couplings of Pati-Salam models $
g_a^2=\alpha g_b^2=\beta \frac{5}{3}g_Y^2=\gamma[\pi e^{\phi_4}]$
with the same gauge coupling ratio yet with different $\gamma$ in their value, (\ref{gabY}) can be related by scaling of the dilaton. Therefore, we propose a new phenomenological relation between the models with filler branes and $d$-stack of D6-branes.


In addition to Models 1-4, special models with either $I_{ac}+I_{ac'}=0, I_{ad}+I_{ad'}=-3$ or $I_{ac}+I_{ac'}=-3, I_{ad}+I_{ad'}=0$ also appear, such as Model 5. 
\begin{table}[h!]
\scriptsize
    \caption{D6-brane configurations and intersection numbers of Model 5, and its MSSM gauge coupling relation is
    $g^2_a=g^2_b=\frac{1}{2}g^2_R=\frac{5}{8} (\frac{5}{3} g^2_Y)=4\sqrt{\frac{2}{3}} \pi  e^{\phi_4}$,
    for which the third torus is tilted.}
	\label{model5}
	\begin{center}
		\begin{tabular}{|c||c|c||c|c|c|c|c|c|c|c|c|c|}
			\hline\rm{Model} 5 & \multicolumn{12}{c|}{$U(4)\times U(2)_L\times U(2)_{R_1}\times U(2)_{R_2}\times USp(2)^2 $}\\
			\hline \hline			\rm{stack} & $N$ & $(n^1,l^1)\times(n^2,l^2)\times(n^3,l^3)$ & $n_{\Ysymm}$& $n_{\Yasymm}$ & $b$ & $b'$ & $c$ & $c'$ &$d$ &$d'$ & 1 & 3\\
			\hline
			$a$ & 8 & $(1,-1)\times (0,-1)\times (-1,1)$ & 0 & 0  & 3 & 0 & -3 & 0 & 0 & 0 & -1 & 1\\
			$b$ & 4 & $(0,1)\times (3,1)\times (-1,-1)$ & 2 & -2  & - & - & 0 & 0 & 0 & -3 & 1 & 0\\
			$c$ & 4 & $(1,0)\times (-3,1)\times (-1,-1)$ & -2 & 2 & - & - & - & - & 0 & 3 & 0 & -1\\
            $d$ & 4 & $(-1,-1)\times (0,-1)\times (-1,-1)$ & 0 & 0 & - & - & - & - & - & - & 1 & -1\\
			\hline
			1 & 2 & $(1, 0)\times (1, 0)\times (2, 0)$& \multicolumn{10}{c|}{$x_A = 3 x_B = x_C= 3 x_D$}\\
            3 & 2 & $(0, -1)\times (1, 0)\times (0, 2)$& \multicolumn{10}{c|}{$\beta^g_1=-2$, \quad\quad$\beta^g_3=-2$}\\
			& & & \multicolumn{10}{c|}{$\chi_1=1$,  $\chi_2=3$, $\chi_3=2$ }\\
			\hline
		\end{tabular}
	\end{center}
\end{table}

\begin{table}[h!]
\scriptsize
    \caption{D6-brane configurations and intersection numbers of Model 6\;(dual to Model 5 in~\cite{He:2021gug}), and its MSSM gauge coupling relation is
    $g^2_a=g^2_b=g^2_c=(\frac{5}{3} g^2_Y)=4\sqrt{\frac{2}{3}} \pi  e^{\phi_4}$.
    for which the third torus is tilted.}
	\label{model6}
	\begin{center}
		\begin{tabular}{|c||c|c||c|c|c|c|c|c|c|c|c|c|}
			\hline\rm{Model} 6  & \multicolumn{12}{c|}{$U(4)\times U(2)_L\times U(2)_{R}\times USp(2)^4 $}\\
			\hline \hline			\rm{stack} & $N$ & $(n^1,l^1)\times(n^2,l^2)\times(n^3,l^3)$ & $n_{\Ysymm}$& $n_{\Yasymm}$ & $b$ & $b'$ & $c$ & $c'$ & 1 & 2 & 3 & 4\\
			\hline
			$a$ & 8 & $(1,-1)\times (0,-1)\times (-1,1)$ & 0 & 0  & 3 & 0 & -3 & 0 & -1 & 0 & 1 & 0 \\
			$b$ & 4 & $(0,1)\times (3,1)\times (-1,-1)$ & 2 & -2  & - & - & 0 & 0 & 1 & -3 & 0 & 0 \\
			$c$ & 4 & $(1,0)\times (-3,1)\times (-1,-1)$ & -2 & 2 & - & - & - & - & 0 & 0 & -1 & 3 \\
			\hline
			1 & 2 & $(1, 0)\times (1, 0)\times (2, 0)$& \multicolumn{10}{c|}{$x_A = 3 x_B = x_C= 3 x_D$}\\
            2 & 2 & $(1, 0)\times (0, -1)\times (0, 2)$& \multicolumn{10}{c|}{ }\\
            3 & 2 & $(0, -1)\times (1, 0)\times (0, 2)$& \multicolumn{10}{c|}{$\beta^g_1=-3$, $\beta^g_2=-3$, $\beta^g_3=-3$, $\beta^g_4=-3$}\\
            4 & 2 & $(0, -1)\times (0, 1)\times (2, 0)$& \multicolumn{10}{c|}{$\chi_1=1$, $\chi_2=3$, $\chi_3=2$}\\
			\hline
		\end{tabular}
	\end{center}
\end{table}

Similarly, as we observed for Model 4 in this work and Model 10 in~\cite{He:2021gug} that there is a hidden relation between the filler brane and the extra $d$-stack of D6-branes introduced, we find that Model 5 and Model 6\;(dual to Model 5 in~\cite{He:2021gug}) share the same intersection numbers. Different from the former case, Model 5 and Model 6 share the same gauge coupling value $g_a^2=4\sqrt{\frac{2}{3}} \pi e^{\phi_4}$ yet  do \emph{not} share the same gauge coupling ratio,  opposite to the former interesting case between Model 4 in this work and Model 10 in~\cite{He:2021gug}. 

From the perspective of gauge couplings unification study, they share the same $g_a$ value while exhibiting completely different ratios of $g_a^2:g_R^2:g_Y^2$.
Therefore, from the perspective of gauge couplings, these models are distinct and will exhibit different behaviors in the gauge coupling unification. 
Their gauge coupling unification at string and GUT scales will be shown latter in Section \ref{G.C.U.}.

\subsubsection{Phenomenological studies}

In this section, we present the full spectra of Models 1 to 4 explicitly and discuss the outcome with the extra $d$-stack of D6-branes introduced.

In particular, different from the standard supersymmetric Pati-Salam models, here Higgs multiplets not only arise from both the intersection of $b$- and $c$-stacks of branes (or its image) but also arise from the intersection of $b$- and $d$-stacks of branes (or its image) simultaneously. This can be read from Table~\ref{spectrum Model 1}-\ref{spectrum Model 4}.
In which, $Q_L$ and $L_L$ represent left-handed quarks and leptons, $Q_R$ and $L_R$ represent right-handed quarks and leptons, respectively.
Besides, $H$ stands for the Higgs particle, $H'$ stands for the Higgs-like particle, and $\Phi$ stands for the bifundamental Higgs.

Firstly, the chiral spectrum of Model 1 is shown in Table \ref{spectrum Model 1}.
From its spectrum, 4 Higgs doublets arise from the ${\cal N}=2$ subsector of the intersection of $b$- and $c$-stacks of D6-branes.
Besides, there is one $USp(2)$ group with $\beta$ function to be zero and one $USp$ group with negative $\beta$ function from Table \ref{model1}.

\begin{table}[htb]
\footnotesize
\renewcommand{\arraystretch}{1.0}
\caption{The chiral spectrum in the open string sector of Model 1}
\label{spectrum Model 1}
\begin{center}
\begin{tabular}{|c||c||c|c|c|c||c|c|c|}\hline
\scriptsize{Model 1} &\scriptsize{$SU(4)\times SU(2)_L\times SU(2)_{R1}$}
& \scriptsize{$Q_4$} & \scriptsize{$Q_{2L}$} & \scriptsize{$Q_{2R_1}$} & \scriptsize{$Q_{2R_2}$} & $Q_{em}$ & $B-L$ & Field \\
 &\scriptsize{$\times SU(2)_{R2}\times USp(2)^2$}
&   &   &   &   &   &   &   \\
\hline\hline
$ab$ & $3 \times (4,\overline{2},1,1,1,1)$ & $1$ & -$1$ & 0 & 0 &  $-\frac 13, \frac 23,-1, 0$ & $\frac 13,-1$ & $Q_L, L_L$\\
$ac$ & $1 \times (\overline{4},1,2,1,1,1)$ & -$1$ & 0 & $1$ & 0 &  $\frac 13, -\frac 23, 1, 0$ & $-\frac 13, 1$ & $Q_R, L_R$\\
$ad$ & $2 \times (\overline{4},1,1,2,1,1)$ & -$1$ & 0 & 0 & $1$  & $\frac 13, -\frac 23, 1, 0$ & $-\frac 13, 1$ & $Q_R, L_R$\\
$bc'$ & $2 \times(1,\overline{2},\overline{2},1,1,1)$ & 0 & -$1$ & -$1$ & 0  & $0,\;\mp 1$ & 0 & $H'$\\
$bd$ & $2 \times(1,2,1,\overline{2},1,1)$ & 0 & $1$ & 0 & -$1$  & $0,\;\pm 1$ & 0 & $H$\\
$bd'$ & $1 \times(1,2,1,2,1,1)$ & 0 & $1$ & 0 & $1$   & $0,\;\pm 1$ & 0 & $H'$\\
$cd$ & $2 \times(1,1,\overline{2},2,1,1)$ & 0 & 0 & -$1$ & $1$   & $0,\;\mp1$ & 0 & $\Phi$\\
$cd'$ & $1 \times(1,1,2,2,1,1)$ & 0 & 0 & $1$ & $1$   & $0,\;\pm1$ & 0 & $\Phi$\\
$b1$ & $3\times (1,\overline{2},1,1,2,1)$ & 0 & -1 & 0 & 0 & $\pm\frac 12$ & 0 & \\
$c4$ & $1\times (1,1,2,1,1,\overline{2})$ & 0 & 0 & 1 & 0 & $\pm\frac 12$ & 0 & \\
$d1$ & $3\times (1,1,1,2,\overline{2},1)$ & 0 & 0 & 0 & 1 & $\pm\frac 12$ & 0 & \\
$d4$ & $1\times (1,1,1,\overline{2},1,2)$ & 0 & 0 & 0 & -1 & $\mp\frac 12$ & 0 & \\
$b_{\Ysymm}$ & $2\times(1,3,1,1,1,1)$ & 0 & 2 & 0 & 0  & $0,\;\pm 1$ & 0 & \\
$b_{\overline{\Yasymm}}$ & $2\times(1,\overline{1},1,1,1,1)$ & 0 & -2 & 0 & 0   & 0 & 0 & \\
$d_{\overline{\Ysymm}}$ & $2\times(1,1,1,\overline{3},1,1)$ & 0 & 0 & 0 & -2   & $0,\;\mp 1$ & 0 & \\
$d_{\Yasymm}$ & $2\times(1,1,1,1,1,1)$ & 0 & 0 & 0 & 2   & 0 & 0 & \\
	\hline\hline
$bc$ & $4\times (1,2,\overline{2},1,1,1)$ & 0 & 1 & -1 & 0   &  $0,\;\pm 1$ &0 &${H}_u^i, {H}_d^i$\\
 & $4\times (1,\overline{2},2,1,1,1)$ & 0 & -1 & 1 & 0   &  & &  \\
\hline
\end{tabular}
\end{center}
\end{table}

\begin{table}[htb]
\scriptsize
    \caption{The composite particle spectrum for Model 1}
	\label{composite model1}
	\begin{center}
		\begin{tabular}{|c|c||c|c|}
			\hline\multicolumn{2}{|c||}{\rm{Model} 1} & \multicolumn{2}{c|}{$U(4)\times U(2)_L\times U(2)_{R1}\times U(2)_{R2}\times USp(2)^2 $}\\
            \hline
            Confining Force & Intersection & Exotic particle Spectrum & Confined Particle Spectrum \\
			\hline \hline
            $USp(2)_1$ & b1 & $3\times(1,\overline{2},1,1,2,1)$ &
            $6\times(1,\overline{2}^2,1,1,1,1)$, $9\times(1,\overline{2},1,2,1,1)$,\\
             & d1 & $3\times(1,1,1,2,\overline{2},1)$ & $6\times(1,1,1,2^2,1,1)$\\
             \hline
            $USp(2)_4$ & c4 & $1\times(1,1,2,1,1,\overline{2})$ &
            $1\times(1,1,2^2,1,1,1)$, $1\times(1,1,2,\overline{2},1,1)$,\\
             & d4 & $1\times(1,1,1,\overline{2},1,2)$ & $1\times(1,1,1,\overline{2}^2,1,1)$\\
			\hline
		\end{tabular}
	\end{center}
\end{table}
Note that a composite particle spectrum also appears for Model 1 as shown in Table \ref{composite model1}.
The $USp(2)_1$ group has two charged intersections, giving rise to three exotic particles $(1,\overline{2},1,1,2,1)$ and three exotic particles $(1,1,1,2,\overline{2},1)$.
The mixing of these two types of exotic particles yields 9 chiral multiplets $(1,\overline{2},1,2,1,1)$.
Besides, their self-confinement results in 6 tensor representations $(1,\overline{2}^2,1,1,1,1)$ and 6 tensor representations $(1,1,1,2^2,1,1)$, which can be further decomposed so that
\beq
\begin{aligned}
    6\times(1,\overline{2}^2,1,1,1,1) &\to 6\times(1,\overline{3},1,1,1,1)+6\times(1,\overline{1},1,1,1,1)~,~\\
    6\times(1,1,1,2^2,1,1) &\to 6\times(1,1,1,3,1,1)+6\times(1,1,1,1,1,1)~,~
\end{aligned}
\eeq
leading to chiral multiplets $(1,\overline{3},1,1,1,1)$ and $(1,1,1,3,1,1)$.
Meanwhile, the $USp(2)_4$ group also has two charged intersections, which give rise to one exotic particle $(1,1,2,1,1,\overline{2})$ and one exotic particle $(1,1,1,\overline{2},1,2)$.
The mixing of these two types of exotic particles yields one chiral multiplet $(1,1,2,2,1,1)$.
Besides, their self-confinement results in a tensor representation $(1,1,2^2,1,1,1)$ and a tensor representation $(1,1,1,2^2,1,1)$, which can be further decomposed into
\beq
\begin{aligned}
    1\times(1,1,2^2,1,1,1) &\to 1\times(1,1,3,1,1,1)+1\times(1,1,1,1,1,1)~,~\\
    1\times(1,1,1,\overline{2}^2,1,1) &\to 1\times(1,1,1,\overline{3},1,1)+1\times(1,1,1,\overline{1},1,1)~,~
\end{aligned}
\eeq
leading to chiral multiplets $(1,1,3,1,1,1)$ and $(1,1,1,\overline{3},1,1)$.

Secondly, the chiral spectrum of Model 2 is shown in Table \ref{spectrum Model 2}, where a bifundamental Higgs doublet arises from the ${\cal N}=2$ subsector of the intersection of the $c$-stack and the $\Zbb_2$ image of the $d$-stack of D6-branes.
Besides, there are two $USp$ groups with negative $\beta$ functions from Table \ref{model2}, where two confining groups could lead to a stable minimum and dynamical supersymmetry breaking.

\begin{table}[htb]
\footnotesize
\renewcommand{\arraystretch}{1.0}
\caption{The chiral spectrum in the open string sector of Model 2}
\label{spectrum Model 2}
\begin{center}
\begin{tabular}{|c||c||c|c|c|c||c|c|c|}\hline
\scriptsize{Model 2} &\scriptsize{$SU(4)\times SU(2)_L\times SU(2)_{R1}$}
& \scriptsize{$Q_4$} & \scriptsize{$Q_{2L}$} & \scriptsize{$Q_{2R_1}$} & \scriptsize{$Q_{2R_2}$} & $Q_{em}$ & $B-L$ & Field \\
 &\scriptsize{$\times SU(2)_{R2}\times USp(2)^2$}
&   &   &   &   &   &   &   \\
\hline\hline
$ab$ & $1 \times (4,\overline{2},1,1,1,1)$ & $1$ & -$1$ & 0 & 0 &  $-\frac 13, \frac 23,-1, 0$ & $\frac 13,-1$ & $Q_L, L_L$\\
$ab'$ & $2 \times (4,2,1,1,1,1)$ & $1$ & $1$ & 0 & 0 &  $-\frac 13, \frac 23,-1, 0$ & $\frac 13,-1$ & $Q_L, L_L$\\
$ac$ & $1 \times (\overline{4},1,2,1,1,1)$ & -$1$ & 0 & $1$ & 0 &  $\frac 13, -\frac 23, 1, 0$ & $-\frac 13, 1$ & $Q_R, L_R$\\
$ad'$ & $2 \times (\overline{4},1,1,\overline{2},1,1)$ & -$1$ & 0 & 0 & -$1$  & $\frac 13, -\frac 23, 1, 0$ & $-\frac 13, 1$ & $Q_R, L_R$\\
$bc'$ & $2 \times(1,\overline{2},\overline{2},1,1,1)$ & 0 & -$1$ & -$1$ & 0  & $0,\;\mp 1$ & 0 & $H$\\
$bd$ & $3 \times(1,\overline{2},1,2,1,1)$ & 0 & -$1$ & 0 & $1$  & $0,\;\mp 1$ & 0 & $H$\\
$bd'$ & $6 \times(1,\overline{2},1,\overline{2},1,1)$ & 0 & -$1$ & 0 & -$1$   & $0,\;\mp1$ & 0 & $H'$\\
$cd$ & $1 \times(1,1,\overline{2},2,1,1)$ & 0 & 0 & -$1$ & $1$   & $0,\;\mp1$ & 0 & $\Phi$\\
$b2$ & $1\times (1,2,1,1,\overline{2},1)$ & 0 & 1 & 0 & 0 & $\pm\frac 12$ & 0 & \\
$c4$ & $1\times (1,1,2,1,1,\overline{2})$ & 0 & 0 & 1 & 0 & $\pm\frac 12$ & 0 & \\
$d2$ & $1\times (1,1,1,2,\overline{2},1)$ & 0 & 0 & 0 & 1 & $\pm\frac 12$ & 0 & \\
$d4$ & $3\times (1,1,1,\overline{2},1,2)$ & 0 & 0 & 0 & -1 & $\mp\frac 12$ & 0 & \\
$b_{\Ysymm}$ & $2\times(1,3,1,1,1,1)$ & 0 & 2 & 0 & 0  & $0,\;\pm 1$ & 0 & \\
$b_{\overline{\Yasymm}}$ & $2\times(1,\overline{1},1,1,1,1)$ & 0 & -2 & 0 & 0   & 0 & 0 & \\
$d_{\Ysymm}$ & $2\times(1,1,1,3,1,1)$ & 0 & 0 & 0 & 2   & $0,\;\pm 1$ & 0 & \\
$d_{\overline{\Yasymm}}$ & $2\times(1,1,1,\overline{1},1,1)$ & 0 & 0 & 0 & -2   & 0 & 0 & \\
	\hline\hline
$cd'$ & $1\times (1,1,2,2,1,1)$ & 0 & 0 & 1 & 1   &  $0,\;\pm 1$ &0 &${\Phi}_u^i, {\Phi}_d^i$\\
& $1\times (1,1,\overline{2},\overline{2},1,1)$ & 0 & 0 & -1 & -1  &  & &  \\
\hline
\end{tabular}
\end{center}
\end{table}

\begin{table}[htb]
\scriptsize
    \caption{The composite particle spectrum for Model 2}
	\label{composite model2}
	\begin{center}
		\begin{tabular}{|c|c||c|c|}
			\hline\multicolumn{2}{|c||}{\rm{Model} 2} & \multicolumn{2}{c|}{$U(4)\times U(2)_L\times U(2)_{R1}\times U(2)_{R2} \times USp(2)^2 $}\\
            \hline
            Confining Force & Intersection & Exotic particle Spectrum & Confined Particle Spectrum \\
			\hline \hline
            $USp(2)_2$ & b2 & $1\times(1,2,1,1,\overline{2},1)$ &
            $1\times(1,2^2,1,1,1,1)$, $1\times(1,2,1,2,1,1)$,\\
             & d2 & $1\times(1,1,1,2,\overline{2},1)$ & $1\times(1,1,1,2^2,1,1)$\\
             \hline
            $USp(2)_4$ & c4 & $1\times(1,1,2,1,1,\overline{2})$ &
            $1\times(1,1,2^2,1,1,1)$, $3\times(1,1,2,\overline{2},1,1)$,\\
             & d4 & $3\times(1,1,1,\overline{2},1,2)$ & $6\times(1,1,1,\overline{2}^2,1,1)$\\
			\hline
		\end{tabular}
	\end{center}
\end{table}

The composite particle spectrum for Model 2 is given in Table \ref{composite model2}.
The $USp(2)_2$ group has two charged intersections, which give rise to an exotic particle $(1,2,1,1,\overline{2},1)$ and an exotic particle $(1,1,1,2,\overline{2},1)$.
The mixing of these two types of exotic particles yields a chiral multiplet $(1,2,1,2,1,1)$.
Besides, their self-confinement results in tensor representations $(1,2^2,1,1,1,1)$ and tensor representations $(1,1,1,2^2,1,1)$, which can be further decomposed
\beq
\begin{aligned}
    1\times(1,2^2,1,1,1,1) &\to 1\times(1,3,1,1,1,1)+1\times(1,1,1,1,1,1)~,~\\
    1\times(1,1,1,2^2,1,1) &\to 1\times(1,1,1,3,1,1)+1\times(1,1,1,1,1,1)~,~
\end{aligned}
\eeq
leading to chiral multipets $(1,3,1,1,1,1)$ and $(1,1,1,3,1,1)$.
Meanwhile, the $USp(2)_4$ group also has two charged intersections, which give rise to one exotic particle $(1,1,2,1,1,\overline{2})$ and three exotic particles $(1,1,1,\overline{2},1,2)$.
The mixing of these two types of exotic particles yields 3 chiral multiplets $(1,1,2,\overline{2},1,1)$.
Besides, their self-confinement results in a tensor representation $(1,1,2^2,1,1,1)$ and 6 tensor representation $(1,1,1,\overline{2}^2,1,1)$, which can be further decomposed,
\beq
\begin{aligned}
    1\times(1,1,2^2,1,1,1) &\to 1\times(1,1,3,1,1,1)+1\times(1,1,1,1,1,1)~,~\\
    6\times(1,1,1,\overline{2}^2,1,1) &\to 6\times(1,1,1,\overline{3},1,1)+6\times(1,1,1,\overline{1},1,1)~,~
\end{aligned}
\eeq
leading to chiral multipets $(1,1,3,1,1,1)$ and $(1,1,1,\overline{3},1,1)$.

\begin{table}[htb]
\footnotesize
\renewcommand{\arraystretch}{1.0}
\caption{The chiral spectrum in the open string sector of Model 3}
\label{spectrum Model 3}
\begin{center}
\begin{tabular}{|c||c||c|c|c|c||c|c|c|}\hline
Model 3 &\scriptsize{$SU(4)\times SU(2)_L\times SU(2)_{R1}\times SU(2)_{R2}$}
& $Q_4$ & $Q_{2L}$ & $Q_{2R_1}$ & $Q_{2R_2}$ & $Q_{em}$ & $B-L$ & Field \\
\hline\hline
$ab$ & $2 \times (4,\overline{2},1,1)$ & $1$ & -$1$ & 0 & 0 &  $-\frac 13,\; \frac 23,\;-1,\; 0$ & $\frac 13,\;-1$ & $Q_L, L_L$\\
$ab'$ & $1 \times (4,2,1,1)$ & $1$ & $1$ & 0 & 0 &  $-\frac 13,\; \frac 23,\;-1,\; 0$ & $\frac 13,\;-1$ & $Q_L, L_L$\\
$ac$ & $1 \times (\overline{4},1,2,1)$ & -$1$ & 0 & $1$ & 0 &  $\frac 13,\; -\frac 23,\; 1,\; 0$ & $-\frac 13,\; 1$ & $Q_R, L_R$\\
$ad$ & $2 \times (\overline{4},1,1,2)$ & -$1$ & 0 & 0 & $1$  & $\frac 13,\; -\frac 23,\; 1,\; 0$ & $-\frac 13,\; 1$ & $Q_R, L_R$\\
$bc$ & $4 \times(1,2,\overline{2},1)$ & 0 & $1$ & -$1$ & 0  & $0,\;\pm 1$ & 0 & $H$\\
$bd$ & $6 \times(1,2,1,\overline{2})$ & 0 & $1$ & 0 & -$1$  & $0,\;\pm 1$ & 0 & $H$\\
$bd'$ & $12 \times(1,2,1,2)$ & 0 & $1$ & 0 & $1$   & $0,\;\mp1$ & 0 & $H'$\\
$cd'$ & $2 \times(1,1,\overline{2},\overline{2})$ & 0 & 0 & -$1$ & -$1$   & $0,\;\mp1$ & 0 & $\Phi$\\
$b_{\overline{\Ysymm}}$ & $5\times(1,\overline{3},1,1)$ & 0 & -2 & 0 & 0  & $0,\;\mp 1$ & 0 & \\
$b_{\Yasymm}$ & $5\times(1,1,1,1)$ & 0 & 2 & 0 & 0   & 0 & 0 & \\
$c_{\Ysymm}$ & $1\times(1,1,3,1)$ & 0 & 0 & 2 & 0   & $0,\;\pm 1$ & 0 & \\
$c_{\overline{\Yasymm}}$ & $1\times(1,1,\overline{1},1)$ & 0 & 0 & -2 & 0   & 0 & 0 & \\
$d_{\overline{\Ysymm}}$ & $2\times(1,1,1,{\overline{3}})$ & 0 & 0 & 0 & -2   & $0,\;\mp 1$ & 0 & \\
$d_{\Yasymm}$ & $2\times(1,1,1,1)$ & 0 & 0 & 0 & 2   & 0 & 0 & \\
	\hline\hline
$cd$ & $2\times (1,1,2,\overline{2})$ & 0 & 0 & 1 & -1   &  $0,\;\pm 1$ &0 &${\Phi}_u^i, {\Phi}_d^i$\\
& $2\times (1,1,\overline{2},2)$ & 0 & 0 & -1 & 1  &  & &  \\
\hline
\end{tabular}
\end{center}
\end{table}

Thirdly, the chiral spectra of Model 3 and 4 are shown in Table \ref{spectrum Model 3} and \ref{spectrum Model 4}, respectively. 
Different from the standard Pati-Salam models found in~\cite{He:2021gug}, there is no $USp$ group in Model 3 and 4.
In Model 3, there are 2 bifundamental Higgs doublets arising from the ${\cal N}=2$ subsector of the intersection of the $c$- and $d$-stacks of the D6-branes.
In Model 4, there are 8 Higgs doublets arising from the ${\cal N}=2$ subsector of the intersection of the $b$- and $c$-stacks of the D6-branes.

\begin{table}[htb]
\footnotesize
\renewcommand{\arraystretch}{1.0}
\caption{The chiral spectrum in the open string sector of Model 4}
\label{spectrum Model 4}
\begin{center}
\begin{tabular}{|c||c||c|c|c|c||c|c|c|}\hline
Model 4 &\scriptsize{$SU(4)\times SU(2)_L\times SU(2)_{R1}\times SU(2)_{R2}$}
& $Q_4$ & $Q_{2L}$ & $Q_{2R_1}$ & $Q_{2R_2}$ & $Q_{em}$ & $B-L$ & Field \\
\hline\hline
$ab$ & $3 \times (4,\overline{2},1,1)$ & 1 & -$1$ & 0 & 0 &  $-\frac 13,\; \frac 23,\;-1,\; 0$ & $\frac 13,\;-1$ & $Q_L, L_L$\\
$ac$ & $1 \times (\overline{4},1,2,1)$ & -$1$ & 0 & $1$ & 0 &  $\frac 13,\; -\frac 23,\; 1,\; 0$ & $-\frac 13,\; 1$ & $Q_R, L_R$\\
$ad'$ & $2 \times (\overline{4},1,1,\overline{2})$ & -$1$ & 0 & 0 &  -$1$  & $\frac 13,\; -\frac 23,\; 1,\; 0$ & $-\frac 13,\; 1$ & $Q_R, L_R$\\
$bc'$ & $4 \times(1,\overline{2},\overline{2},1)$ & 0 & -$1$ & -$1$ & 0  & $0,\;\pm 1$ & 0 & $H'$\\
$bd$ & $2 \times(1,2,1,\overline{2})$ & 0 & $1$ & 0 & -$1$  & $0,\;\pm 1$ & 0 & $H'$\\
$bd'$ & $4 \times(1,2,1,2)$ & 0 & $1$ & 0 & $1$   & $0,\;\mp1$ & 0 & $H$\\
$cd$ & $2 \times(1,1,2,\overline{2})$ & 0 & 0 & $1$ & -$1$   & $0,\;\pm1$ & 0 & $\Phi$\\
$cd'$ & $4 \times(1,1,\overline{2},\overline{2})$ & 0 & 0 & -$1$ & -$1$   & $0,\;\mp1$ & 0 & $\Phi$\\
$b_{\Ysymm}$ & $1\times(1,3,1,1)$ & 0 & 2 & 0 & 0  & $0,\;\pm 1$ & 0 & \\
$b_{\overline{\Yasymm}}$ & $1\times(1,\overline{1},1,1)$ & 0 & -2 & 0 & 0   & 0 & 0 & \\
$c_{\Ysymm}$ & $1\times(1,1,{3},1)$ & 0 & 0 & 2 & 0   & $0,\;\pm 1$ & 0 & \\
$c_{\overline{\Yasymm}}$ & $1\times(1,1,\overline{1},1)$ & 0 & 0 & -2 & 0   & 0 & 0 & \\
$d_{\Ysymm}$ & $2\times(1,1,1,3)$ & 0 & 0 & 0 & 2   & $0,\;\pm 1$ & 0 & \\
$d_{\overline{\Yasymm}}$ & $2\times(1,1,\overline{1},1)$ & 0 & 0 & 0 & -2   & 0 & 0 & \\
	\hline\hline
$bc$ & $8\times (1,2,\overline{2},1)$ & 0 & 1 & -1 & 0   &  $0,\;\pm 1$ &0 &${H}_u^i, {H}_d^i$\\
& $8\times (1,\overline{2},2,1)$ & 0 & -1 & 1 & 0   &  & &  \\
\hline
\end{tabular}
\end{center}
\end{table}

These four classes of extended Pati-Salam models are constructed with two confining groups (see Model 1 and 2) and without confining group (see Model 3 and 4).
The supersymmetry breaking can be realized via gaugino condensation when there are four confining gauge groups in the hidden sector, which means that all $\beta$ functions are negative.
However, as discussed in~\cite{Cvetic:2003yd,Brustein:1992nk,Taylor:1990wr}, there should be at least two confining $USp$ groups with negative $\beta$ functions for gaugino condensation in order to stabilize the complex structure moduli.
For this reason, among these four classes of models, models in the same class of Model 2 with two negative $\beta$ functions are the most likely to break the supersymmetry via gaugino condensation
\footnote{Note that only when the moduli stabilization is achieved at the stable extremum supersymmetry breaking appears, while at the saddle points, such supersymmetry breaking cannot be staggered that alternative mechanism need to be searched for.}.

\begin{table}[htb]
\footnotesize
\renewcommand{\arraystretch}{1.0}
\caption{The chiral spectrum in the open string sector of Model 10 from~\cite{He:2021gug}}
\label{spectrum Comp M10}
\begin{center}
\begin{tabular}{|c||c||c|c|c|c||c|c|}\hline
Model 10~\cite{He:2021gug} &\scriptsize{$SU(4)\times SU(2)_L\times SU(2)_{R}\times USp(2)^3 $}
& $Q_4$ & $Q_{2L}$ & $Q_{2R}$ & $Q_{em}$ & $B-L$ & Field \\
\hline\hline
$ab$ & $3 \times (4,\overline{2},1,1,1,1)$ & 1 & -1 & 0 &  $-\frac 13,\; \frac 23,\;-1,\; 0$ & $\frac 13,\;-1$ & $Q_L, L_L$\\
$ac'$ & $3 \times (\overline{4},1,\overline{2},1,1,1)$ & -1 & 0 & -1 &  $\frac 13,\; -\frac 23,\;1,\; 0$ & $-\frac 13,\;1$ & $Q_R, L_R$\\
$bc$ & $3 \times (1,2,\overline{2},1,1,1)$ & 0 & 1 & -1 & $0,\pm 1$ & 0 & $H'$\\
$a2$ & $1 \times (4,1,1,1,\overline{2},1)$ & 1 & 0 & 0 & $\frac16,-\frac12$ & $\frac13,-1$ & \\
$a4$ & $1 \times (\overline{4},1,1,1,1,2)$ & -1 & 0 & 0 & $-\frac16,\frac12$ & $-\frac13,1$ & \\
$b3$ & $3 \times (1,\overline{2},1,1,2,1)$ & 0 & -1 & 0 & $\mp\frac12$ & 0 & \\
$b4$ & $2 \times (1,2,1,1,1,\overline{2})$ & 0 & 1 & 0 & $\pm\frac12$ & 0 & \\
$c2$ & $1 \times (1,1,2,\overline{2},1,1)$ & 0 & 0 & 1 & $\pm\frac12$ & 0 & \\
$b_{\Ysymm}$ & $1 \times (1,3,1,1,1,1)$ & 0 & 2 & 0 & $0,\pm 1$ & 0 & \\
$b_{\overline{\Yasymm}}$ & $1 \times (1,\overline{1},1,1,1,1)$ & 0 & -2 & 0 & $0$ & 0 & \\
$c_{\Ysymm}$ & $2 \times (1,1,3,1,1,1)$ & 0 & 0 & 2 & $0,\pm 1$ & 0 & \\
$c_{\overline{\Yasymm}}$ & $2 \times (1,1,\overline{1},1,1,1)$ & 0 & 0 & -2 & $0$ & 0 & \\
\hline\hline
$bc'$ & $9 \times (1,2,2,1,1,1)$ & 0 & 1 & 1 & $0,\pm 1$ & 0 & $H_u^i,H_d^i$ \\
 & $9\times (1,\overline{2},\overline{2},1,1,1)$ & 0 & -1 & -1 & & & \\
\hline
\end{tabular}
\end{center}
\end{table}

Recall that Model 4 in this work and Model 10 in~\cite{He:2021gug} share the same gauge coupling ratio $g_a^2:g_b^2:g_R^2:(\frac{5}{3}g_Y^2)$  with different $g_a$ value and $\gamma$ in (\ref{gabY}). It is obvious that with an extra 
$d$-stack of D6-branes, the number of filler branes turns to be ``replaced'' by the additional stack of branes that carry $SU(2)$ gauge symmetry. 
From the spectra shown in Table \ref{spectrum Model 4} and \ref{spectrum Comp M10}, it is obvious that with an extra $d$-stack of brane introduced, the Higgs multiplets have two sources:  
$bd'$-brane and $bc$-brane intersections, although without extra $d$-stack of branes there is only one source for Higgs doublets to appear from $bc'$-brane intersection.

The chiral spectrum of Model 10 from~\cite{He:2021gug} is shown in Table \ref{spectrum Comp M10}.
This spectrum is quite different from the chiral spectrum of Model 4 with $d$-stack of $SU(2)$ gauge, although they have the same gauge coupling ration.
In Model 10 from~\cite{He:2021gug}, there are 9 Higgs doublets arising from the ${\cal N}=2$ subsector of the intersection of the $b$-stack and the $\Zbb_2$ image of $c$-stack of D6-branes only, while with $d$-stack, Model 4 has 8 Higgs multiplets from  $bc$-brane intersections, and 4 Higgs multiplets from $bd'$-brane  intersections.
Besides, there are three $USp$ groups in Model 10 from~\cite{He:2021gug}, yielding charged exotic particles by intersecting with branes in the visible sector, while there is no $USp$ group in Model 4 with $d$-stack of $SU(2)$ gauge branes.

From the perspective that they share the same gauge coupling ratio (and therefore behave similarly in the RGE rolling), it indicates that by introducing an extra stack of D6-branes carrying $SU(2)$ gauge will provide more possibilities/origins  for Higgs particles.

\subsection{$SU(2)_L$ gauge from extra symmetry breaking}
\label{secL}

Secondly, note that different with the 3-stack brane construction in\cite{Li:2019miw}, with $d$-stack of D6-branes introduced, non-trivial interchange appears to the three-family chiral fermion constraints appears and gauge coupling relations are different as well.

\subsubsection{$SU(2)_L$ gauge from $b/d$-stack of D6-branes}

Analogously to the configuration with an extra $d$-stack of D6-branes introduced to the right-handed $SU(2)_R$,  the extra $d$-stack of D6-branes can also be introduced to the left-handed $SU(2)_L$ symmetry, with $SU(2)_{L_1} \times SU(2)_{L_2} \rightarrow SU(2)_L$. 
The $SU(2)_L$ gauge coupling $g_L^{-2}$ is then realized through the symmetry breaking of $SU(2)_{L_1}$ and $SU(2)_{L_2}$ which can be written in terms of $g_b^{-2}$ and $g_d^{-2}$ as
\beq
g_L^{-2}= g_b^{-2}+g_d^{-2}=|\mathrm{Re}(f_b)|+|\mathrm{Re}(f_d)|~.~
\eeq
The gauge kinetic function related to $SU(2)_L$ is then written in
\beq
f_L=f_b+f_d~,~
\eeq
while the kinetic function of $U(1)_Y$ gauge symmetry remains the same as that in~\cite{He:2021gug}, simply
\beq
f_Y =\frac35(\frac23 f_a+f_c)~.~
\eeq
The corresponding gauge coupling relations are read as follows
\beq
g_a^2=\alpha g_L^2=\beta \frac{5}{3}g_Y^2=\gamma[\pi e^{\phi_4}]~,~
\label{gabYL}
\eeq
with $g_L$ replacing $g_b$ as the new weak gauge coupling.

The three-family of chiral fermions is then obtained through (\ref{3g})  as
\beq
I_{ab}+I_{ab'}+I_{ad}+I_{ad'}=3 ~,~
I_{ac}+I_{ac'}=-3~.~
\label{3gL}
\eeq
  
In this configuration, we found there are 14 classes of independent models with left-handed extension as shown in the Appendix.
Here in this section, we discuss on four independent representative  Models 7-10 (Table \ref{model7}-\ref{model10}).
In particular, the tadpole cancellation conditions in Model 9 and 10 are  satisfied without a filler brane as in Model 3 and 4 in the right-handed extension.

In Model 10, the ratio of gauge couplings $g_a^2:g_b^2:g_R^2:(\frac{5}{3}g_Y^2)$ is the same as that of Model 18 in the previous complete search for supersymmetric three-family Pati-Salam models with only $a$-, $b$- and $c$-stacks~\cite{He:2021gug}.
The difference in gauge couplings is that with extra $d$-stack of $SU(2)_L$ we have $g_a^2=4\times3^{1/4} \pi e^{\phi_4}$ while in ~\cite{He:2021gug} $g_a^2=\frac{16\times2^{1/4}}{3\sqrt{3}} \pi e^{\phi_4}$. Again, these two different values of $\gamma$ in (\ref{gabYL}) are supposed to be related by the scaling of dilaton.

Moreover, the gauge coupling relations of  Model 1 and 7, Model 2 and 8, Model 3 and 9, and Model 4 and 10, respectively, share the same values of $\gamma$ from (\ref{gabY}) and (\ref{gabYL}), which shows a left-right symmetry in Pati-Salam models.

\begin{table}[h!]
\scriptsize
    \caption{D6-brane configurations and intersection numbers of Model 7, and its MSSM gauge coupling relation is
    $g^2_a=\frac{1}{2} g^2_L=g^2_c=(\frac{5}{3} g^2_Y)=2\sqrt{2}\ 3^{1/4} \pi  e^{\phi_4}$,
    for which the first torus is tilted.}
	\label{model7}
	\begin{center}
		\begin{tabular}{|c||c|c||c|c|c|c|c|c|c|c|c|c|}
			\hline\rm{Model} 7 & \multicolumn{12}{c|}{$U(4)\times U(2)_{L_1}\times U(2)_R\times U(2)_{L_2}\times USp(2)^2 $}\\
			\hline \hline			\rm{stack} & $N$ & $(n^1,l^1)\times(n^2,l^2)\times(n^3,l^3)$ & $n_{\Ysymm}$& $n_{\Yasymm}$ & $b$ & $b'$ & $c$ & $c'$ &$d$ &$d'$ & 2 & 4\\
			\hline
			$a$ & 8 & $(1,-1)\times (0,-1)\times (-1,1)$ & 0 & 0  & 1 & 0 & -3 & 0 & 0 & 2 & 0 & 0\\
			$b$ & 4 & $(-1,-1)\times (-1,-1)\times (0,-1)$ & 0 & 0  & - & - & 0 & 2 & -1 & 2 & 0 & -1\\
			$c$ & 4 & $(-1,-1)\times (-3,1)\times (1,0)$ & -2 & 2 & - & - & - & - & -1 & -2 & 3 & 0\\
            $d$ & 4 & $(1,3)\times (1,0)\times (1,-1)$ & -2 & 2  & - & - & - & - & - & - & 3 & -1\\
			\hline
			2 & 2 & $(2, 0)\times (0, -1)\times (0, 1)$& \multicolumn{10}{c|}{$x_A = 3 x_B = x_C=x_D$}\\
            4 & 2 & $(0, -2)\times (0, 1)\times (1, 0)$& \multicolumn{10}{c|}{$\beta^g_2=0$, \quad\quad$\beta^g_4=-4$}\\
			& & & \multicolumn{10}{c|}{$\chi_1=\frac{2}{\sqrt{3}}$,  $\chi_2=\sqrt{3}$, $\chi_3=\sqrt{3}$ }\\
			\hline
		\end{tabular}
	\end{center}
\end{table}

\begin{table}[h!]
\scriptsize
    \caption{D6-brane configurations and intersection numbers of Model 8, and its MSSM gauge coupling relation is
    $g^2_a=\frac{3}{4}g^2_L= 3 g^2_c=\frac{5}{3} (\frac{5}{3} g^2_Y)=2\sqrt{2}\ 3^{3/4} \pi  e^{\phi_4}$,
    for which the first torus is tilted.}
	\label{model8}
	\begin{center}
		\begin{tabular}{|c||c|c||c|c|c|c|c|c|c|c|c|c|}
			\hline\rm{Model} 8 & \multicolumn{12}{c|}{$U(4)\times U(2)_{L_1}\times U(2)_R\times U(2)_{L_2}\times USp(2)^2 $}\\
			\hline \hline			\rm{stack} & $N$ & $(n^1,l^1)\times(n^2,l^2)\times(n^3,l^3)$ & $n_{\Ysymm}$& $n_{\Yasymm}$ & $b$ & $b'$ & $c$ & $c'$ &$d$ &$d'$ & 2 & 4\\
			\hline
			$a$ & 8 & $(1,-1)\times (0,-1)\times (-1,1)$ & 0 & 0  & 1 & 0 & -1 & -2 & 2 & 0 & 0 & 0\\
			$b$ & 4 & $(-1,-1)\times (-1,-1)\times (0,-1)$ & 0 & 0  & - & - & 0 & 2 & 0 & 1 & 0 & -1\\
			$c$ & 4 & $(-3,1)\times (-1,-1)\times (1,0)$ & -2 & 2 & - & - & - & - & 6 & 3 & -1 & 0\\
            $d$ & 4 & $(1,1)\times (1,0)\times (1,-3)$ & 2 & -2  & - & - & - & - & - & - & 1 & -3\\
			\hline
			2 & 2 & $(2, 0)\times (0, -1)\times (0, 1)$& \multicolumn{10}{c|}{$x_A = \frac{1}{3} x_B = x_C= x_D$}\\
            4 & 2 & $(0, -2)\times (0, 1)\times (1, 0)$& \multicolumn{10}{c|}{$\beta^g_2=-4$, \quad$\beta^g_4=-2$}\\
			& & & \multicolumn{10}{c|}{$\chi_1=2\sqrt{3}$,  $\chi_2=\frac{1}{\sqrt{3}}$, $\chi_3=\frac{1}{\sqrt{3}}$ }\\
			\hline
		\end{tabular}
	\end{center}
\end{table}

\begin{table}[h!]
\scriptsize
    \caption{D6-brane configurations and intersection numbers of Model 9, and its MSSM gauge coupling relation is
    $g^2_a=\frac{3}{2}g^2_L=6 g^2_c=2 (\frac{5}{3} g^2_Y)=4\times3^{3/4} \pi  e^{\phi_4}$,
    for which the first torus is tilted.}
	\label{model9}
	\begin{center}
		\begin{tabular}{|c||c|c||c|c|c|c|c|c|c|c|}
			\hline\rm{Model} 9 & \multicolumn{10}{c|}{$U(4)\times U(2)_{L_1}\times U(2)_R\times U(2)_{L_2} $}\\
			\hline \hline			\rm{stack} & $N$ & $(n^1,l^1)\times(n^2,l^2)\times(n^3,l^3)$ & $n_{\Ysymm}$& $n_{\Yasymm}$ & $b$ & $b'$ & $c$ & $c'$ &$d$ &$d'$ \\
			\hline
			$a$ & 8 & $(1,-1)\times (0,-1)\times (-1,1)$ & 0 & 0  & 1 & 0 & -2 & -1 & 2 & 0\\
			$b$ & 4 & $(-1,-1)\times (-1,-2)\times (0,-1)$ & -1 & 1  & - & - & 4 & 0 & 0 & 2\\
			$c$ & 4 & $(-3,-1)\times (-1,2)\times (1,0)$ & 5 & -5 & - & - & - & - & -6 & -12\\
            $d$ & 4 & $(1,1)\times (1,0)\times (1,-3)$ & 2 & -2  & - & - & - & - & - & -\\
			\hline
			   &  &  & \multicolumn{8}{c|}{$x_A = \frac{1}{6} x_B = x_C= \frac{1}{2}x_D$}\\
			& & & \multicolumn{8}{c|}{$\chi_1=2\sqrt{3}$,  $\chi_2=\frac{1}{2\sqrt{3}}$, $\chi_3=\frac{1}{\sqrt{3}}$ }\\
			\hline
		\end{tabular}
	\end{center}
\end{table}

\begin{table}[h!]
\scriptsize
    \caption{D6-brane configurations and intersection numbers of Model 10, and its MSSM gauge coupling relation is
    $g^2_a=g^2_L=2g^2_c=\frac{10}{7}(\frac{5}{3} g^2_Y)=4\times 3^{1/4} \pi  e^{\phi_4}$,
    for which the first torus is tilted.}
	\label{model10}
	\begin{center}
		\begin{tabular}{|c||c|c||c|c|c|c|c|c|c|c|}
			\hline\rm{Model} 10 & \multicolumn{10}{c|}{$U(4)\times U(2)_{L_1}\times U(2)_R\times U(2)_{L_2}  $}\\
			\hline \hline			\rm{stack} & $N$ & $(n^1,l^1)\times(n^2,l^2)\times(n^3,l^3)$ & $n_{\Ysymm}$& $n_{\Yasymm}$ & $b$ & $b'$ & $c$ & $c'$ &$d$ &$d'$ \\
			\hline
			$a$ & 8 & $(1,-1)\times (0,-1)\times (-1,1)$ & 0 & 0  & 1 & 0 & -3 & 0 & 0 & 2\\
			$b$ & 4 & $(-1,-1)\times (-1,-2)\times (0,-1)$ & -1 & 1  & - & - & 0 & 4 & -2 & 4\\
			$c$ & 4 & $(-1,-1)\times (-3,2)\times (1,0)$ & -1 & 1 & - & - & - & - & -2 & -4\\
            $d$ & 4 & $(1,3)\times (1,0)\times (1,-1)$ & -2 & 2  & - & - & - & - & - & -\\
			\hline
			   &  &  & \multicolumn{8}{c|}{$x_A = \frac{3}{2} x_B = x_C= \frac{1}{2}x_D$}\\
			& & & \multicolumn{8}{c|}{$\chi_1=\frac{2}{\sqrt{3}}$,  $\chi_2=\frac{\sqrt{3}}{2}$, $\chi_3=\sqrt{3}$ }\\
			\hline
		\end{tabular}
	\end{center}
\end{table}

\subsubsection{Phenomenological studies}

The particle spectra of Models 7-10, are shown in Table \ref{spectrum Model 7}, \ref{spectrum Model 8}, \ref{spectrum Model 9}, and \ref{spectrum Model 10} respectively.
From the spectrum of Model 7, 4 Higgs doublets arise from the ${\cal N}=2$ subsector of the intersection of $b$- and $c$-stacks of D6-branes.
Besides, there is one $USp$ group with negative $\beta$ function and one $USp(2)$ group with $\beta$ function to be zero from Table \ref{model7}.
This $\beta$ with zero value might result in difficulty in dynamical supersymmetry breaking.

\begin{table}[htb]
\footnotesize
\renewcommand{\arraystretch}{1.0}
\caption{The chiral spectrum in the open string sector of Model 7}
\label{spectrum Model 7}
\begin{center}
\begin{tabular}{|c||c||c|c|c|c||c|c|c|}\hline
\scriptsize{Model 7} &\scriptsize{$SU(4)\times SU(2)_{L1}\times SU(2)_{R}$}
& \scriptsize{$Q_4$} & \scriptsize{$Q_{2L_1}$} & \scriptsize{$Q_{2R}$} & \scriptsize{$Q_{2L_2}$} & $Q_{em}$ & $B-L$ & Field \\
 &\scriptsize{$\times SU(2)_{L2}\times USp(2)^2$}
&   &   &   &   &   &   &   \\
\hline\hline
$ab$ & $1 \times (4,\overline{2},1,1,1,1)$ & $1$ & -$1$ & 0 & 0 &  $-\frac 13, \frac 23,-1, 0$ & $\frac 13,-1$ & $Q_L, L_L$\\
$ac$ & $3 \times (\overline{4},1,2,1,1,1)$ & -$1$ & 0 & $1$ & 0 &  $\frac 13, -\frac 23, 1, 0$ & $-\frac 13, 1$ & $Q_R, L_R$\\
$ad'$ & $2 \times (4,1,1,2,1,1)$ & $1$ & 0 & 0 & $1$  & $-\frac 13, \frac 23, -1, 0$ & $\frac 13, -1$ & $Q_L, L_L$\\
$bc'$ & $2 \times(1,2,2,1,1,1)$ & 0 & $1$ & $1$ & 0  & $0,\;\pm 1$ & 0 & $H'$\\
$bd$ & $1 \times(1,\overline{2},1,2,1,1)$ & 0 & -$1$ & 0 & $1$  & $0,\;\mp 1$ & 0 & $\Phi$\\
$bd'$ & $2 \times(1,2,1,2,1,1)$ & 0 & $1$ & 0 & $1$   & $0,\;\pm 1$ & 0 & $\Phi$\\
$cd$ & $1 \times(1,1,\overline{2},2,1,1)$ & 0 & 0 & -$1$ & $1$   & $0,\;\mp1$ & 0 & $H'$\\
$cd'$ & $2 \times(1,1,\overline{2},\overline{2},1,1)$ & 0 & 0 & -$1$ & -$1$   & $0,\;\mp 1$ & 0 & $H$\\
$b4$ & $1\times (1,\overline{2},1,1,1,2)$ & 0 & -1 & 0 & 0 & $\mp\frac 12$ & 0 & \\
$c2$ & $3\times (1,1,2,1,\overline{2},1)$ & 0 & 0 & 1 & 0 & $\pm\frac 12$ & 0 & \\
$d2$ & $3\times (1,1,1,2,\overline{2},1)$ & 0 & 0 & 0 & 1 & $\pm\frac 12$ & 0 & \\
$d4$ & $1\times (1,1,1,\overline{2},1,2)$ & 0 & 0 & 0 & -1 & $\mp\frac 12$ & 0 & \\
$c_{\overline{\Ysymm}}$ & $2\times(1,1,\overline{3},1,1,1)$ & 0 & 0 & -2 & 0  & $0,\;\mp 1$ & 0 & \\
$c_{\Yasymm}$ & $2\times(1,1,1,1,1,1)$ & 0 & 0 & 2 & 0   & 0 & 0 & \\
$d_{\overline{\Ysymm}}$ & $2\times(1,1,1,\overline{3},1,1)$ & 0 & 0 & 0 & -2   & $0,\;\mp 1$ & 0 & \\
$d_{\Yasymm}$ & $2\times(1,1,1,1,1,1)$ & 0 & 0 & 0 & 2   & 0 & 0 & \\
	\hline\hline
$bc$ & $4\times (1,2,\overline{2},1,1,1)$ & 0 & 1 & -1 & 0   &  $0,\;\pm 1$ &0 &${H}_u^i, {H}_d^i$\\
 & $4\times (1,\overline{2},2,1,,11)$ & 0 & -1 & 1 & 0   &  & &  \\
\hline
\end{tabular}
\end{center}
\end{table}

\begin{table}[htb]
\scriptsize
    \caption{The composite particle spectrum for Model 7}
	\label{composite model7}
	\begin{center}
		\begin{tabular}{|c|c||c|c|}
			\hline\multicolumn{2}{|c||}{\rm{Model} 7} & \multicolumn{2}{c|}{$U(4)\times U(2)_{L1}\times U(2)_{R}\times U(2)_{L2} \times USp(2)^2 $}\\
            \hline
            Confining Force & Intersection & Exotic particle Spectrum & Confined Particle Spectrum \\
			\hline \hline
            $USp(2)_2$ & c2 & $3\times(1,1,2,1,\overline{2},1)$ &
            $6\times(1,1,2^2,1,1,1)$, $9\times(1,1,2,2,1,1)$,\\
             & d2 & $3\times(1,1,1,2,\overline{2},1)$ & $6\times(1,1,1,2^2,1,1)$\\
             \hline
            $USp(2)_4$ & b4 & $1\times(1,\overline{2},1,1,1,2)$ &
            $1\times(1,\overline{2}^2,1,1,1,1)$, $1\times(1,\overline{2},1,\overline{2},1,1)$,\\
             & d4 & $1\times(1,1,1,\overline{2},1,2)$ & $1\times(1,1,1,\overline{2}^2,1,1)$\\
			\hline
		\end{tabular}
	\end{center}
\end{table}

The composite particle spectrum for Model 7 is given in Table \ref{composite model7}.
The $USp(2)_2$ group has two charged intersections, which give rise to three exotic particles with quantum number $(1,1,2,1,\overline{2},1)$ and three exotic particles with quantum number $(1,1,1,2,\overline{2},1)$.
The mixing of these two kinds of exotic particles yields 9 chiral multiplets $(1,1,2,2,1,1)$.
Their self-confinement results in 6 tensor representations $(1,2^2,1,1,1,1)$ and 6 tensor representations $(1,1,1,2^2,1,1)$, which can be further decomposed into
\beq
\begin{aligned}
    6\times(1,1,2^2,1,1,1) &\to 6\times(1,1,3,1,1,1)+6\times(1,1,1,1,1,1)~,~\\
    6\times(1,1,1,2^2,1,1) &\to 6\times(1,1,1,3,1,1)+6\times(1,1,1,1,1,1)~,~
\end{aligned}
\eeq
leading to chiral multiplets $(1,1,3,1,1,1)$ and $(1,1,1,3,1,1)$.
Meanwhile, the $USp(2)_4$ group also has two charged intersections, which give rise to one exotic particle $(1,\overline{2},1,1,1,2)$ and one exotic particle $(1,1,1,\overline{2},1,2)$.
The mixing of these two kinds of exotic particles yields one chiral multiplet $(1,\overline{2},1,\overline{2},1,1)$.
Besides, their self-confinement results in a tensor representation $(1,\overline{2}^2,1,1,1,1)$ and a tensor representation $(1,1,1,\overline{2}^2,1,1)$, which can be further decomposed
\beq
\begin{aligned}
    1\times(1,\overline{2}^2,1,1,1,1) &\to 1\times(1,\overline{3},1,1,1,1)+1\times(1,\overline{1},1,1,1,1)~,~\\
    1\times(1,1,1,\overline{2}^2,1,1) &\to 1\times(1,1,1,\overline{3},1,1)+1\times(1,1,1,\overline{1},1,1)~,~
\end{aligned}
\eeq
leading to chiral multiplets $(1,\overline{3},1,1,1,1)$ and $(1,1,1,\overline{3},1,1)$.

The chiral spectrum of Model 8 is shown in Table \ref{spectrum Model 8}, where a bifundamental Higgs doublet arises from the ${\cal N}=2$ subsector of the intersection of $b$-stack and $d$-stack of D6-branes.
Besides, there are two $USp$ groups with negative $\beta$ functions from Table \ref{model8}.
These two possible confining groups may lead to stable minimum and dynamical supersymmetry breaking.

\begin{table}[htb]
\footnotesize
\renewcommand{\arraystretch}{1.0}
\caption{The chiral spectrum in the open string sector of Model 8}
\label{spectrum Model 8}
\begin{center}
\begin{tabular}{|c||c||c|c|c|c||c|c|c|}\hline
\scriptsize{Model 8} &\scriptsize{$SU(4)\times SU(2)_{L1}\times SU(2)_{R}$}
& \scriptsize{$Q_4$} & \scriptsize{$Q_{2L_1}$} & \scriptsize{$Q_{2R}$} & \scriptsize{$Q_{2L_2}$} & $Q_{em}$ & $B-L$ & Field \\
 &\scriptsize{$\times SU(2)_{L2}\times USp(2)^2$}
&   &   &   &   &   &   &   \\
\hline\hline
$ab$ & $1 \times (4,\overline{2},1,1,1,1)$ & $1$ & -$1$ & 0 & 0 &  $-\frac 13, \frac 23,-1, 0$ & $\frac 13,-1$ & $Q_L, L_L$\\
$ac$ & $1 \times (\overline{4},1,2,1,1,1)$ & -$1$ & 0 & $1$ & 0 &  $\frac 13, -\frac 23, 1, 0$ & $-\frac 13, 1$ & $Q_R, L_R$\\
$ac'$ & $2 \times (\overline{4},1,\overline{2},1,1,1)$ & -$1$ & 0 & -$1$ & 0 &  $\frac 13, -\frac 23, 1, 0$ & $-\frac 13, 1$ & $Q_R, L_R$\\
$ad$ & $2 \times (4,1,1,\overline{2},1,1)$ & $1$ & 0 & 0 & -$1$  & $-\frac 13, \frac 23, -1, 0$ & $\frac 13, -1$ & $Q_L, L_L$\\
$bc'$ & $2 \times(1,2,2,1,1,1)$ & 0 & $1$ & $1$ & 0  & $0,\;\pm 1$ & 0 & $H$\\
$bd'$ & $1 \times(1,2,1,2,1,1)$ & 0 & $1$ & 0 & $1$   & $0,\;\pm 1$ & 0 & $\Phi$\\
$cd$ & $6 \times(1,1,2,\overline{2},1,1)$ & 0 & 0 & $1$ & -$1$   & $0,\;\pm 1$ & 0 & $H'$\\
$cd'$ & $3 \times(1,1,2,2,1,1)$ & 0 & 0 & $1$ & $1$   & $0,\;\pm 1$ & 0 & $H$\\
$b4$ & $1\times (1,\overline{2},1,1,1,2)$ & 0 & -1 & 0 & 0 & $\mp\frac 12$ & 0 & \\
$c2$ & $1\times (1,1,\overline{2},1,2,1)$ & 0 & 0 & -1 & 0 & $\mp\frac 12$ & 0 & \\
$d2$ & $1\times (1,1,1,2,\overline{2},1)$ & 0 & 0 & 0 & 1 & $\pm\frac 12$ & 0 & \\
$d4$ & $3\times (1,1,1,\overline{2},1,2)$ & 0 & 0 & 0 & -1 & $\mp\frac 12$ & 0 & \\
$c_{\overline{\Ysymm}}$ & $2\times(1,1,\overline{3},1,1,1)$ & 0 & 0 & -2 & 0  & $0,\;\mp 1$ & 0 & \\
$c_{\Yasymm}$ & $2\times(1,1,1,1,1,1)$ & 0 & 0 & 2 & 0   & 0 & 0 & \\
$d_{\Ysymm}$ & $2\times(1,1,1,3,1,1)$ & 0 & 0 & 0 & 2   & $0,\;\pm 1$ & 0 & \\
$d_{\overline{\Yasymm}}$ & $2\times(1,1,1,\overline{1},1,1)$ & 0 & 0 & 0 & -2   & 0 & 0 & \\
	\hline\hline
$bd$ & $1\times (1,2,1,\overline{2},1,1)$ & 0 & 1 & 0 & -1   &  $0,\;\pm 1$ &0 &${\Phi}_u^i, {\Phi}_d^i$\\
& $1\times (1,\overline{2},1,2,1,1)$ & 0 & -1 & 0 & 1  &  & &  \\
\hline
\end{tabular}
\end{center}
\end{table}

\begin{table}[htb]
\scriptsize
    \caption{The composite particle spectrum for Model 8}
	\label{composite model8}
	\begin{center}
		\begin{tabular}{|c|c||c|c|}
			\hline\multicolumn{2}{|c||}{\rm{Model} 8} & \multicolumn{2}{c|}{$U(4)\times U(2)_{L1}\times U(2)_{R}\times U(2)_{L2} \times USp(2)^2 $}\\
            \hline
            Confining Force & Intersection & Exotic particle Spectrum & Confined Particle Spectrum \\
			\hline \hline
            $USp(2)_2$ & c2 & $1\times(1,1,\overline{2},1,2,1)$ &
            $1\times(1,1,\overline{2}^2,1,1,1)$, $1\times(1,1,\overline{2},2,1,1)$,\\
             & d2 & $1\times(1,1,1,2,\overline{2},1)$ & $1\times(1,1,1,2^2,1,1)$\\
             \hline
            $USp(2)_4$ & b4 & $1\times(1,\overline{2},1,1,1,2)$ &
            $1\times(1,\overline{2}^2,1,1,1,1)$, $3\times(1,\overline{2},1,\overline{2},1,1)$,\\
             & d4 & $3\times(1,1,1,\overline{2},1,2)$ & $6\times(1,1,1,\overline{2}^2,1,1)$\\
			\hline
		\end{tabular}
	\end{center}
\end{table}

The composite particle spectrum for Model 8 is given in Table \ref{composite model8}.
The $USp(2)_2$ group has two charged intersections, which give rise to an exotic particle $(1,1,\overline{2},1,2,1)$ and an exotic particle $(1,1,1,2,\overline{2},1)$.
The mixing of these two kinds of exotic particles yields a chiral multiplet $(1,1,\overline{2},2,1,1)$.
Besides, their self-confinement results in tensor representations $(1,1,\overline{2}^2,1,1,1)$ and tensor representations $(1,1,1,2^2,1,1)$, which can be further decomposed
\beq
\begin{aligned}
    1\times(1,1,\overline{2}^2,1,1,1) &\to 1\times(1,1,\overline{3},1,1,1)+1\times(1,1,\overline{1},1,1,1)~,~\\
    1\times(1,1,1,2^2,1,1) &\to 1\times(1,1,1,3,1,1)+1\times(1,1,1,1,1,1)~,~
\end{aligned}
\eeq
leading to chiral multipets $(1,1,\overline{3},1,1,1)$ and $(1,1,1,3,1,1)$.
Meanwhile, the $USp(2)_4$ group also has two charged intersections that give rise to one exotic particle $(1,\overline{2},1,1,1,2)$ and three exotic particles $(1,1,1,\overline{2},1,2)$.
The mixing of these two kinds of exotic particles yields 3 chiral multiplets $(1,\overline{2},1,\overline{2},1,1)$.
Besides, their self-confinement results in a tensor representation $(1,\overline{2}^2,1,1,1,1)$ and a 6 tensor representation $(1,1,1,\overline{2}^2,1,1)$, which can be further decomposed, with
\beq
\begin{aligned}
    1\times(1,\overline{2}^2,1,1,1,1) &\to 1\times(1,\overline{3},1,1,1,1)+1\times(1,\overline{1},1,1,1,1)~,~\\
    6\times(1,1,1,\overline{2}^2,1,1) &\to 6\times(1,1,1,\overline{3},1,1)+6\times(1,1,1,\overline{1},1,1)~,~
\end{aligned}
\eeq
leading to chiral multipets $(1,\overline{3},1,1,1,1)$ and $(1,1,1,\overline{3},1,1)$.

\begin{table}[h!]
\footnotesize
\renewcommand{\arraystretch}{1.0}
\caption{The chiral spectrum in the open string sector of Model 9}
\label{spectrum Model 9}
\begin{center}
\begin{tabular}{|c||c||c|c|c|c||c|c|c|}\hline
Model 9 &\scriptsize{$SU(4)\times SU(2)_{L_1}\times SU(2)_{R}\times SU(2)_{L_2}$}
& $Q_4$ & $Q_{2L_1}$ & $Q_{2R}$ & $Q_{2L_2}$ & $Q_{em}$ & $B-L$ & Field \\
\hline\hline
$ab$ & $1 \times (4,\overline{2},1,1)$ & $1$ & -$1$ & 0 & 0 &  $-\frac 13,\; \frac 23,\;-1,\; 0$ & $\frac 13,\;-1$ & $Q_L, L_L$\\
$ac$ & $2 \times (\overline{4},1,2,1)$ & -$1$ & 0 & $1$ & 0 &  $\frac 13,\; -\frac 23,\; 1,\; 0$ & $-\frac 13,\; 1$ & $Q_R, L_R$\\
$ac'$ & $1 \times (\overline{4},1,\overline{2},1)$ & -$1$ & 0 & -$1$ & 0 &  $\frac 13,\; -\frac 23,\; 1,\; 0$ & $-\frac 13,\; 1$ & $Q_R, L_R$\\
$ad$ & $2 \times (4,1,1,\overline{2})$ & $1$ & 0 & 0 & -$1$  & $-\frac 13,\; \frac 23,\; -1,\; 0$ & $\frac 13,\; -1$ & $Q_L, L_L$\\
$bc$ & $4 \times(1,2,\overline{2},1)$ & 0 & $1$ & -$1$ & 0  & $0,\;\pm 1$ & 0 & $H$\\
$bd'$ & $2 \times(1,2,1,2)$ & 0 & $1$ & 0 & $1$  & $0,\;\pm 1$ & 0 & $\Phi$\\
$cd$ & $6 \times(1,1,\overline{2},2)$ & 0 & 0 & -$1$ & $1$   & $0,\;\mp1$ & 0 & $H$\\
$cd'$ & $12 \times(1,1,\overline{2},\overline{2})$ & 0 & 0 & -$1$ & -$1$   & $0,\;\mp1$ & 0 & $H'$\\
$b_{\overline{\Ysymm}}$ & $1 \times(1,\overline{3},1,1)$ & 0 & -2 & 0 & 0  & $0,\;\mp 1$ & 0 & \\
$b_{\Yasymm}$ & $1\times(1,1,1,1)$ & 0 & 2 & 0 & 0   & 0 & 0 & \\
$c_{\Ysymm}$ & $5\times(1,1,3,1)$ & 0 & 0 & 2 & 0   & $0,\;\pm 1$ & 0 & \\
$c_{\overline{\Yasymm}}$ & $5\times(1,1,\overline{1},1)$ & 0 & 0 & -2 & 0   & 0 & 0 & \\
$d_{\Ysymm}$ & $2\times(1,1,1,3)$ & 0 & 0 & 0 & 2   & $0,\;\pm 1$ & 0 & \\
$d_{\overline{\Yasymm}}$ & $2\times(1,1,1,\overline{1})$ & 0 & 0 & 0 & -2   & 0 & 0 & \\
	\hline\hline
$bd$ & $2\times (1,2,1,\overline{2})$ & 0 & 1 & 0 & -1   &  $0,\;\pm 1$ &0 &${\Phi}_u^i, {\Phi}_d^i$\\
& $2\times (1,\overline{2},1,2)$ & 0 & -1 & 0 & 1  &  & &  \\
\hline
\end{tabular}
\end{center}
\end{table}

The full spectra of Model 9 and 10 are shown in Table \ref{spectrum Model 9} and \ref{spectrum Model 10}, respectively.
In Model 9, there are 2 bifundamental Higgs doublets arising from the ${\cal N}=2$ subsector of the intersection of the $b$- and $d$-stacks of the D6-branes. 
In Model 10, there are 8 Higgs doublets arising from the ${\cal N}=2$ subsector of the intersection of the $b$- and $c$-stacks of the D6-branes.
In addition, since Model 9 and 10 have no $USp$ group, one cannot break the supersymmetry through gaugino condensation.

\begin{table}[h!]
\footnotesize
\renewcommand{\arraystretch}{1.0}
\caption{The chiral spectrum in the open string sector of Model 10}
\label{spectrum Model 10}
\begin{center}
\begin{tabular}{|c||c||c|c|c|c||c|c|c|}\hline
Model 10 &\scriptsize{$SU(4)\times SU(2)_{L_1}\times SU(2)_{R}\times SU(2)_{L_2}$}
& $Q_4$ & $Q_{2L_1}$ & $Q_{2R}$ & $Q_{2L_2}$ & $Q_{em}$ & $B-L$ & Field \\
\hline\hline
$ab$ & $1 \times (4,\overline{2},1,1)$ & 1 & -$1$ & 0 & 0 &  $-\frac 13,\; \frac 23,\;-1,\; 0$ & $\frac 13,\;-1$ & $Q_L, L_L$\\
$ac$ & $3 \times (\overline{4},1,2,1)$ & -$1$ & 0 & $1$ & 0 &  $\frac 13,\; -\frac 23,\; 1,\; 0$ & $-\frac 13,\; 1$ & $Q_R, L_R$\\
$ad'$ & $2 \times (4,1,1,2)$ & $1$ & 0 & 0 &  $1$  & $-\frac 13,\; \frac 23,\; -1,\; 0$ & $\frac 13,\; -1$ & $Q_L, L_L$\\
$bc'$ & $4 \times(1,2,2,1)$ & 0 & $1$ & $1$ & 0  & $0,\;\pm 1$ & 0 & $H'$\\
$bd$ & $2 \times(1,\overline{2},1,2)$ & 0 & -$1$ & 0 & $1$  & $0,\;\mp 1$ & 0 & $\Phi$\\
$bd'$ & $4 \times(1,2,1,2)$ & 0 & $1$ & 0 & $1$   & $0,\;\pm1$ & 0 & $\Phi$\\
$cd$ & $2 \times(1,1,\overline{2},2)$ & 0 & 0 & -$1$ & $1$   & $0,\;\mp1$ & 0 & $H'$\\
$cd'$ & $4 \times(1,1,\overline{2},\overline{2})$ & 0 & 0 & -$1$ & -$1$   & $0,\;\mp1$ & 0 & $H$\\
$b_{\overline{\Ysymm}}$ & $1\times(1,\overline{3},1,1)$ & 0 & -2 & 0 & 0  & $0,\;\mp 1$ & 0 & \\
$b_{\Yasymm}$ & $1\times(1,1,1,1)$ & 0 & 2 & 0 & 0   & 0 & 0 & \\
$c_{\overline{\Ysymm}}$ & $1\times(1,1,\overline{3},1)$ & 0 & 0 & -2 & 0   & $0,\;\pm 1$ & 0 & \\
$c_{\Yasymm}$ & $1\times(1,1,1,1)$ & 0 & 0 & 2 & 0   & 0 & 0 & \\
$d_{\overline{\Ysymm}}$ & $2\times(1,1,1,\overline{3})$ & 0 & 0 & 0 & -2   & $0,\;\pm 1$ & 0 & \\
$d_{\Yasymm}$ & $2\times(1,1,1,1)$ & 0 & 0 & 0 & 2   & 0 & 0 & \\
	\hline\hline
$bc$ & $8\times (1,2,\overline{2},1)$ & 0 & 1 & -1 & 0   &  $0,\;\pm 1$ &0 &${H}_u^i, {H}_d^i$\\
& $8\times (1,\overline{2},2,1)$ & 0 & -1 & 1 & 0   &  & &  \\
\hline
\end{tabular}
\end{center}
\end{table}
\FloatBarrier

\begin{table}[htb]
\footnotesize
\renewcommand{\arraystretch}{1.0}
\caption{The chiral spectrum in the open string sector of Model 18 from~\cite{He:2021gug}}
\label{spectrum Comp M18}
\begin{center}
\begin{tabular}{|c||c||c|c|c|c||c|c|}\hline
Model 18~\cite{He:2021gug} &\scriptsize{$SU(4)\times SU(2)_L\times SU(2)_{R}\times USp(2)^3 $}
& $Q_4$ & $Q_{2L}$ & $Q_{2R}$ & $Q_{em}$ & $B-L$ & Field \\
\hline\hline
$ab$ & $3 \times (4,\overline{2},1,1,1,1)$ & 1 & -1 & 0 &  $-\frac 13,\; \frac 23,\;-1,\; 0$ & $\frac 13,\;-1$ & $Q_L, L_L$\\
$ac'$ & $3 \times (\overline{4},1,\overline{2},1,1,1)$ & -1 & 0 & -1 &  $\frac 13,\; -\frac 23,\;1,\; 0$ & $-\frac 13,\;1$ & $Q_R, L_R$\\
$bc$ & $3 \times (1,\overline{2},2,1,1,1)$ & 0 & -1 & 1 & $0,\mp 1$ & 0 & $H'$\\
$a2$ & $1 \times (4,1,1,\overline{2},1,1)$ & 1 & 0 & 0 & $\frac16,-\frac12$ & $\frac13,-1$ & \\
$a4$ & $1 \times (\overline{4},1,1,1,1,2)$ & -1 & 0 & 0 & $-\frac16,\frac12$ & $-\frac13,1$ & \\
$b4$ & $1 \times (1,2,1,1,1,\overline{2})$ & 0 & 1 & 0 & $\pm\frac12$ & 0 & \\
$c1$ & $3 \times (1,1,\overline{2},2,1,1)$ & 0 & 0 & -1 & $\mp\frac12$ & 0 & \\
$c2$ & $2 \times (1,1,2,\overline{2},1,1)$ & 0 & 0 & 1 & $\pm\frac12$ & 0 & \\
$b_{\Ysymm}$ & $2 \times (1,3,1,1,1,1)$ & 0 & 2 & 0 & $0,\pm 1$ & 0 & \\
$b_{\overline{\Yasymm}}$ & $2 \times (1,\overline{1},1,1,1,1)$ & 0 & -2 & 0 & $0$ & 0 & \\
$c_{\Ysymm}$ & $1 \times (1,1,3,1,1,1)$ & 0 & 0 & 2 & $0,\pm 1$ & 0 & \\
$c_{\overline{\Yasymm}}$ & $1 \times (1,1,\overline{1},1,1,1)$ & 0 & 0 & -2 & $0$ & 0 & \\
\hline\hline
$bc'$ & $9 \times (1,2,2,1,1,1)$ & 0 & 1 & 1 & $0,\pm 1$ & 0 & $H_u^i,H_d^i$ \\
 & $9\times (1,\overline{2},\overline{2},1,1,1)$ & 0 & -1 & -1 & & & \\
\hline
\end{tabular}
\end{center}
\end{table}
\FloatBarrier

Recall that Model 10 shares the same gauge coupling ratio with 
 Model 18 from~\cite{He:2021gug} with spectrum shown in Table \ref{spectrum Comp M18}.
In Model 18 from~\cite{He:2021gug}, there are 9 Higgs doublets arising from the ${\cal N}=2$ subsector of the intersection of the $b$-stack and the $\Zbb_2$ image of the $c$-stack of D6-branes only, while in Model 10,  Higgs multiplets and Higgs-like particles arise from more intersecting origins.
Again, similar to the $d$-stack of D6-branes introduced into the $SU(2)_R$ sector, extra $d$-stack turns to reduce the number of filler branes.  
From the perspective of phenomenology, these two models are considered to be different as their values of gauge couplings $g_a$ are different from $\gamma$ in (\ref{gabY}).

\section{String and GUT scale gauge coupling relation}
\label{G.C.U.}

With an extra $d$-stack of D6-branes introduced, we have  phenomenologically discussed eight representative ${\cal N}=1$ supersymmetric Pati-Salam models from their spectrum. In this section, we discuss on their behavior of  string and GUT scale gauge coupling unification with RGE evolution.  

In particular, the new  models constructed by introducing new $d$-stack D6-branes do not have chiral multiplets in the symmetric representation of $SU(4)_C$ appearing from $aa$ sector, and thus the exotic particles can be decoupled~\cite{Li:2022cqk}.
In principle,  the gauge coupling unification can be performed at the string scale ($M_{string}\simeq5\times 10^{17}\text{GeV}$) or the GUT scale($M_{GUT}\simeq2\times 10^{16}\text{GeV}$) as shown in~\cite{Li:2022cqk}.
In which, through the method of two-loop evolution and the addition of vector-like particles, gauge coupling unification can be achieved. 
These vector-like particles come from ${\cal N}=2$ subsectors and are chiral multiplets from adjoint representations of $SU(4)_C$ and $SU(2)_L$.

Recall that the gauge coupling relations can be expressed in (\ref{gabY}), and at the unification scale, we have
\beq
k_1 g_a^2=k_2 g_b^2=k_Yg_Y^2=g_U^2~,~
\label{uniR}
\eeq
\beq
k_1 g_a^2=k_2 g_L^2=k_Yg_Y^2=g_U^2~,~
\label{uniL}
\eeq
for models with $SU(2)_{R_1}\times SU(2)_{R_2}\rightarrow SU(2)_R$ and $SU(2)_{L_1}\times SU(2)_{L_2}\rightarrow SU(2)_L$ respectively.
In these two cases, one can take $k_1=1$ for convenience, and the constants $k_2$ and $k_Y=\frac53 k_y$ can be obtained from the gauge coupling relations of the models shown in Sections \ref{models} and \ref{secL}.
By defining $\alpha_1\equiv \frac{k_Y g_Y^2}{4\pi}$, $\alpha_2\equiv \frac{k_2 g_L^2}{4\pi}$ and $\alpha_3\equiv \frac{g_a^2}{4\pi}$, the gauge coupling relations at the unification scale can be represented as~\cite{Li:2022cqk}
\beq
\alpha_U ^{-1}\equiv\alpha_1^{-1}=\frac{\alpha_2^{-1}+\alpha_3^{-1}}{2}~,~
\eeq
and such the accuracy can be defined by
\beq
\Delta =\frac{|\alpha_1^{-1}-\alpha_2^{-1}|}{\alpha_1^{-1}}~,~
\eeq
which is usually required to be less than $1\%$.

Unlike Model 5 in~\cite{He:2021gug} with $k_2=1$ and $k_Y=\frac53$, the new models found in our work cannot naturally achieve unification without introducing additional vector-like particles or chiral multiplets.
Thus, in order to obtain the string-scale or GUT scale gauge coupling relation, two-loop RGEs must be performed~\cite{Li:2022cqk,He:2021kbj,Chen:2017rpn,Chen:2018ucf,Barger:2005qy,Barger:2007qb}
\beq
\frac{{\rm d} g_i}{{\rm d} \ln{\mu}} =\frac{b_i}{(4\pi)^2}g_i^3+\sum_j\frac{g_i^3}{(4\pi)^4}\left[\sum_{j=1}^3 B_{ij}g_j^2-\sum_{\alpha=u,d,e}d_i^{\alpha}Tr(h^{\alpha\dagger}h^{\alpha})\right]
~, ~
\eeq
where $g_i(i=1,2,3)$ and $h^\alpha(\alpha=u,d,e)$ stand for the SM gauge couplings and the Yukawa couplings, respectively.
The coefficients for the beta functions in SM~\cite{Machacek:1983tz,Machacek:1983fi,Machacek:1984zw,Cvetic:1998uw,Gogoladze:2010in} and supersymmetric models~\cite{Barger:1992ac,Barger:1993gh,Martin:1993zk} are given by
\beq
b_{\rm SM}=\left( \frac{41}{6}\frac{1}{k_Y},-\frac{19}{6}\frac{1}{k_2},-7 \right)~,~
B_{\rm SM}=\left(
\begin{matrix}
    \frac{199}{18}\frac{1}{k_Y^2} & \frac{27}{6}\frac{1}{k_Y k_2} & \frac{44}{3}\frac{1}{k_Y}\\
    \frac{3}{2}\frac{1}{k_Y k_2} & \frac{35}{6}\frac{1}{k_2^2} & 12\frac{1}{k_2^2}\\
    \frac{11}{6}\frac{1}{k_Y} & \frac{9}{2}\frac{1}{k_2} & -26
\end{matrix}\right)~,~
\eeq
\beq
d_{\rm SM}^u=\left(\frac{17}{6} \frac{1}{k_Y},\frac{3}{2}\frac{1}{k_2},2 \right)~,~
d_{\rm SM}^d=0~,~d_{\rm SM}^e=0~,~
\eeq

\beq
b_{\rm SUSY}=\left( 11\frac{1}{k_Y},\frac{1}{k_2},-3 \right)~,~
B_{\rm SUSY}=\left(
\begin{matrix}
    \frac{199}{9}\frac{1}{k_Y^2} & 9\frac{1}{k_Y k_2} & \frac{88}{3}\frac{1}{k_Y}\\
    3\frac{1}{k_Y k_2} & 25\frac{1}{k_2^2} & 24\frac{1}{k_2^2}\\
    \frac{11}{3}\frac{1}{k_Y} & 9\frac{1}{k_2} & 14
\end{matrix}\right)~,~
\eeq
\beq
d_{\rm SUSY}^u=\left(\frac{26}{3} \frac{1}{k_Y},6\frac{1}{k_2},4 \right)~,~
d_{\rm SUSY}^d=0~,~d_{\rm SUSY}^e=0~.~
\eeq
The two-loop RGEs for the SM gauge couplings can be numerically solved, and the
calculations of one-loop RGEs can be performed taking into account of the new physics contributions and threshold.
The general one-loop RGEs for Yukawa couplings are given in~\cite{Gogoladze:2010in}.
The gauge couplings run from $M_Z$, the mass scale of the $Z$ boson, with the corresponding initial conditions
\beq
g_1(M_Z)=\sqrt{k_Y}\frac{g_{em}}{\cos{\theta_W}}~,~
g_2(M_Z)=\sqrt{k_2}\frac{g_{em}}{\sin{\theta_W}}~,~
g_3(M_Z)=\sqrt{4\pi\alpha_s}~,~
\eeq
in which $g_{em}$ is the effective coupling strength of the electromagnetic interaction in electroweak theory, $\theta_W$ is the mixing angle (or the Weinberg angle) and $\alpha_s$ is the strong coupling constant.
The analysis considers the non-supersymmetric SM spectrum with the top quark pole mass set to $m_t = 173.34$ GeV, over the range of $M_Z$ to a supersymmetry breaking scale $M_S$.
Guided by the experimental lower limits for preserving supersymmetry and the gauge hierarchy, one can adopt a supersymmetry breaking scale $M_S\simeq3.0$ TeV~\cite{Li:2022cqk}.
In addition, one can choose the mass of the $Z$ boson, the pole mass of the top quark, the Higgs Vacuum Expectation Value (VEV), the strong coupling constant, the fine structure constant and the relation of the mixing angle as follows~\cite{ParticleDataGroup:2018ovx,ParticleDataGroup:2020ssz},
\beq
\begin{aligned}
M_Z=91.1876\;{\rm{GeV}}~,~
m_t=173.34\pm 0.27(stat)\pm0.71(syst)\;{\rm{GeV}}~,~
v=174.10\;{\rm{GeV}}~,~\\
\alpha_S(M_Z)=0.1181\pm 0.0011~,~
\alpha_{em}^{-1}(M_Z)=128.91\pm 0.02~,~
\sin^2\theta_W(M_Z)=0.23122~.~
\end{aligned}
\eeq
And string and GUT scales are usually applied to be $M_{string}\simeq 5\times10^{17}\rm{GeV}$ and $M_{GUT}\simeq 2\times10^{16}\rm{GeV}$.
Under $SU(3)_C\times SU(2)_L\times U(1)_Y$, the quantum numbers of vector-like particles that can be introduced to RGE rolling and their chiral multiplets can be explicitly written down as~\cite{Jiang:2006hf,Jiang:2008xrg}
\begin{align} \label{subsector-Q}
    &(XQ+\overline{XQ})=(\mathbf{3},\mathbf{2},\mathbf{\frac{1}{6}})+(\mathbf{\bar{3}},\mathbf{2},\mathbf{-\frac{1}{6}})~,\\ \label{subsector-D}
    &(XD+\overline{XD})=(\mathbf{3},\mathbf{1},\mathbf{-\frac{1}{3}})+(\mathbf{\bar{3}},\mathbf{1},\mathbf{\frac{1}{3}})~,\\ \label{subsector-U}
    &(XU+\overline{XU})=(\mathbf{3},\mathbf{1},\mathbf{\frac{2}{3}})+(\mathbf{\bar{3}},\mathbf{1},\mathbf{-\frac{2}{3}})~,\\ \label{subsector-E}
    &(XE+\overline{XE})=(\mathbf{1},\mathbf{1},\mathbf{1})+(\mathbf{1},\mathbf{1},\mathbf{-1})~,\\ \label{subsector-G}
    &XG=(\mathbf{8},\mathbf{1},\mathbf{0})~,\\ \label{subsector-W}
    &XW=(\mathbf{1},\mathbf{3},\mathbf{0})~. 
\end{align}
And their contributions to the one-loop beta functions are simply
\begin{align}
    &\Delta b^{(XQ+\overline{XQ})}=\left(\frac{1}{5},3,2\right)~,~~\Delta b^{(XD+\overline{XD})}= \left(\frac{2}{5},0,1\right)~,~~
    \Delta b^{(XU+\overline{XU})}=\left(\frac{8}{5},0,1\right)~,\\
    &\Delta b^{(XE+\overline{XE})}=\left(\frac{6}{5},0,0\right)~,~~
    \Delta b^{(XG)}= \left(0,0,3\right)~,~~
    \Delta b^{(XW)}=\left(0,2,0\right)~, 
\end{align}
with their contributions to the two-loop beta function to be
\begin{align}  
    &\Delta B^{(XQ+\overline{XQ})}=
    \begin{pmatrix}
        \frac{1}{75} & \frac{3}{5} & \frac{16}{15} \\
        \frac{1}{5}  & 21          & 16 \\
        \frac{2}{15} & 6           & \frac{68}{3}
    \end{pmatrix}~,~
    \Delta B^{(XD+\overline{XD})}=
    \begin{pmatrix}
        \frac{8}{75}& 0& \frac{32}{15} \\
             0& 0& 0\\
        \frac{4}{15}& 0& \frac{34}{3}
    \end{pmatrix}~,~
    \Delta B^{(XU+\overline{XU})}=
    \begin{pmatrix}
        \frac{128}{75}& 0& \frac{128}{15} \\
        0& 0& 0\\
        \frac{16}{15}& 0& \frac{34}{3}
    \end{pmatrix}~,~\\
    &\Delta B^{(XE+\overline{XE})}=
    \begin{pmatrix}
        \frac{72}{25} & 0 & 0 \\
        0  & 0          & 0 \\
        0 & 0           & 0
    \end{pmatrix}~,~
    \Delta B^{(XG)}=
    \begin{pmatrix}
        0& 0& 0 \\
             0& 0& 0\\
        0& 0& 54
    \end{pmatrix}~,~
    \Delta B^{(XW)}=
    \begin{pmatrix}
        0& 0& 0\\
        0& 24& 0\\
        0& 0& 0
    \end{pmatrix}.
\end{align}

By introducing additional vector-like particles that arise from ${\cal N}=2$ subsectors, both string and GUT scale gauge coupling unification can be expected to be achieved via two-loop RGE evolution as in~\cite{He:2021kbj, Li:2022cqk}.
For Model 1 to Model 4, the gauge coupling unification are expected to achieve
at the two-loop level with additionally  vector-like particles from the ${\cal N}=2$ subsector introduced,
whose quantum numbers and their Hermitian conjugates are shown in \eqref{subsector-Q}-\eqref{subsector-W}.
The number of these vector-like particles, $n_v$, shall be completely determined by the intersections of the D6-branes from model building.
And the gauge coupling unification can be finally be realized by fine-tuning the masses of the additional vector-like particles.
As for Models 7-10, in addition to vector-like particles,  chiral multiplets are also required from adjoint representations of $SU(4)_C$, $SU(2)_{L_1}$ and $SU(2)_{L_2}$.
The chiral multiplet $XG$ comes from the $aa$ sector, while the chiral multiplet $XW$ comes from the $bb$ and the $dd$ sectors, as shown in Table \ref{tab2}.
Since there are three adjoint chiral multiplets arising from each one of $aa$, $bb$, $cc$ and $dd$ sectors, the number of chiral multiplets $XG$ is restricted to be less than or equal to 3, and the number of chiral multiplets $XW$ is restricted to be less than or equal to 6.
With the extra $d$-stack of D6-brane introduced to $SU(2)_R$ and $SU(2)_L$ gauge sector, we take Model 4 and Model 10 as examples respectively, to discuss their string and GUT scale gauge coupling unification in the following.

First, while the extra $d$-stack of D6-branes are introduced to the $SU(2)_R$ sector, we take Model 4 as example. 
For Model 4, the extra $d$-stack of D6-branes are introduced to the $SU(2)_R$ configuration, for which the first torus is tilted.
Its MSSM gauge coupling relation is
$g^2_a=2 g^2_b=g^2_R=(\frac{5}{3} g^2_Y)=4\times3^{1/4} \pi  e^{\phi_4}$. To realize gauge coupling unification at string or GUT scale, the contributions of introduced vector-like particles to the one-loop beta functions play the essential role.
Recall that the vector-like particles from $XD+\overline{XD}$ and $XU+\overline{XU}$ contribute to the one-loop beta functions with $\Delta b^{(XD+\overline{XD})}= \left(\frac{2}{5},0,1\right),
\Delta b^{(XU+\overline{XU})}=\left(\frac{8}{5},0,1\right)$. With such introduced particle, the gauge coupling $g_a$ and $g_Y$ will be lifted while $g_b$ remain the same till the gauge unification.
Therefore, it can be expected that by introducing vector-like particles from ${\cal N}=2$ subsector $XD+\overline{XD}$ 
with quantum number $(\mathbf{3},\mathbf{1},\mathbf{-\frac{1}{3}})+(\mathbf{\bar{3}},\mathbf{1},\mathbf{\frac{1}{3}})$ in \eqref{subsector-D} and $XU+\overline{XU}$ with quantum number $(\mathbf{3},\mathbf{1},\mathbf{\frac{2}{3}})+(\mathbf{\bar{3}},\mathbf{1},\mathbf{-\frac{2}{3}})$ in \eqref{subsector-U} from the ${\cal N}=2$ subsector, 
the gauge coupling unification at both string and GUT scales can be obtained.
As these vector-like particles  naturally  arise from the ${\cal N}=2$ subsector, the number of these particles are completely determined from the intersection of D6-branes with\footnote{Here the upper indices of the intersection number denote which two of the three tori are wrapped in the ${\cal N}=2$ subsectors, {\emph e.g.} $(2,3)$ in $I_{ac'}^{(2,3)}$.} $|I_{ac'}^{(2,3)}|=|(0+1)(-1+0)|=1$ and $|I_{ad}^{(1,2)}|=|\frac12 (1+3)(0+1 )|=2$.
We verify that the gauge coupling unification at both string scale and GUT scale can be achieved by adding $2(XD+\overline{XD})+2(XU+\overline{XU})$ and $3(XD+\overline{XD})+3(XU+\overline{XU})$, respectively.
    
Second, while the extra $d$-stack of D6-branes are introduced to the $SU(2)_L$ sector, we take Model 10 as an example. 
For Model 10, its MSSM gauge coupling relation is
$g^2_a=g^2_L=2g^2_c=\frac{10}{7}(\frac{5}{3} g^2_Y)=4\times 3^{1/4} \pi  e^{\phi_4}$, for which the first torus is tilted.
To suppress the gauge coupling $g_Y$ in the RGE evolution, 
with the aforementioned definition 
$\alpha_1\equiv \frac{k_Y g_Y^2}{4\pi}$, $\alpha_2\equiv \frac{k_2 g_L^2}{4\pi}$ and $\alpha_3\equiv \frac{g_a^2}{4\pi}$,
the vector-like particle  $(XQ+\overline{XQ})$ and chiral multiplets from adjoint representations of $SU(4)_C$ $XG$ can be introduced.
Such that, their contributions to the one-loop beta functions
$\Delta b^{(XQ+\overline{XQ})}=\left(\frac{1}{5},3,2\right)$ and 
$\Delta b^{(XG)}= \left(0,0,3\right)$ can finally lead to the gauge coupling unification.
In precise, the introduced vector-like particles are $XQ+\overline{XQ}$ with quantum number $(\mathbf{3},\mathbf{2},\mathbf{\frac{1}{6}})+(\mathbf{\bar{3}},\mathbf{2},\mathbf{-\frac{1}{6}})$ in \eqref{subsector-Q} from ${\cal N}=2$ subsector and $XG$ with quantum number $(\mathbf{8},\mathbf{1},\mathbf{0})$ in \eqref{subsector-G} from $aa$ sector with $n_{XG}=3$,
while the number $n_{XQ}$ determined by the intersection with $|I_{ab'}^{(2,3)}|=1$ and $|I_{ad}^{(1,2)}|=2$ or $n_{XQ}=|I_{ab'}^{(2,3)}|+|I_{ad}^{(1,2)}|=3$.
It can be verified that in such construction the gauge coupling unification can be achieved at both string scale and GUT scale as in the former  studies~\cite{He:2021kbj, Li:2022cqk}.

While Models 1-3 and 6-9 give rise to completely different gauge coupling relations from the standard Pati-Salam construction,
Model 4 in this work and Model 10 in~\cite{He:2021gug} share the same gauge coupling ratio $g_a^2:g_b^2:g_R^2:(\frac{5}{3}g_Y^2)$,
and our Model 10  with $d$-stack configuration has the same gauge coupling ratio with the standard Model 18 of~\cite{He:2021gug}. 
This suggests that for a certain gauge coupling relation, a possible ``degeneracy'' behavior of $k_y$ and $k_2$ with extra $d$-stack of D6-branes exist. Namely, with the same gauge coupling ratio, there are models that phenomenologically independent from the gauge coupling unification aspect, yet with different values of $\gamma$ in (\ref{gabY}) or (\ref{gabYL}).
It is obvious that the models in the same gauge coupling degeneracy class shall behave similarly in the RGE rolling with vector-like particles  introduced from ${\cal N}=2$ subsector. 

As an advantage, while the extra $d$-stack of D6-brane introduced to $SU(2)_R$ and $SU(2)_L$ gauge sector, more D6-brane intersections are introduced and therefore more origins of vector-like particles arise from ${\cal N}=2$ subsector as well, such as $I_{ad}$ for Model 4.
Thus, string and GUT scale gauge coupling unification can be realized in  more efficient manner with $d$-stack of D6-brane introduced.

\section{String Landscape of ${\cal N}=1$ supersymmetric Pati-Salam models with extra $d$-stack of D6-branes and other model buildings}
\label{landscape}

To study the string landscape of intersecting branes with four stacks of D6-branes, we generalize the deterministic algorithm constructed in \cite{He:2021gug} which completed the standard ${\cal N}=1$ supersymmetric Pati-Salam models with three stacks of D6-branes.

\subsection{Deterministic Algorithm with $d$-stack extension}

\subsubsection*{Supersymmetry System}
\label{secsusy}
When there are $a$-, $b$-, and $c$-stacks of D6-branes, the supersymmetry system with three stacks of D6-branes
\beq
\left\{ \begin{array}{ll}
x_{A}\tilde{A}_{a} + x_{B}\tilde{B}_{a} + x_{C}\tilde{C}_{a} + x_{D}\tilde{D}_{a} = 0,\\ 
x_{A}\tilde{A}_{b} + x_{B}\tilde{B}_{b} + x_{C}\tilde{C}_{b} + x_{D}\tilde{D}_{b} = 0,\\ 
x_{A}\tilde{A}_{c} + x_{B}\tilde{B}_{c} + x_{C}\tilde{C}_{c} + x_{D}\tilde{D}_{c} = 0 ,
\end{array} 
\right. 
\eeq
with the linear system has rank 3. Cramer's rule indicates that equation $M\vec{x}=\vec{0}$ has infinitely many solutions with only one fundamental set of solutions.
Therefore, we must first identify a row in matrix $M$ where all cofactors are non-zeros, then assign its four corresponding algebraic cofactors as $y_A$, $y_B$, $y_C$, and $y_D$.
Without loss of generality, we can take the second row as an example and have
{\footnotesize 
\beq \label{cofactor}
y_A=\left| \begin{matrix}
    \tilde{B}_a && \tilde{C}_a && \tilde{D}_a \\
\tilde{B}_b && \tilde{C}_b && \tilde{D}_b \\
\tilde{B}_c && \tilde{C}_c && \tilde{D}_c
    \end{matrix}\right|~,~\,
y_B=-\left| \begin{matrix}
    \tilde{A}_a && \tilde{C}_a && \tilde{D}_a \\
\tilde{A}_b && \tilde{C}_b && \tilde{D}_b \\
\tilde{A}_c && \tilde{C}_c && \tilde{D}_c
    \end{matrix}\right|~,~\,
y_C=\left| \begin{matrix}
    \tilde{A}_a && \tilde{B}_a && \tilde{D}_a \\
\tilde{A}_b && \tilde{B}_b && \tilde{D}_b \\
\tilde{A}_c && \tilde{B}_c && \tilde{D}_c
    \end{matrix}\right|~,~\,
y_D=-\left| \begin{matrix}
    \tilde{A}_a && \tilde{B}_a && \tilde{C}_a \\
\tilde{A}_b && \tilde{B}_b && \tilde{C}_b \\
\tilde{A}_c && \tilde{B}_c && \tilde{C}_c
    \end{matrix}\right|~,~\,
\eeq
}
which are all non-zeros. Such that, the set of solutions to the supersymmetry equations~\eqref{susyconditions} can be rewritten as $x_A=\lambda$, $x_B=\lambda\, y_B/y_A$, $x_C=\lambda\, y_C/y_A$ and $x_D=\lambda\, y_D/y_A$ as for the three stacks of D6-branes construction in \cite{He:2021gug}.

While there are $a$-, $b$-, $c$- and $d$-stacks of D6-branes, the supersymmetry system 
are extended with the extra $d$-stack of contribution and becomes
\beq \label{4susy}
\left\{ \begin{array}{ll}
x_{A}\tilde{A}_{a} + x_{B}\tilde{B}_{a} + x_{C}\tilde{C}_{a} + x_{D}\tilde{D}_{a} = 0,\\ 
x_{A}\tilde{A}_{b} + x_{B}\tilde{B}_{b} + x_{C}\tilde{C}_{b} + x_{D}\tilde{D}_{b} = 0,\\ 
x_{A}\tilde{A}_{c} + x_{B}\tilde{B}_{c} + x_{C}\tilde{C}_{c} + x_{D}\tilde{D}_{c} = 0 ,\\ 
x_{A}\tilde{A}_{d} + x_{B}\tilde{B}_{d} + x_{C}\tilde{C}_{d} + x_{D}\tilde{D}_{d} = 0 .
\end{array} 
\right. 
\eeq
This extended linear system  shall still has rank 3, which is  constrained by the supersymmetry constraints~\eqref{susyconditions}.
This is because if the rank  is 4, equation $M\vec{x} = \vec{0}$ admits only the trivial solution $x_A = x_B = x_C = x_D = 0$, and if the rank of $M$ is less than three, the complex moduli of the tori will not be determined.
However, note that while the rank of the linear system is 3, there must have two stacks of supersymmetry equations that are linearly related. 

Considering that there are 3 and only 3 independent supersymmetry equations, this provides 4 subsystems of supersymmetry equations with distinct types of algebraic cofactor as \eqref{cofactor}. This can be categorized by which stack of D6-brane is missing in the three independent linear system of supersymmetry equations.
To complete the searching of string landscape with $d$-stack extension, we extended the searching algorithm with 4 types of subsystems of supersymmetry equations fully considered.
For example, while the independent stacks of D6-branes are a-, c-, $d$-stacks, the solution of ($x_A , x_B , x_C , x_D$) is proportional to
$y_A$, $y_B$, $y_C$, and $y_D$ with
{\footnotesize 
\beq \label{4cofactor}
y_A=\left| \begin{matrix}
    \tilde{B}_a && \tilde{C}_a && \tilde{D}_a \\
\tilde{B}_c && \tilde{C}_c && \tilde{D}_c \\
\tilde{B}_d && \tilde{C}_d && \tilde{D}_d
    \end{matrix}\right|~,~\,
y_B=-\left| \begin{matrix}
    \tilde{A}_a && \tilde{C}_a && \tilde{D}_a \\
\tilde{A}_c && \tilde{C}_c && \tilde{D}_c \\
\tilde{A}_d && \tilde{C}_d && \tilde{D}_d
    \end{matrix}\right|~,~\,
y_C=\left| \begin{matrix}
    \tilde{A}_a && \tilde{B}_a && \tilde{D}_a \\
\tilde{A}_c && \tilde{B}_c && \tilde{D}_c \\
\tilde{A}_d && \tilde{B}_d && \tilde{D}_d
    \end{matrix}\right|~,~\,
y_D=-\left| \begin{matrix}
    \tilde{A}_a && \tilde{B}_a && \tilde{C}_a \\
\tilde{A}_c && \tilde{B}_c && \tilde{C}_c \\
\tilde{A}_d && \tilde{B}_d && \tilde{C}_d
    \end{matrix}\right|~.~\,
\eeq
}
Again positive parameter $\lambda$ can be introduced to make  the variables $(A_a , A_b , A_c, A_d)$ in ~\eqref{susyconditions} at equivalent position. The set of solutions to the supersymmetry equations can be written in terms of $y_A$, $y_B$, $y_C$, and $y_D$.
And as follows, the D6-branes configurations that preserve four-dimensional ${\cal N}=1$ supersymmetry can be classified into three categories~\cite{Cvetic:2002pj}.

First, there are D6-branes whose wrapping numbers are the same as O6-planes (shown in Table \ref{orientifold1}). 
Among $A_a$, $B_a$, $C_a$ and $D_a$, where the subscript $a$ denotes any stack, there are one negative and three zeros, represented as $(-, 0, 0, 0)$.
This type of D6-branes carrying $USp(n)$ gauge symmetry can be identified with $USp(n)_1$, $USp(n)_2$, $USp(n)_3$ and $USp(n)_4$ groups according to which one of $A_a$, $B_a$, $C_a$ and $D_a$ takes zero value, respectively.

Second, this case corresponds to wrapping numbers that are all non-zero, leading to one positive number and three negative numbers $(+,-,-,-)$ in $A_a$, $B_a$, $C_a$ and $D_a$, which is called NZ-type D6-branes.

Third, if there is one and only one zero wrapping number, there will be two negative numbers and two zeros in $A_a$, $B_a$, $C_a$ and $D_a$ $(0,0,-,-)$ and this case is called Z-type D6-branes.
Since both Z- and NZ-type branes carry $U(n)$ gauge symmetry, they are sometimes called U-branes.
The combinations of wrapping numbers of NZ-type D6-branes can be classified into
\begin{eqnarray}
	A1: (-,-)\times(+,+)\times(+,+),~& A2:(-,+)\times(-,+)\times(-,+);\\
	B1: (+,-)\times(+,+)\times(+,+),~& B2:(+,+)\times(-,+)\times(-,+);\\
	C1: (+,+)\times(+,-)\times(+,+),~& C2:(-,+)\times(+,+)\times(-,+);\\
	D1: (+,+)\times(+,+)\times(+,-),~& D2:(-,+)\times(-,+)\times(+,+).
\end{eqnarray}
To conclude, $A_a$, $B_a$, $C_a$ and $D_a$ in (\ref{tadpole}) each corresponds to one of the types of U-branes.
This leads to 10 types of possible sign for ($A_a$, $B_a$, $C_a$, $D_a$), \emph{i.e.,} $(0,0,-1,-1)$, $(0,-1,0,-1)$, $(0,-1,-1,0)$, $(-1,0,0,-1)$, $(-1,0,-1,0)$, $(-1,-1,0,0)$, $(-1,-1,-1,1)$, $(-1,-1,1,-1)$, $(-1,1,-1,-1)$, $(1,-1,-1,-1)$, where the first six possible signs are from the Z-type of branes and the later four possible signs are from the NZ-type branes.

\subsubsection*{Tadpole Constraint}

Based on the possible sign of ($A_a$, $B_a$, $C_a$, $D_a$) , we consider the tadpole condition \eqref{tadpole}, where we have the constraints that the number of filler brane must be equal or greater than zero, such that 
\begin{eqnarray}
N^{(1)}= 4+2 A_a + A_b +A_c+A_d \geq 0 ,\\
N^{(2)}= 4+2 B_a + B_b +B_c+B_d \geq 0 ,\\
N^{(3)}= 4+2 C_a + C_b +C_c+C_d \geq 0 ,\\
N^{(4)}= 4+2 D_a + D_b +D_c+D_d \geq 0 .
\end{eqnarray}
The total 10 types of possible sign for ($A_a$, $B_a$, $C_a$, $D_a$) provides $10^4=10000 \to 1948$ choices by imposing the tadpole constraints. With $N^{(1)},N^{(2)},N^{(3)},N^{(4)} \geq 0$ satisfied simultaneously in the same system, the number of possible sign for ($A_a$, $B_a$, $C_a$, $D_a$) can be reduced to 1948 types. 
In the three-stack configuration this was $10^3=1000 \to 468$ types of possible signs with tadpole constraints considered. 
These constraints to ($A_a$, $B_a$, $C_a$, $D_a$) reduce the moduli space of the wrapping numbers essentially, and play a similar role as it was carefully discussed for the scanning bounds from tadpole and supersymmetry constraints.
in~\cite{Douglas:2003um, Gmeiner:2005vz,Loges:2022mao}.
It was shown that among the billion number of intersecting brane models there shall be countable number of MSSM-like D-brane models, which is consistent with the 33 independent classes of Pati-Salam models found in~\cite{He:2021gug} as well.

\subsubsection*{Three-generation Condition}

Inspired by the earlier investigations in~\cite{Douglas:2003um, Gmeiner:2005vz,He:2021gug, Loges:2022mao, Marchesano:2024gul}, we observe that under the tadpole and supersymmetry constraints the MSSM-like D-brane models are rather rare with a frequency of ``one in a billion''. Although it shall be expected that there is an upper bound for the range of the wrapping number $n, l$, the exact number is not obvious. Therefore, categorized by the 1948 type of possible signs for ($A_a$, $B_a$, $C_a$, $D_a$),  we choose to solve with the \emph{Three-generation Condition} as the first step in the deterministic algorithm. 

As it was proven by statics in~\cite{Douglas:2003um, Gmeiner:2005vz,Loges:2022mao}, solving the \emph{Three-generation Condition} as the first step will dramatically reduce the integer set of wrapping number when searching for MSSM-like D-brane models. And although it is hard to provide a strict mathematical proof with the upper bound of the wrapping number, it was statically proven in \cite{He:2021gug}, to satisfy the \emph{Three-generation Condition} with three family of chiral fermion, there appears a digital upper bound(5 in the three-stack of D-brane case) for the value of $n, l$ from the integer system. This can be easily seen from the products $\tilde{A}_{a}$ and $A_{a}$.
When extended three-generation conditions (\ref{3g}) under an extra group for $SU(2)_{R}$ is rewritten in terms of products of wrapping numbers (\ref{variables}), it becomes
\beq
I_{ab}+I_{ab'}=-\sum_{A,B,C,D}\tilde{A}_{a}A_{b}=3 ~,~
\label{3gPRb}
\eeq

\beq
I_{ac}+I_{ac'}+I_{ad}+I_{ad'}=-\sum_{A,B,C,D}(\tilde{A}_{a}A_{c}+\tilde{A}_{a}A_{d})=-3~.~
\label{3gPRcd}
\eeq
And in the case with $SU(2)_{L}$ extension, it becomes
\beq
I_{ab}+I_{ab'}+I_{ad}+I_{ad'}=-\sum_{A,B,C,D}(\tilde{A}_{a}A_{b}+\tilde{A}_{a}A_{d})=3 ~,~
\label{3gPLbd}
\eeq

\beq
I_{ac}+I_{ac'}=-\sum_{A,B,C,D}\tilde{A}_{a}A_{c}=-3~,~
\label{3gPLc}
\eeq
where $\sum_{A,B,C,D}\tilde{A}_{a}A_{b} = \tilde{A}_{a}A_{b}+\tilde{B}_{a}B_{b}+\tilde{C}_{a}C_{b}+\tilde{D}_{a}D_{b}$.

Based on these consideration, we have the strategy to complete the string landscape of MSSM-like D-brane models with $d$-stack extension as follows:

{\bf{Step one.}} Generate all the possible signs of ($A_a$, $B_a$, $C_a$, $D_a$) and constraint them with the tadpole condition 
$N^{(1)},N^{(2)},N^{(3)},N^{(4)} \geq 0$. This produce a 1948 sign combinations which produce a sign matrix that required by $N^{(1)},N^{(2)},N^{(3)},N^{(4)} \geq 0$ simultaneously. 

{\bf{Step two.}} In each ($A_a$, $B_a$, $C_a$, $D_a$) sign set of the 1948  sign matrix, we solve the three-generation condition by factorizing and enumerating. Because of the bound with absolute value 3, this will significantly reduce the scanning range of wrapping numbmer $n, l$. In other words, this is precisely dealing with the ``{\emph{one}}'' in ``{\emph{one in a billion}}''.

{\bf{Step three.}} In the second step, with the three-generation condition and wrapping number solved as much as possible, we have a significantly smaller integer set. In this step, we solve the supersymmetry equations with algebraic cofactor given in \eqref{4cofactor}. 

Note that there are four classes of cofactor solutions from the degeneracy of \eqref{4susy} as given in the discussion of {\bf Supersymmetry System}~\ref{secsusy}. Therefore, the moduli solution $x_A, x_B, x_C, x_D $ can be explicitly given by the wrapping numbers. The supersymmetry in-equality conditions~\eqref{susyconditions}  are also utilized to  further restrict the string landscape. 

Till here, the tadpole conditions, supersymmetry constraints, and three-generation condition for MSSM-like models are solved as a {\bf Diophantine equation system}, and thus the landscape of supersymmetric Pati-Salam models with $d$-stack extension is completed through the deterministic algorithm. 

\emph{It is worthwhile to emphasize that, distinct with the standard string landscape search, we divide the tadpole landscape into certain sign categories, and in each category directly solve the three generation condition as the first step to solve the Diophantine system.
 As was shown in~\cite{Douglas:2003um, Gmeiner:2005vz,Loges:2022mao}, the three-generation condition significantly constrain the moduli space and thus impose it at an earlier stage will make it more efficient to complete the MSSM-like string landscape. }

\subsection{Landscape of ${\cal N}=1$ supersymmetric Pati-Salam models with extra $d$-stack extension}

As the solution from the deterministic algorithm search, we have the full list of 14 independent classes of ${\cal N}=1$ supersymmetric Pati-Salam models with $d$-stack extension in $SU(2)_L$ and $SU(2)_R$ respectively in the Appendix, while the first torus to be tilted. 
In particular, there are in total 897024 models in the right‑handed $SU(2)_R$ extension and 694272 models in the left‑handed $SU(2)_L$ extension.

Recall that without $d$-stack extension, the standard ${\cal N}=1$ supersymmetric Pati-Salam landscape was completed with 202752 number of models that with 33 classes of gauge coupling relations~\cite{He:2021gug}.
In each class, the models are symmetrically related, such as by type I/II T-dualities, DSEP.
With the additional $d$-stack of D6-branes introduced to the $SU(2)_R$ and $SU(2)_L$ sectors, the symmetries of  ${\cal N}=1$ supersymmetric Pati-Salam models get further extended, and therefore we can extend the landscape with an additional $d$-stack of D6-branes. 
While $d$-stack of D6-branes introduced, the symmetry in this configuration is discussed as follows.

Firstly, recall that the models are all constructed on $\Tbb^{6} = \Tbb^{2} \times \Tbb^{2} \times \Tbb^{2}$ with a two-torus to be tilted.
These three tori are all at democratic positions, and thus we have three equal options of the tilted one of three, which accounts for a symmetry factor 3.
One can certainly choose the first or the third torus to be tilted without changing the gauge couplings relation.

Secondly,  the three-generation condition in the $SU(2)_R$ extension \eqref{3g} and \eqref{3gL} in the $SU(2)_L$ extension indicate that $I_{ab}$ and $I_{ab'}$ are in equal positions, which implies that equivalent models can be produced by performing $b\leftrightarrow b'$. The symmetry transformations $c\leftrightarrow c'$ and $d\leftrightarrow d'$ can be derived by following the procedure used for $b\leftrightarrow b'$. This provides a symmetry factor with $2\times 2\times 2=8$\footnote{
At the first sight, {\emph{e.g.},} in the $SU(2)_R$ extension, one could also consider the symmetry interchange of the $c$- and $d$-stacks of D6-branes as they both contribute to the $SU(2)_R$ gauge with the kinetic function  $f_R=f_c+f_d$. However, when the c- and $d$-stack of branes contribute to the three generation with an unequal value(which is often the case), this symmetry shall not be counted generically.}.

Thirdly, recall that in three-stack brane configuration, the DSEP~\cite{Cvetic:2004ui} shows two models are equivalent if, for any stack of brane, the wrapping numbers on any two tori have opposite signs, while those on the remaining torus are identical.
In other words, the transformation $(n_x^i,l_x^i)\to(-n_x^i,-l_x^i)$ together with $(n_x^j,l_x^j)\to(-n_x^j,-l_x^j)$ leads to an equivalent model, where $x$ can be any one of $a,~b,~c$ and $d$, and $i\neq j$ are chosen from $1,2$ and $3$. Consider the original model and the alternative 3 variations, there are $(1+3)^3=4^3$ symmetric models counted in~\cite{He:2021gug}. 
The base $(1+3)$ is derived as follows: ``1" represents the case without applying DSEP, while ``3" corresponds to the number of combinations for selecting two tori (out of three) to change the wrapping numbers, given by $C_3^2=3$. 
With $d$-stack extension, one might naively think that the DSEP symmetry factor shall be generalized to  $(1+3)^4=4^4$. However, this is improper recalling that the DSEP shall not change anything about the D-brane model except interchanging among the four pairs of 
wrapping numbers  products $(A,\tilde{A})$, $(B,\tilde{B})$, $(C,\tilde{C})$
and $(D,\tilde{D})$~\cite{Cvetic:2004ui}. However, the $d$-stack extension combined with the $b$/$c$-stack contribute to the three-generation condition together as in \eqref{3gPRcd} and \eqref{3gPLbd}. Therefore, their contribution under DSEP symmetry shall be counted together and therefore it still contributes a $(1+3)^3=4^3$ symmetry.

Fourthly, as  T-duality transformation fully depends on the three tori. The T-duality and the extended T-duality shall remain the same.  In which, type I T-duality provides a transformation on the $i$-th and the $j$-th tori and the wrapping numbers are governed by the transformation rule as below~\cite{Cvetic:2004ui}
\beq
(n_x^i,l_x^i)\to (-l_x^i,n_x^i)~\,{\rm and}~\,
(n_x^j,l_x^j)\to (l_x^j,-n_x^j)~,~
\eeq
where $x$ runs over all $a,b,c$ and $d$ stacks simultaneously, and  contributes a symmetry factor with $1+3=4$.

Moreover, if we combine DSEP with type I T-duality, it will double the number of such models and provide a symmetry factor of extra $2$.
This so-called extended type I T-duality as a variation of type II T-duality works on the rectangular $i$- and $j$-th tori as transformation by
\beq
(n_x^i,l_x^i)\to (l_x^j,-n_x^j)~\,{\rm and}~\,
(n_x^j,l_x^j)\to (-l_x^i,n_x^i)~.~
\label{extT}
\eeq
In total, considering all the solid symmetries given above, there are 
\beq
3\times 2\times 2\times 2\times 2\times 4^3\times 4\times 2
=12288=6144\times 2
\eeq
basic equivalent models with the same gauge coupling relation for each class of models\footnote{
Note that models with the same $\alpha$ and $\beta$ but different $\gamma$, such as Model 4 in this work and Model 10 in~\cite{He:2021gug}, are not considered equivalent, since it is impossible to relate these two models via any symmetry and duality discussed above.}.
This explains the ``Factor/6144'' of 2 in the Multiplicity of equivalent models table \ref{R} and \ref{L}. Because for particular model(such as Class 2 in the $SU(2)_R$ configuration), extra symmetry of $b$- and $d$-stack of brane exchanging provides a same gauge coupling relation while $g_b=g_d$, extra symmetry may arise and contributes with greater ``Factor/6144''.
On the other hand, because of the extra $d$-stack extension, the three-family of chiral fermions has a more complicated source from brane intersection, and certain hidden degeneracies exist when the symmetry permutation provides with the same models.

\subsection{String Landscape with other model building}

For the connection with other String landscape searches, let's briefly review the earlier work first. 
Considering generic RR tadpole cancellation conditions and supersymmetry condition to be satisfied, the entire ensemble of distinct vacua for $\Tbb^6/(\Zbb_2\times\Zbb_2)$ configurations were exactly exhausted in~\cite{Loges:2022mao}, with a total count to be ${\cal O}(10^{11})$.

By fully solving the Diophantine system constituting of supersymmetry constraint, tadpole condition and MSSM gauge requirements with three chiral fermion families, the landscape of ${\cal N}=1$ supersymmetric Pati-Salam models was completed in~\cite{He:2021gug} with 33 independent(\emph{i.e.} gauge invariant) particle physics vacua aer obtained. This precisely matches the search of~\cite{Loges:2022mao} and \cite{Gmeiner:2005vz, Douglas:2003um} that while Standard Model features are required, the frequency of MSSM-like gauge group with three generations is around ``{\emph{one in a billion}}''.

In particular,  numerical hints towards an analytic proof of finiteness of the considered space of models(under supersymmetry and tadpole constraints) were given in ~\cite{Gmeiner:2005vz} as well. This analysis was also given consistently with a finiteness conjecture in~\cite{Douglas:2003um}.
It was also analyzed in~\cite{Blumenhagen:2004xx} that the tadpole
equations admit only a \emph{finite} number of supersymmetric
configurations for a fixed choice of rational complex structures. 
This was also observed in~\cite{He:2021gug} that the linear inequalities($N^{(1)},N^{(2)},N^{(3)},N^{(4)} \geq 0$) given by the tadpole conditions over the real numbers corresponds to a polyhedron. There are well-known algorithms (e.g. Fourier-Motzkin elimination) that can determine the volume of the polyhedron.
By cross checking with the supersymmetry constraint, especially the inequality constraint, we find the system is finite. Actually, the supersymmetry inequality condition can be checked through the requirements of supersymmetry equality condition with positive complex structures moduli.

With $d$-stack extension to the supersymmetric Pati-Salam models, we implemented the deterministic algorithm~\cite{He:2021gug} with extra $d$-stack. By generating all the possible signs for the wrapping number products ($A_a$, $B_a$, $C_a$, $D_a$) as the tadpole system's coefficients, we reduce the rough possible sign from $10^4=10000$ to $1948$ within a finite space. 
Recall that three generation condition of MSSM-like gauge requirements is a much stronger condition as ``one in a billion'' than the tadpole and supersymmetry conditions. Therefore, by imposing the three generation condition, as soon as the possible signs of ($A_a$, $B_a$, $C_a$, $D_a$) is generated, will helps to zoom into the MSSM-like brane models more efficiently. As an advantage, this will also give the bound of the wrapping number as the three-family fermion is bound with 3 for the sum of the intersection number. Here the algorithm does not rely on a
pre-estimated wrapping number bound but given from the fact of fulfilling three-generation condition.

On the other hand, it would also be interesting to generalize the brane intersecting landscape search algorithm to other string phenomenology model building, as reviewed in~\cite{Marchesano:2024gul}.
In different corners of the string landscape, the construction of Standard Model–like vacua typically involves scanning large sets of consistent compactifications and imposing increasingly restrictive phenomenological conditions.
However, if we utilize the strategy of imposing the three-family fermion condition at an earlier stage in the landscape searching, the moduli space of the vacua will be significantly smaller and accelerate the process of completing the MSSM-like particle vacua dramatically.
For example, in the context of heterotic compactifications, systematic searches over large classes of internal gauge bundles have been carried out, leading to a statistical picture in which realistic spectra appear in a small subset of the allowed configurations~\cite{Lebedev:2006kn}.
Similar features also arise in F-theory constructions, where scans over large ensembles of elliptically fibered Calabi–Yau fourfolds indicate that obtaining the Standard Model gauge group and chiral spectrum requires imposing a series of highly nontrivial constraints on the geometry and flux data, as discussed for instance in~\cite{Anderson:2012yf,Cvetic:2019gnh}.
It emerges from these studies that the requirement of realistic phenomenology also drastically reduces the number of Standard Model vacua. 

Besides the complete search of Pati-Salam models in~\cite{He:2021gug}, our investigation completed with 14 independent classes of $d$-stack brane extension(with one tilted torus) to the Pati-Salam models provides another illustration of ``{\emph {one in a billion}}'' in the Standard model landscape search. 
Starting from the vast space of consistent intersecting brane configurations, the simultaneous requirements of realizing the extended Pati–Salam gauge symmetry, three chiral families, and accommodating the necessary Higgs sector, strongly restrict the allowed number of solutions/vacua.
The complete search of 14 independent classes of models obtained in the $d$-stack extension landscape is again consistent with the order of general results observed across different string model buildings.

\section{Conclusions and outlook}

Realizing that the chiral fermion multiplets can arise from various intersection of branes, we introduced extra $d$-stack of D6-branes to the $SU(2)_R$ and $SU(2)_L$ gauge constructions.
The standard Pati-Salam gauge symmetry $SU(4)_C\times SU(2)_L\times SU(2)_R$ is then extended to
$SU(4)_C\times SU(2)_L\times SU(2)_{R_1}\times SU(2)_{R_2}$ and $SU(4)_C\times SU(2)_{L_1}\times SU(2)_{R}\times SU(2)_{L_2}$, while $d$-stack of branes contributes to the symmetries $SU(2)_{R_2}$ and $SU(2)_{L_2}$, respectively.
Based on the duality and symmetry of supersymmetric Pati-Salam models, the extra $d$-stack of branes introduced an extra factor of 2 from $d\leftrightarrow d'$ to the variation of symmetric transformation compared to the complete search of the landscape in~\cite{He:2021gug}.

Based on the model building with extra $d$-stack of D6-branes, we complete the landscape of Pati-Salam models with gauge symmetry $SU(4)_C\times SU(2)_L\times SU(2)_{R_1}\times SU(2)_{R_2}$ with 14 independent classes of models. This is represented and discussed with Model 1-4 (see Table \ref{model1} to \ref{model4}) with $g_a^2=\alpha g_b^2=\beta \frac53 g_Y^2=\gamma[\pi e^{\phi_4}]$. 
In addition, in the left handed extension, we also completed the landscape with 14 independent classes of models with gauge symmetry $SU(4)_C\times SU(2)_{L_1}\times SU(2)_{R}\times SU(2)_{L_2}$, represented by Model 7-10 (see Table \ref{model7} to \ref{model10}) with $g_a^2=\alpha g_L^2=\beta \frac53 g_Y^2=\gamma[\pi e^{\phi_4}]$.
In precise, the multiplicity of 
equivalent models with the same gauge coupling relation for each  class of models are presented in the Appendix in detail.
These equivalent models are all related via the permutation of tori, the transformation $b\leftrightarrow b'$, $c\leftrightarrow c'$, $d\leftrightarrow d'$ and $c\leftrightarrow d$/$b\leftrightarrow d$(in certain classes), DSEP and T-duality. This greatly extended the landscape of supersymmetric Pati-Salam models.

From the spectra of the representative models, we found that introducing extra $d$-stack of D6-branes tends to reduce the number of filler branes that carry $USp(N)$ gauge symmetries. For Model 1, 2, 7 and 8, only two $USp$ groups are observed in the hidden sector;
while no $USp$ group appears in Model 3, 4, 9 and 10, which is rare compared with all the previous studies of three-family supersymmetric Pati-Salam models.
However, as an advantage, it provides more origins to arise Higgs/Higgs-like multiplets that would be interesting phenomenologically. 
In particular, for Model 1 to 4, Higgs multiplets arise from the intersection of $b$- and $c$-stacks of branes (or its $\Zbb_2$ image) and  $b$- and $d$-stacks of branes (or its $\Zbb_2$ image) appear simultaneously;
while for Model 7 to 10, Higgs multiplets arise from the intersection of $b$- and $c$-stacks of branes (or its $\Zbb_2$ image) and intersection of $d$- and $c$-stacks of branes (or its $\Zbb_2$ image).

At the string/GUT gauge unification scale, with $d$-stack of D6-branes introduced, the gauge coupling relations become
$
    g_a^2=k_2 g_b^2=k_Y g_Y^2=g_U^2,~~
    g_a^2=k_2 g_L^2=k_Y g_Y^2=g_U^2,~
$
respectively.
The coefficients $k_2$ and $k_Y$ rely on different models and will be utilized to achieve gauge coupling unification.
Among these models, we find that there are newly constructed models that share the same gauge coupling ratio as standard supersymmetric Pati-Salam models in~\cite{He:2021gug} but with a different value of gauge coupling, such as Model 4 in this work and Model 10 in~\cite{He:2021gug}. 
This shows that a ``degeneracy'' behavior of gauge coupling ratio exists, which also appears in Model 10 and Model 18 in~\cite{He:2021gug}.
It is also inspiring that these models with different values of $\gamma$ (such as in (\ref{gabY}) or (\ref{gabYL})) but the same gauge coupling ratios can be unified at the target scale with different numbers of additional particles from various origins, for string scale and GUT scale unification, respectively.

For future investigation, it would be interesting to further study the ``degeneracy'' class with the same gauge coupling relation ratio, from both the phenomenology and the model building aspects. 
Moreover, consider that the $d$-stack extended landscape is completed with both  14 independent classes of models from the right-handed and left-handed $SU(2)$ gauge sector respectively, yet with different number of models in each class. This indicates non-trivial extra symmetries exist besides the basic 12288 symmetries explained by simple permutation, DSEP and T-duality when extra stack of brane introduced.

In the landscape search, we observed that it will be more efficient to search the MSSM-like brane models by solving the three-generation condition first in the brane intersecting Diophantine system.  This strategy will dramatically reduce the  moduli space of particle physics vacua, and leads to complete the brane intersecting landscape as in~\cite{He:2021gug}.
Therefore, we found it would be interesting to study other brane intersecting extensions,  and string landscapes utilizing the deterministic algorithm, especially while the Standard Model gauge group and three-family chiral spectrum are required.

\begin{acknowledgments}
   
	TL is supported in part by the National Key Research and Development Program of China Grant No. 2020YFC2201504, by the Projects No. 11875062, No. 11947302, No. 12047503, and No. 12275333 supported by the National Natural Science Foundation of China, by the Key Research Program of the Chinese Academy of Sciences, Grant No. XDPB15, by the Scientific Instrument Developing Project of the Chinese Academy of Sciences, Grant No. YJKYYQ20190049, by the International Partnership Program of Chinese Academy of Sciences for Grand Challenges, Grant No. 112311KYSB20210012, and by the Henan Province Outstanding Foreign Scientist Studio Project, No.GZS2025008.	 
    RS is supported by the Fundamental Research Funds for the Central Universities Grant No.~E4EQ0102X2.

\end{acknowledgments}

\bibliography{reference}

\appendix

\section{Complete list of independent classes of models with $SU(2)_R$ extension}

\begin{table}[h!]
\scriptsize
\caption{D6-brane configurations and intersection numbers of Class 1, and its MSSM gauge coupling relation is $g_a^2 = \frac{2}{3}g_b^2 = \frac{4}{3}g_R^2 = \frac{6}{5} (\frac{5}{3}g_Y^2) = \frac{16}{45}14^{\frac{3}{4}}\pi e^{\phi_4}$, with $12288\,(6144\times 2)$ dual models.}
\label{Rmodel7}
\begin{center}
\begin{tabular}{|c||c|c||c|c|c|c|c|c|c|c|c|c|}
\hline
\rm{Model} 1 & \multicolumn{12}{c|}{$U(4)\times U(2)_L\times U(2)_{R_1}\times U(2)_{R_2}\times USp(2)^2 $}\\
\hline \hline
\rm{stack} & $N$ & $(n^1,l^1)\times(n^2,l^2)\times(n^3,l^3)$ & $n_{\Ysymm}$ & $n_{\Yasymm}$ & $b$ & $b'$ & $c$ & $c'$ & $d$ & $d'$ & 1 & 2\\
\hline
$a$ & 8 & $(-1, -1) \times (-1, -1) \times (-1, -1)$ & 0 & -4  & 0 & 3 & 0 & 1 & 0 & -4 & 1 & -1\\
$b$ & 4 & $(-1, -1) \times (4, -1) \times (-1, 0)$ & -3 & 3  & - & - & 0 & -1 & 0 & 3 & 0 & 4\\
$c$ & 4 & $(-1, -1) \times (-1, 0) \times (2, -1)$ & -1 & 1  & - & - & - & - & 0 & 1 & 0 & 2\\
$d$ & 4 & $(-1, -1) \times (-1, -1) \times (-1, -1)$ & 0 & -4  & - & - & - & - & - & - & 1 & -1\\
\hline
1 & 2 & $(2, 0) \times (1, 0) \times (1, 0)$ & \multicolumn{10}{c|}{$x_A = 7x_B = \frac{7}{4}x_C = \frac{7}{2}x_D$}\\
2 & 2 & $(2, 0) \times (0, -1) \times (0, 1)$ & \multicolumn{10}{c|}{$\beta^g_1=-3$, $\beta^g_2=3$}\\
 & & & \multicolumn{10}{c|}{$\chi_1=\sqrt{\frac{7}{2}}$, $\chi_2=\sqrt{14}$, $\chi_3=\sqrt{\frac{7}{2}}$}\\
\hline
\end{tabular}
\end{center}
\end{table}

\begin{table}[h!]
\scriptsize
\caption{D6-brane configurations and intersection numbers of Class 2, and its MSSM gauge coupling relation is $g_a^2 = g_b^2 = \frac{4}{3}g_R^2 = \frac{6}{5} (\frac{5}{3}g_Y^2) = 4\sqrt{\frac{2}{3}}\pi e^{\phi_4}$, with $24576\,(6144\times 4)$ dual models.}
\label{Rmodel2}
\begin{center}
\begin{tabular}{|c||c|c||c|c|c|c|c|c|c|c|c|c|}
\hline
\rm{Model} 2 & \multicolumn{12}{c|}{$U(4)\times U(2)_L\times U(2)_{R_1}\times U(2)_{R_2}\times USp(2)^2 $}\\
\hline \hline
\rm{stack} & $N$ & $(n^1,l^1)\times(n^2,l^2)\times(n^3,l^3)$ & $n_{\Ysymm}$ & $n_{\Yasymm}$ & $b$ & $b'$ & $c$ & $c'$ & $d$ & $d'$ & 1 & 3\\
\hline
$a$ & 8 & $(-1, 1) \times (-1, 0) \times (1, 1)$ & 0 & 0  & 3 & 0 & 0 & 0 & -3 & 0 & 0 & 0\\
$b$ & 4 & $(-1, -1) \times (-1, 3) \times (1, 0)$ & 2 & -2  & - & - & 0 & 1 & 0 & 0 & 0 & -3\\
$c$ & 4 & $(-1, -1) \times (0, -1) \times (-1, -1)$ & 0 & 0  & - & - & - & - & 0 & -1 & 1 & -1\\
$d$ & 4 & $(-1, -1) \times (-1, -3) \times (0, -1)$ & -2 & 2  & - & - & - & - & - & - & 3 & 0\\
\hline
1 & 2 & $(2, 0) \times (1, 0) \times (1, 0)$ & \multicolumn{10}{c|}{$x_A = \frac{1}{3}x_B = 1x_C = \frac{1}{3}x_D$}\\
3 & 2 & $(0, -2) \times (1, 0) \times (0, 1)$ & \multicolumn{10}{c|}{$\beta^g_1=-2$, $\beta^g_3=-2$}\\
 & & & \multicolumn{10}{c|}{$\chi_1=2$, $\chi_2=\frac{1}{3}$, $\chi_3=1$}\\
\hline
\end{tabular}
\end{center}
\end{table}

\begin{table}[h!]
\scriptsize
\caption{D6-brane configurations and intersection numbers of Class 3, and its MSSM gauge coupling relation is $g_a^2 = 10g_b^2 = 16g_R^2 = 10 (\frac{5}{3}g_Y^2) = 8\sqrt{3}\pi e^{\phi_4}$, with $24576\,(6144\times 4)$ dual models.}
\label{Rmodel5}
\begin{center}
\begin{tabular}{|c||c|c||c|c|c|c|c|c|c|c|}
\hline
\rm{Model} 3 & \multicolumn{10}{c|}{$U(4)\times U(2)_L\times U(2)_{R_1}\times U(2)_{R_2} $}\\
\hline \hline
\rm{stack} & $N$ & $(n^1,l^1)\times(n^2,l^2)\times(n^3,l^3)$ & $n_{\Ysymm}$ & $n_{\Yasymm}$ & $b$ & $b'$ & $c$ & $c'$ & $d$ & $d'$ \\
\hline
$a$ & 8 & $(-1, 1) \times (-1, 0) \times (1, 1)$ & 0 & 0  & 2 & 1 & 0 & 0 & -2 & -1\\
$b$ & 4 & $(-1, -3) \times (-2, 1) \times (1, 0)$ & -5 & 5  & - & - & 2 & 4 & 16 & 0\\
$c$ & 4 & $(-1, -1) \times (0, -1) \times (-1, -1)$ & 0 & 0  & - & - & - & - & 2 & -4\\
$d$ & 4 & $(-3, -1) \times (-2, -1) \times (0, -1)$ & 5 & -5  & - & - & - & - & - & -\\
\hline
 & & & \multicolumn{8}{c|}{$x_A = 6x_B = x_C = 6x_D$}\\
 & & & \multicolumn{8}{c|}{$\chi_1=2$, $\chi_2=6$, $\chi_3=1$}\\
\hline
\end{tabular}
\end{center}
\end{table}

\begin{table}[h!]
\scriptsize
\caption{D6-brane configurations and intersection numbers of Class 4, and its MSSM gauge coupling relation is $g_a^2 = 2g_b^2 = 4g_R^2 = \frac{14}{5} (\frac{5}{3}g_Y^2) = 4\times3^{\frac{1}{4}}\pi e^{\phi_4}$, with $24576\,(6144\times 4)$ dual models.}
\label{Rmodel8}
\begin{center}
\begin{tabular}{|c||c|c||c|c|c|c|c|c|c|c|}
\hline
\rm{Model} 4 & \multicolumn{10}{c|}{$U(4)\times U(2)_L\times U(2)_{R_1}\times U(2)_{R_2} $}\\
\hline \hline
\rm{stack} & $N$ & $(n^1,l^1)\times(n^2,l^2)\times(n^3,l^3)$ & $n_{\Ysymm}$ & $n_{\Yasymm}$ & $b$ & $b'$ & $c$ & $c'$ & $d$ & $d'$ \\
\hline
$a$ & 8 & $(-1, 1) \times (-1, 0) \times (1, 1)$ & 0 & 0  & 3 & 0 & 0 & -2 & -1 & 0\\
$b$ & 4 & $(-1, -1) \times (-2, 3) \times (1, 0)$ & 1 & -1  & - & - & 2 & 4 & 0 & -4\\
$c$ & 4 & $(-3, -1) \times (0, -1) \times (-1, -1)$ & 2 & -2  & - & - & - & - & -2 & -4\\
$d$ & 4 & $(-1, -1) \times (-2, -1) \times (0, -1)$ & 1 & -1  & - & - & - & - & - & -\\
\hline
 & & & \multicolumn{8}{c|}{$x_A = 2x_B = 3x_C = 2x_D$}\\
 & & & \multicolumn{8}{c|}{$\chi_1=2\sqrt{3}$, $\chi_2=\frac{2}{\sqrt{3}}$, $\chi_3=\sqrt{3}$}\\
\hline
\end{tabular}
\end{center}
\end{table}

\begin{table}[h!]
\scriptsize
\caption{D6-brane configurations and intersection numbers of Class 5, and its MSSM gauge coupling relation is $g_a^2 = 2g_b^2 = \frac{8}{3}g_R^2 = 2 (\frac{5}{3}g_Y^2) = \frac{8}{\sqrt{3}}\pi e^{\phi_4}$, with $24576\,(6144\times 4)$ dual models.}
\label{Rmodel9}
\begin{center}
\begin{tabular}{|c||c|c||c|c|c|c|c|c|c|c|}
\hline
\rm{Model} 5 & \multicolumn{10}{c|}{$U(4)\times U(2)_L\times U(2)_{R_1}\times U(2)_{R_2} $}\\
\hline \hline
\rm{stack} & $N$ & $(n^1,l^1)\times(n^2,l^2)\times(n^3,l^3)$ & $n_{\Ysymm}$ & $n_{\Yasymm}$ & $b$ & $b'$ & $c$ & $c'$ & $d$ & $d'$ \\
\hline
$a$ & 8 & $(-1, 1) \times (-1, 0) \times (1, 1)$ & 0 & 0  & 3 & 0 & 0 & 0 & -3 & 0\\
$b$ & 4 & $(-1, -1) \times (-2, 3) \times (1, 0)$ & 1 & -1  & - & - & 0 & 2 & 0 & 0\\
$c$ & 4 & $(-1, -1) \times (0, -1) \times (-1, -1)$ & 0 & 0  & - & - & - & - & 0 & -2\\
$d$ & 4 & $(-1, -1) \times (-2, -3) \times (0, -1)$ & -1 & 1  & - & - & - & - & - & -\\
\hline
 & & & \multicolumn{8}{c|}{$x_A = \frac{2}{3}x_B = x_C = \frac{2}{3}x_D$}\\
 & & & \multicolumn{8}{c|}{$\chi_1=2$, $\chi_2=\frac{2}{3}$, $\chi_3=1$}\\
\hline
\end{tabular}
\end{center}
\end{table}

\begin{table}[h!]
\scriptsize
\caption{D6-brane configurations and intersection numbers of Class 6, and its MSSM gauge coupling relation is $g_a^2 = 3g_b^2 = 4g_R^2 = \frac{14}{5} (\frac{5}{3}g_Y^2) = 2\sqrt{2}\times3^{\frac{3}{4}}\pi e^{\phi_4}$, with $24576\,(6144\times 4)$ dual models.}
\label{Rmodel11}
\begin{center}
\begin{tabular}{|c||c|c||c|c|c|c|c|c|c|c|c|c|}
\hline
\rm{Model} 6 & \multicolumn{12}{c|}{$U(4)\times U(2)_L\times U(2)_{R_1}\times U(2)_{R_2}\times USp(2)^2 $}\\
\hline \hline
\rm{stack} & $N$ & $(n^1,l^1)\times(n^2,l^2)\times(n^3,l^3)$ & $n_{\Ysymm}$ & $n_{\Yasymm}$ & $b$ & $b'$ & $c$ & $c'$ & $d$ & $d'$ & 1 & 3\\
\hline
$a$ & 8 & $(-1, 1) \times (-1, 0) \times (1, 1)$ & 0 & 0  & 2 & 1 & -2 & 0 & -1 & 0 & 0 & 0\\
$b$ & 4 & $(-1, -3) \times (-1, 1) \times (1, 0)$ & -2 & 2  & - & - & 3 & 6 & 2 & 0 & 0 & -1\\
$c$ & 4 & $(-1, -1) \times (0, -1) \times (-1, -3)$ & -2 & 2  & - & - & - & - & 0 & -1 & 3 & -1\\
$d$ & 4 & $(-1, -1) \times (-1, -1) \times (0, -1)$ & 0 & 0  & - & - & - & - & - & - & 1 & 0\\
\hline
1 & 2 & $(2, 0) \times (1, 0) \times (1, 0)$ & \multicolumn{10}{c|}{$x_A = x_B = \frac{1}{3}x_C = x_D$}\\
3 & 2 & $(0, -2) \times (1, 0) \times (0, 1)$ & \multicolumn{10}{c|}{$\beta^g_1=-2$, $\beta^g_3=-4$}\\
 & & & \multicolumn{10}{c|}{$\chi_1=\frac{2}{\sqrt{3}}$, $\chi_2=\sqrt{3}$, $\chi_3=\frac{1}{\sqrt{3}}$}\\
\hline
\end{tabular}
\end{center}
\end{table}

\begin{table}[h!]
\scriptsize
\caption{D6-brane configurations and intersection numbers of Class 7, and its MSSM gauge coupling relation is $g_a^2 = 5g_b^2 = 8g_R^2 = \frac{26}{5} (\frac{5}{3}g_Y^2) = 4\sqrt{6}\pi e^{\phi_4}$, with $24576\,(6144\times 4)$ dual models.4}
\label{Rmodel13}
\begin{center}
\begin{tabular}{|c||c|c||c|c|c|c|c|c|c|c|c|c|}
\hline
\rm{Model} 7 & \multicolumn{12}{c|}{$U(4)\times U(2)_L\times U(2)_{R_1}\times U(2)_{R_2}\times USp(2)^2 $}\\
\hline \hline
\rm{stack} & $N$ & $(n^1,l^1)\times(n^2,l^2)\times(n^3,l^3)$ & $n_{\Ysymm}$ & $n_{\Yasymm}$ & $b$ & $b'$ & $c$ & $c'$ & $d$ & $d'$ & 1 & 3\\
\hline
$a$ & 8 & $(-1, 1) \times (-1, 0) \times (1, 1)$ & 0 & 0  & 2 & 1 & 0 & 0 & -2 & -1 & 0 & 0\\
$b$ & 4 & $(-1, -3) \times (-1, 1) \times (1, 0)$ & -2 & 2  & - & - & 1 & 2 & 8 & 0 & 0 & -1\\
$c$ & 4 & $(-1, -1) \times (0, -1) \times (-1, -1)$ & 0 & 0  & - & - & - & - & 1 & -2 & 1 & -1\\
$d$ & 4 & $(-3, -1) \times (-1, -1) \times (0, -1)$ & 2 & -2  & - & - & - & - & - & - & 1 & 0\\
\hline
1 & 2 & $(2, 0) \times (1, 0) \times (1, 0)$ & \multicolumn{10}{c|}{$x_A = 3x_B = x_C = 3x_D$}\\
3 & 2 & $(0, -2) \times (1, 0) \times (0, 1)$ & \multicolumn{10}{c|}{$\beta^g_1=-4$, $\beta^g_3=-4$}\\
 & & & \multicolumn{10}{c|}{$\chi_1=2$, $\chi_2=3$, $\chi_3=1$}\\
\hline
\end{tabular}
\end{center}
\end{table}

\begin{table}[h!]
\scriptsize
\caption{D6-brane configurations and intersection numbers of Class 8, and its MSSM gauge coupling relation is $g_a^2 = 6g_b^2 = 8g_R^2 = \frac{26}{5} (\frac{5}{3}g_Y^2) = 4\times3^{\frac{3}{4}}\pi e^{\phi_4}$, with $24576\,(6144\times 4)$ dual models.}
\label{Rmodel14}
\begin{center}
\begin{tabular}{|c||c|c||c|c|c|c|c|c|c|c|}
\hline
\rm{Model} 8 & \multicolumn{10}{c|}{$U(4)\times U(2)_L\times U(2)_{R_1}\times U(2)_{R_2} $}\\
\hline \hline
\rm{stack} & $N$ & $(n^1,l^1)\times(n^2,l^2)\times(n^3,l^3)$ & $n_{\Ysymm}$ & $n_{\Yasymm}$ & $b$ & $b'$ & $c$ & $c'$ & $d$ & $d'$ \\
\hline
$a$ & 8 & $(-1, 1) \times (-1, 0) \times (1, 1)$ & 0 & 0  & 2 & 1 & -2 & 0 & -1 & 0\\
$b$ & 4 & $(-1, -3) \times (-2, 1) \times (1, 0)$ & -5 & 5  & - & - & 6 & 12 & 4 & 0\\
$c$ & 4 & $(-1, -1) \times (0, -1) \times (-1, -3)$ & -2 & 2  & - & - & - & - & 0 & -2\\
$d$ & 4 & $(-1, -1) \times (-2, -1) \times (0, -1)$ & 1 & -1  & - & - & - & - & - & -\\
\hline
 & & & \multicolumn{8}{c|}{$x_A = 2x_B = \frac{1}{3}x_C = 2x_D$}\\
 & & & \multicolumn{8}{c|}{$\chi_1=\frac{2}{\sqrt{3}}$, $\chi_2=\sqrt{3}$, $\chi_3=\frac{1}{\sqrt{3}}$}\\
\hline
\end{tabular}
\end{center}
\end{table}

\begin{table}[h!]
\scriptsize
\caption{D6-brane configurations and intersection numbers of Class 9, and its MSSM gauge coupling relation is $g_a^2 = \frac{5}{3}g_b^2 = \frac{4}{3}g_R^2 = \frac{6}{5} (\frac{5}{3}g_Y^2) = 4\sqrt{\frac{2}{3}}\pi e^{\phi_4}$, with $61440\,(6144\times 10)$ dual models.}
\label{Rmodel12}
\begin{center}
\begin{tabular}{|c||c|c||c|c|c|c|c|c|c|c|c|c|}
\hline
\rm{Model} 9 & \multicolumn{12}{c|}{$U(4)\times U(2)_L\times U(2)_{R_1}\times U(2)_{R_2}\times USp(2)^2 $}\\
\hline \hline
\rm{stack} & $N$ & $(n^1,l^1)\times(n^2,l^2)\times(n^3,l^3)$ & $n_{\Ysymm}$ & $n_{\Yasymm}$ & $b$ & $b'$ & $c$ & $c'$ & $d$ & $d'$ & 1 & 3\\
\hline
$a$ & 8 & $(-1, -1) \times (0, -1) \times (-1, -1)$ & 0 & 0  & 0 & 3 & 0 & -3 & 0 & 0 & 1 & -1\\
$b$ & 4 & $(-1, -1) \times (-1, 1) \times (2, 1)$ & 2 & 6  & - & - & 0 & -4 & 0 & -1 & 1 & -2\\
$c$ & 4 & $(-1, -1) \times (-3, -1) \times (0, -1)$ & 2 & -2  & - & - & - & - & 0 & 1 & 1 & 0\\
$d$ & 4 & $(-1, -1) \times (-1, 0) \times (1, -1)$ & 0 & 0  & - & - & - & - & - & - & 0 & 0\\
\hline
1 & 2 & $(2, 0) \times (1, 0) \times (1, 0)$ & \multicolumn{10}{c|}{$x_A = 3x_B = x_C = 3x_D$}\\
3 & 2 & $(0, -2) \times (1, 0) \times (0, 1)$ & \multicolumn{10}{c|}{$\beta^g_1=-2$, $\beta^g_3=-2$}\\
 & & & \multicolumn{10}{c|}{$\chi_1=2$, $\chi_2=3$, $\chi_3=1$}\\
\hline
\end{tabular}
\end{center}
\end{table}

\begin{table}[h!]
\scriptsize
\caption{D6-brane configurations and intersection numbers of Class 10, and its MSSM gauge coupling relation is $g_a^2 = \frac{1}{3}g_b^2 = 2g_R^2 = \frac{8}{5} (\frac{5}{3}g_Y^2) = \frac{16}{5\sqrt{3}}\pi e^{\phi_4}$, with $73728\,(6144\times 12)$ dual models.}
\label{Rmodel6}
\begin{center}
\begin{tabular}{|c||c|c||c|c|c|c|c|c|c|c|}
\hline
\rm{Model} 10 & \multicolumn{10}{c|}{$U(4)\times U(2)_L\times U(2)_{R_1}\times U(2)_{R_2} $}\\
\hline \hline
\rm{stack} & $N$ & $(n^1,l^1)\times(n^2,l^2)\times(n^3,l^3)$ & $n_{\Ysymm}$ & $n_{\Yasymm}$ & $b$ & $b'$ & $c$ & $c'$ & $d$ & $d'$ \\
\hline
$a$ & 8 & $(-1, -1) \times (0, -1) \times (-1, -1)$ & 0 & 0  & 0 & 3 & 0 & -3 & 0 & 0\\
$b$ & 4 & $(-1, -1) \times (-1, 0) \times (4, -1)$ & -3 & 3  & - & - & 0 & 4 & 0 & 3\\
$c$ & 4 & $(-1, -1) \times (-3, -1) \times (0, -1)$ & 2 & -2  & - & - & - & - & 0 & -3\\
$d$ & 4 & $(-1, -1) \times (0, -1) \times (-1, -1)$ & 0 & 0  & - & - & - & - & - & -\\
\hline
 & & & \multicolumn{8}{c|}{$x_A = \frac{1}{12}x_B = 1x_C = \frac{1}{3}x_D$}\\
 & & & \multicolumn{8}{c|}{$\chi_1=1$, $\chi_2=\frac{1}{12}$, $\chi_3=3$}\\
\hline
\end{tabular}
\end{center}
\end{table}

\begin{table}[h!]
\scriptsize
\caption{D6-brane configurations and intersection numbers of Class 11, and its MSSM gauge coupling relation is $g_a^2 = 3g_b^2 = \frac{18}{5}g_R^2 = \frac{64}{25} (\frac{5}{3}g_Y^2) = \frac{16}{5}\sqrt{3}\pi e^{\phi_4}$, with $73728\,(6144\times 12)$ dual models.}
\label{Rmodel10}
\begin{center}
\begin{tabular}{|c||c|c||c|c|c|c|c|c|c|c|}
\hline
\rm{Model} 11 & \multicolumn{10}{c|}{$U(4)\times U(2)_L\times U(2)_{R_1}\times U(2)_{R_2} $}\\
\hline \hline
\rm{stack} & $N$ & $(n^1,l^1)\times(n^2,l^2)\times(n^3,l^3)$ & $n_{\Ysymm}$ & $n_{\Yasymm}$ & $b$ & $b'$ & $c$ & $c'$ & $d$ & $d'$ \\
\hline
$a$ & 8 & $(-1, -1) \times (0, -1) \times (-1, -1)$ & 0 & 0  & 0 & 3 & -1 & -2 & 0 & 0\\
$b$ & 4 & $(-1, -1) \times (-1, 0) \times (4, -1)$ & -3 & 3  & - & - & -4 & 8 & 0 & 3\\
$c$ & 4 & $(-1, -3) \times (-1, -1) \times (0, -1)$ & -2 & 2  & - & - & - & - & 1 & -2\\
$d$ & 4 & $(-1, -1) \times (0, -1) \times (-1, -1)$ & 0 & 0  & - & - & - & - & - & -\\
\hline
 & & & \multicolumn{8}{c|}{$x_A = \frac{4}{3}x_B = 1x_C = \frac{1}{3}x_D$}\\
 & & & \multicolumn{8}{c|}{$\chi_1=1$, $\chi_2=\frac{2}{3}$, $\chi_3=2$}\\
\hline
\end{tabular}
\end{center}
\end{table}

\begin{table}[h!]
\scriptsize
\caption{D6-brane configurations and intersection numbers of class 12, and its MSSM gauge coupling relation is $g_a^2 = g_b^2 = 2g_R^2 = \frac{8}{5} (\frac{5}{3}g_Y^2) = 4\sqrt{\frac{2}{3} } \pi e^{\phi_4}$, with $98304\,(6144\times 16)$ dual models.}
\label{Rmodel1}
\begin{center}
\begin{tabular}{|c||c|c||c|c|c|c|c|c|c|c|c|c|}
\hline
\rm{Model} 12 & \multicolumn{12}{c|}{$U(4)\times U(2)_L\times U(2)_{R_1}\times U(2)_{R_2}\times USp(2)^2 $}\\
\hline \hline
\rm{stack} & $N$ & $(n^1,l^1)\times(n^2,l^2)\times(n^3,l^3)$ & $n_{\Ysymm}$ & $n_{\Yasymm}$ & $b$ & $b'$ & $c$ & $c'$ & $d$ & $d'$ & 1 & 3\\
\hline
$a$ & 8 & $(1, -1) \times (0, -1) \times (-1, 1)$ & 0 & 0  & 3 & 0 & 0 & 0 & -3 & 0 & -1 & 1\\
$b$ & 4 & $(-1, -1) \times (-3, -1) \times (0, -1)$ & 2 & -2  & - & - & 0 & 1 & 0 & -4 & 1 & 0\\
$c$ & 4 & $(-1, -1) \times (-1, 0) \times (1, -1)$ & 0 & 0  & - & - & - & - & 0 & -1 & 0 & 0\\
$d$ & 4 & $(-1, -1) \times (-1, 1) \times (2, 1)$ & 2 & 6  & - & - & - & - & - & - & 1 & -2\\
\hline
1 & 2 & $(2, 0) \times (1, 0) \times (1, 0)$ & \multicolumn{10}{c|}{$x_A = 3x_B = x_C = 3x_D$}\\
3 & 2 & $(0, -2) \times (1, 0) \times (0, 1)$ & \multicolumn{10}{c|}{$\beta^g_1=-2$, $\beta^g_3=-2$}\\
 & & & \multicolumn{10}{c|}{$\chi_1=2$, $\chi_2=3$, $\chi_3=1$}\\
\hline
\end{tabular}
\end{center}
\end{table}

\begin{table}[h!]
\scriptsize
\caption{D6-brane configurations and intersection numbers of Class 13, and its MSSM gauge coupling relation is $g_a^2 = \frac{5}{3}g_b^2 = \frac{5}{3}g_R^2 = \frac{7}{5} (\frac{5}{3}g_Y^2) = 4\sqrt{\frac{2}{3}}\pi e^{\phi_4}$, with $184320\,(6144\times 30)$ dual models.}
\label{Rmodel4}
\begin{center}
\begin{tabular}{|c||c|c||c|c|c|c|c|c|c|c|}
\hline
\rm{Model} 13 & \multicolumn{10}{c|}{$U(4)\times U(2)_L\times U(2)_{R_1}\times U(2)_{R_2} $}\\
\hline \hline
\rm{stack} & $N$ & $(n^1,l^1)\times(n^2,l^2)\times(n^3,l^3)$ & $n_{\Ysymm}$ & $n_{\Yasymm}$ & $b$ & $b'$ & $c$ & $c'$ & $d$ & $d'$ \\
\hline
$a$ & 8 & $(-1, -1) \times (0, -1) \times (-1, -1)$ & 0 & 0  & 0 & 3 & 0 & -3 & 0 & 0\\
$b$ & 4 & $(-1, -1) \times (-1, 1) \times (2, 1)$ & 2 & 6  & - & - & 0 & -4 & 0 & -2\\
$c$ & 4 & $(-1, -1) \times (-3, -1) \times (0, -1)$ & 2 & -2  & - & - & - & - & 0 & 2\\
$d$ & 4 & $(-2, -2) \times (-1, 0) \times (1, -1)$ & 0 & 0  & - & - & - & - & - & -\\
\hline
 & & & \multicolumn{8}{c|}{$x_A = 3x_B = x_C = 3x_D$}\\
 & & & \multicolumn{8}{c|}{$\chi_1=2$, $\chi_2=3$, $\chi_3=1$}\\
\hline
\end{tabular}
\end{center}
\end{table}

\begin{table}[h!]
\scriptsize
\caption{D6-brane configurations and intersection numbers of Class 14, and its MSSM gauge coupling relation is $g_a^2 = g_b^2 = \frac{7}{3}g_R^2 = \frac{9}{5} (\frac{5}{3}g_Y^2) = 4\sqrt{\frac{2}{3}}\pi e^{\phi_4}$, with $221184\,(6144\times 36)$ dual models.36}
\label{Rmodel3}
\begin{center}
\begin{tabular}{|c||c|c||c|c|c|c|c|c|c|c|}
\hline
\rm{Model} 14 & \multicolumn{10}{c|}{$U(4)\times U(2)_L\times U(2)_{R_1}\times U(2)_{R_2} $}\\
\hline \hline
\rm{stack} & $N$ & $(n^1,l^1)\times(n^2,l^2)\times(n^3,l^3)$ & $n_{\Ysymm}$ & $n_{\Yasymm}$ & $b$ & $b'$ & $c$ & $c'$ & $d$ & $d'$ \\
\hline
$a$ & 8 & $(1, -1) \times (0, -1) \times (-1, 1)$ & 0 & 0  & 3 & 0 & 0 & 0 & -3 & 0\\
$b$ & 4 & $(-1, -1) \times (-3, -1) \times (0, -1)$ & 2 & -2  & - & - & 0 & 2 & 0 & -4\\
$c$ & 4 & $(-2, -2) \times (-1, 0) \times (1, -1)$ & 0 & 0  & - & - & - & - & 0 & -2\\
$d$ & 4 & $(-1, -1) \times (-1, 1) \times (2, 1)$ & 2 & 6  & - & - & - & - & - & -\\
\hline
 & & & \multicolumn{8}{c|}{$x_A = 3x_B = x_C = 3x_D$}\\
 & & & \multicolumn{8}{c|}{$\chi_1=2$, $\chi_2=3$, $\chi_3=1$}\\
\hline
\end{tabular}
\end{center}
\end{table}

\FloatBarrier

The full list of 14 independent classes of models with extra $d$-stack in the $SU(2)_R$ gauge extension can be summarized into Table~\ref{R} as follows.
\begin{table}[h]
\centering
\caption{Multiplicity of equivalent models in each gauge coupling class of the right‑handed $SU(2)_R$ extension. The basic factor $6144$ originates from torus permutations, orientifold image, T‑duality and the D6‑brane sign equivalence principle.}
\label{R}
\begin{tabular}{|c||c||c||c|}
\hline
Class & Total models & Factor $/6144$ & Gauge coupling relation \\
\hline
\hline
1 & 12288 & 2 & - \\
2 & 24576 & 4 & $g_b = g_d$ \\
3 & 24576 & 4 &  $g_b = g_d$ \\
4 & 24576 & 4 & $g_b = g_c = g_d$ \\
5 & 24576 & 4 & $g_b = g_d$ \\
6 & 24576 & 4 & $g_b = g_c$ \\
7 & 24576 & 4 & $g_b = g_d$ \\
8 & 24576 & 4 & $g_b = g_c$ \\
9 & 61440 & 10 & - \\
10 & 73728 & 12&  $g_c = g_d$ \\
11 & 73728 & 12 & - \\
12 & 98304 & 16 & $g_b = g_c = g_d$ \\
13 & 184320 & 30 & - \\
14 & 221184 & 36 & - \\
\hline
\end{tabular}
\end{table}


\section{Complete list of independent classes of models with $SU(2)_L$ extension}

\begin{table}[h!]
\scriptsize
\caption{D6-brane configurations and intersection numbers of Class 1, and its MSSM gauge coupling relation is $g_a^2 = 16g_L^2 = 10g_c^2 = \frac{32}{5} (\frac{5}{3}g_Y^2) = 8\sqrt{3}\pi e^{\phi_4}$, with $12288\,(6144\times 2)$ dual models.}
\label{Lmodel2}
\begin{center}
\begin{tabular}{|c||c|c||c|c|c|c|c|c|c|c|}
\hline
\rm{Model} 1 & \multicolumn{10}{c|}{$U(4)\times U(2)_{L_1}\times U(2)_R\times U(2)_{L_2} $}\\
\hline \hline
\rm{stack} & $N$ & $(n^1,l^1)\times(n^2,l^2)\times(n^3,l^3)$ & $n_{\Ysymm}$ & $n_{\Yasymm}$ & $b$ & $b'$ & $c$ & $c'$ & $d$ & $d'$ \\
\hline
$a$ & 8 & $(-1, -1) \times (-1, -1) \times (0, -1)$ & 0 & 0  & 0 & 0 & -1 & -2 & 1 & 2\\
$b$ & 4 & $(-1, -1) \times (-1, 1) \times (1, 0)$ & 0 & 0  & - & - & -2 & 4 & 2 & -4\\
$c$ & 4 & $(-1, -3) \times (0, -1) \times (-1, -2)$ & -5 & 5  & - & - & - & - & 16 & 0\\
$d$ & 4 & $(-3, -1) \times (-1, 0) \times (1, -2)$ & 5 & -5  & - & - & - & - & - & -\\
\hline
 & & & \multicolumn{8}{c|}{$x_A = \frac{1}{6}x_B = \frac{1}{6}x_C = 1x_D$}\\
 & & & \multicolumn{8}{c|}{$\chi_1=2$, $\chi_2=1$, $\chi_3=\frac{1}{6}$}\\
\hline
\end{tabular}
\end{center}
\end{table}
\FloatBarrier

\begin{table}[h!]
\scriptsize
\caption{D6-brane configurations and intersection numbers of Class 2, and its MSSM gauge coupling relation is $g_a^2 = \frac{4}{3}g_L^2 = g_c^2 = (\frac{5}{3}g_Y^2) = 4\sqrt{\frac{2}{3}}\pi e^{\phi_4}$, with $12288\,(6144\times 2)$ dual models.}
\label{Lmodel6}
\begin{center}
\begin{tabular}{|c||c|c||c|c|c|c|c|c|c|c|c|c|}
\hline
\rm{Model} 2 & \multicolumn{12}{c|}{$U(4)\times U(2)_{L_1}\times U(2)_R\times U(2)_{L_2}\times USp(2)^2 $}\\
\hline \hline
\rm{stack} & $N$ & $(n^1,l^1)\times(n^2,l^2)\times(n^3,l^3)$ & $n_{\Ysymm}$ & $n_{\Yasymm}$ & $b$ & $b'$ & $c$ & $c'$ & $d$ & $d'$ & 2 & 3\\
\hline
$a$ & 8 & $(-1, -1) \times (-1, -1) \times (0, -1)$ & 0 & 0  & 0 & 0 & 0 & -3 & 0 & 3 & 0 & 0\\
$b$ & 4 & $(-1, -1) \times (-1, 1) \times (1, 0)$ & 0 & 0  & - & - & 0 & 1 & 0 & -1 & 1 & -1\\
$c$ & 4 & $(-1, -1) \times (0, -1) \times (-3, -1)$ & 2 & -2  & - & - & - & - & 0 & 0 & 0 & -3\\
$d$ & 4 & $(-1, -1) \times (-1, 0) \times (3, -1)$ & -2 & 2  & - & - & - & - & - & - & 3 & 0\\
\hline
2 & 2 & $(2, 0) \times (0, -1) \times (0, 1)$ & \multicolumn{10}{c|}{$x_A = 3x_B = 3x_C = x_D$}\\
3 & 2 & $(0, -2) \times (1, 0) \times (0, 1)$ & \multicolumn{10}{c|}{$\beta^g_2=-2$, $\beta^g_3=-2$}\\
 & & & \multicolumn{10}{c|}{$\chi_1=2$, $\chi_2=1$, $\chi_3=3$}\\
\hline
\end{tabular}
\end{center}
\end{table}

\begin{table}[h!]
\scriptsize
\caption{D6-brane configurations and intersection numbers of Class 3, and its MSSM gauge coupling relation is $g_a^2 = 4g_L^2 = 2g_c^2 = \frac{8}{5} (\frac{5}{3}g_Y^2) = 4\times3^{\frac{1}{4}}\pi e^{\phi_4}$, with $12288\,(6144\times 2)$ dual models.}
\label{Lmodel9}
\begin{center}
\begin{tabular}{|c||c|c||c|c|c|c|c|c|c|c|}
\hline
\rm{Model} 3 & \multicolumn{10}{c|}{$U(4)\times U(2)_{L_1}\times U(2)_R\times U(2)_{L_2} $}\\
\hline \hline
\rm{stack} & $N$ & $(n^1,l^1)\times(n^2,l^2)\times(n^3,l^3)$ & $n_{\Ysymm}$ & $n_{\Yasymm}$ & $b$ & $b'$ & $c$ & $c'$ & $d$ & $d'$ \\
\hline
$a$ & 8 & $(-1, -1) \times (-1, -1) \times (0, -1)$ & 0 & 0  & 2 & 0 & 0 & -3 & 0 & 1\\
$b$ & 4 & $(-3, -1) \times (-1, 1) \times (1, 0)$ & 2 & -2  & - & - & -2 & 4 & -2 & -4\\
$c$ & 4 & $(-1, -1) \times (0, -1) \times (-3, -2)$ & 1 & -1  & - & - & - & - & 0 & -4\\
$d$ & 4 & $(-1, -1) \times (-1, 0) \times (1, -2)$ & 1 & -1  & - & - & - & - & - & -\\
\hline
 & & & \multicolumn{8}{c|}{$x_A = \frac{1}{2}x_B = \frac{3}{2}x_C = x_D$}\\
 & & & \multicolumn{8}{c|}{$\chi_1=2\sqrt{3}$, $\chi_2=\frac{1}{\sqrt{3}}$, $\chi_3=\frac{\sqrt{3}}{2}$}\\
\hline
\end{tabular}
\end{center}
\end{table}

\begin{table}[h!]
\scriptsize
\caption{D6-brane configurations and intersection numbers of Class 4, and its MSSM gauge coupling relation is $g_a^2 = 4g_L^2 = 3g_c^2 = \frac{11}{5} (\frac{5}{3}g_Y^2) = 2\sqrt{2}\times3^{\frac{3}{4}}\pi e^{\phi_4}$, with $12288\,(6144\times 2)$ dual models.}
\label{Lmodel10}
\begin{center}
\begin{tabular}{|c||c|c||c|c|c|c|c|c|c|c|c|c|}
\hline
\rm{Model} 4 & \multicolumn{12}{c|}{$U(4)\times U(2)_{L_1}\times U(2)_R\times U(2)_{L_2}\times USp(2)^2 $}\\
\hline \hline
\rm{stack} & $N$ & $(n^1,l^1)\times(n^2,l^2)\times(n^3,l^3)$ & $n_{\Ysymm}$ & $n_{\Yasymm}$ & $b$ & $b'$ & $c$ & $c'$ & $d$ & $d'$ & 2 & 3\\
\hline
$a$ & 8 & $(-1, -1) \times (-1, -1) \times (0, -1)$ & 0 & 0  & 0 & 2 & -1 & -2 & 0 & 1 & 0 & 0\\
$b$ & 4 & $(-1, -1) \times (-3, 1) \times (1, 0)$ & -2 & 2  & - & - & -3 & 6 & 0 & -1 & 3 & -1\\
$c$ & 4 & $(-1, -3) \times (0, -1) \times (-1, -1)$ & -2 & 2  & - & - & - & - & 2 & 0 & 0 & -1\\
$d$ & 4 & $(-1, -1) \times (-1, 0) \times (1, -1)$ & 0 & 0  & - & - & - & - & - & - & 1 & 0\\
\hline
2 & 2 & $(2, 0) \times (0, -1) \times (0, 1)$ & \multicolumn{10}{c|}{$x_A = x_B = \frac{1}{3}x_C = x_D$}\\
3 & 2 & $(0, -2) \times (1, 0) \times (0, 1)$ & \multicolumn{10}{c|}{$\beta^g_2=-2$, $\beta^g_3=-4$}\\
 & & & \multicolumn{10}{c|}{$\chi_1=\frac{2}{\sqrt{3}}$, $\chi_2=\sqrt{3}$, $\chi_3=\frac{1}{\sqrt{3}}$}\\
\hline
\end{tabular}
\end{center}
\end{table}

\begin{table}[h!]
\scriptsize
\caption{D6-brane configurations and intersection numbers of Class 5, and its MSSM gauge coupling relation is $g_a^2 = \frac{8}{3}g_L^2 = 2g_c^2 = \frac{8}{5} (\frac{5}{3}g_Y^2) = \frac{8}{\sqrt{3}}\pi e^{\phi_4}$, with $12288\,(6144\times 2)$ dual models.}
\label{Lmodel12}
\begin{center}
\begin{tabular}{|c||c|c||c|c|c|c|c|c|c|c|}
\hline
\rm{Model} 5 & \multicolumn{10}{c|}{$U(4)\times U(2)_{L_1}\times U(2)_R\times U(2)_{L_2} $}\\
\hline \hline
\rm{stack} & $N$ & $(n^1,l^1)\times(n^2,l^2)\times(n^3,l^3)$ & $n_{\Ysymm}$ & $n_{\Yasymm}$ & $b$ & $b'$ & $c$ & $c'$ & $d$ & $d'$ \\
\hline
$a$ & 8 & $(-1, -1) \times (-1, -1) \times (0, -1)$ & 0 & 0  & 0 & 0 & 0 & -3 & 0 & 3\\
$b$ & 4 & $(-1, -1) \times (-1, 1) \times (1, 0)$ & 0 & 0  & - & - & 0 & 2 & 0 & -2\\
$c$ & 4 & $(-1, -1) \times (0, -1) \times (-3, -2)$ & 1 & -1  & - & - & - & - & 0 & 0\\
$d$ & 4 & $(-1, -1) \times (-1, 0) \times (3, -2)$ & -1 & 1  & - & - & - & - & - & -\\
\hline
 & & & \multicolumn{8}{c|}{$x_A = \frac{3}{2}x_B = \frac{3}{2}x_C = x_D$}\\
 & & & \multicolumn{8}{c|}{$\chi_1=2$, $\chi_2=1$, $\chi_3=\frac{3}{2}$}\\
\hline
\end{tabular}
\end{center}
\end{table}

\begin{table}[h!]
\scriptsize
\caption{D6-brane configurations and intersection numbers of Class 6, and its MSSM gauge coupling relation is $g_a^2 = 8g_L^2 = 5g_c^2 = \frac{17}{5} (\frac{5}{3}g_Y^2) = 4\sqrt{6}\pi e^{\phi_4}$, with $12288\,(6144\times 2)$ dual models.}
\label{Lmodel13}
\begin{center}
\begin{tabular}{|c||c|c||c|c|c|c|c|c|c|c|c|c|}
\hline
\rm{Model} 6 & \multicolumn{12}{c|}{$U(4)\times U(2)_{L_1}\times U(2)_R\times U(2)_{L_2}\times USp(2)^2 $}\\
\hline \hline
\rm{stack} & $N$ & $(n^1,l^1)\times(n^2,l^2)\times(n^3,l^3)$ & $n_{\Ysymm}$ & $n_{\Yasymm}$ & $b$ & $b'$ & $c$ & $c'$ & $d$ & $d'$ & 2 & 3\\
\hline
$a$ & 8 & $(-1, -1) \times (-1, -1) \times (0, -1)$ & 0 & 0  & 0 & 0 & -1 & -2 & 1 & 2 & 0 & 0\\
$b$ & 4 & $(-1, -1) \times (-1, 1) \times (1, 0)$ & 0 & 0  & - & - & -1 & 2 & 1 & -2 & 1 & -1\\
$c$ & 4 & $(-1, -3) \times (0, -1) \times (-1, -1)$ & -2 & 2  & - & - & - & - & 8 & 0 & 0 & -1\\
$d$ & 4 & $(-3, -1) \times (-1, 0) \times (1, -1)$ & 2 & -2  & - & - & - & - & - & - & 1 & 0\\
\hline
2 & 2 & $(2, 0) \times (0, -1) \times (0, 1)$ & \multicolumn{10}{c|}{$x_A = \frac{1}{3}x_B = \frac{1}{3}x_C = x_D$}\\
3 & 2 & $(0, -2) \times (1, 0) \times (0, 1)$ & \multicolumn{10}{c|}{$\beta^g_2=-4$, $\beta^g_3=-4$}\\
 & & & \multicolumn{10}{c|}{$\chi_1=2$, $\chi_2=1$, $\chi_3=\frac{1}{3}$}\\
\hline
\end{tabular}
\end{center}
\end{table}

\begin{table}[h!]
\scriptsize
\caption{D6-brane configurations and intersection numbers of Class 7, and its MSSM gauge coupling relation is $g_a^2 = 8g_L^2 = 6g_R^2 = 4 (\frac{5}{3}g_Y^2) = 4\times3^{\frac{3}{4}}\pi e^{\phi_4}$, with $12288\,(6144\times 2)$ dual models.}
\label{Lmodel14}
\begin{center}
\begin{tabular}{|c||c|c||c|c|c|c|c|c|c|c|}
\hline
\rm{Model} 7 & \multicolumn{10}{c|}{$U(4)\times U(2)_{L_1}\times U(2)_R\times U(2)_{L_2} $}\\
\hline \hline
\rm{stack} & $N$ & $(n^1,l^1)\times(n^2,l^2)\times(n^3,l^3)$ & $n_{\Ysymm}$ & $n_{\Yasymm}$ & $b$ & $b'$ & $c$ & $c'$ & $d$ & $d'$ \\
\hline
$a$ & 8 & $(-1, -1) \times (-1, -1) \times (0, -1)$ & 0 & 0  & 0 & 2 & -1 & -2 & 0 & 1\\
$b$ & 4 & $(-1, -1) \times (-3, 1) \times (1, 0)$ & -2 & 2  & - & - & -6 & 12 & 0 & -2\\
$c$ & 4 & $(-1, -3) \times (0, -1) \times (-1, -2)$ & -5 & 5  & - & - & - & - & 4 & 0\\
$d$ & 4 & $(-1, -1) \times (-1, 0) \times (1, -2)$ & 1 & -1  & - & - & - & - & - & -\\
\hline
 & & & \multicolumn{8}{c|}{$x_A = \frac{1}{2}x_B = \frac{1}{6}x_C = x_D$}\\
 & & & \multicolumn{8}{c|}{$\chi_1=\frac{2}{\sqrt{3}}$, $\chi_2=\sqrt{3}$, $\chi_3=\frac{1}{2\sqrt{3}}$}\\
\hline
\end{tabular}
\end{center}
\end{table}

\begin{table}[h!]
\scriptsize
\caption{D6-brane configurations and intersection numbers of Class 8, and its MSSM gauge coupling relation is $g_a^2 = \frac{4}{3}g_L^2 = \frac{2}{3}g_c^2 = \frac{4}{5} (\frac{5}{3}g_Y^2) = \frac{16}{45}14^{\frac{3}{4}}\pi e^{\phi_4}$, with $30720\,(6144\times 5)$ dual models.}
\label{Lmodel7}
\begin{center}
\begin{tabular}{|c||c|c||c|c|c|c|c|c|c|c|c|c|}
\hline
\rm{Model} 8 & \multicolumn{12}{c|}{$U(4)\times U(2)_{L_1}\times U(2)_R\times U(2)_{L_2}\times USp(2)^2 $}\\
\hline \hline
\rm{stack} & $N$ & $(n^1,l^1)\times(n^2,l^2)\times(n^3,l^3)$ & $n_{\Ysymm}$ & $n_{\Yasymm}$ & $b$ & $b'$ & $c$ & $c'$ & $d$ & $d'$ & 1 & 2\\
\hline
$a$ & 8 & $(-1, 1) \times (1, -1) \times (1, -1)$ & 0 & 4  & -1 & 0 & -3 & 0 & 4 & 0 & -1 & 1\\
$b$ & 4 & $(-1, -1) \times (-2, 1) \times (1, 0)$ & -1 & 1  & - & - & 0 & -1 & 0 & 1 & 0 & 2\\
$c$ & 4 & $(-1, -1) \times (-1, 0) \times (4, -1)$ & -3 & 3  & - & - & - & - & 0 & 3 & 0 & 4\\
$d$ & 4 & $(-1, -1) \times (-1, -1) \times (-1, -1)$ & 0 & -4  & - & - & - & - & - & - & 1 & -1\\
\hline
1 & 2 & $(2, 0) \times (1, 0) \times (1, 0)$ & \multicolumn{10}{c|}{$x_A = 7x_B = \frac{7}{2}x_C = \frac{7}{4}x_D$}\\
2 & 2 & $(2, 0) \times (0, -1) \times (0, 1)$ & \multicolumn{10}{c|}{$\beta^g_1=-3$, $\beta^g_2=3$}\\
 & & & \multicolumn{10}{c|}{$\chi_1=\sqrt{\frac{7}{2}}$, $\chi_2=\sqrt{\frac{7}{2}}$, $\chi_3=\sqrt{14}$}\\
\hline
\end{tabular}
\end{center}
\end{table}

\begin{table}[h!]
\scriptsize
\caption{D6-brane configurations and intersection numbers of Class 9, and its MSSM gauge coupling relation is $g_a^2 = \frac{18}{5}g_L^2 = 3g_c^2 = \frac{11}{5} (\frac{5}{3}g_Y^2) = \frac{16}{5}\sqrt{3}\pi e^{\phi_4}$, with $61440\,(6144\times 10)$ dual models.}
\label{Lmodel3}
\begin{center}
\begin{tabular}{|c||c|c||c|c|c|c|c|c|c|c|}
\hline
\rm{Model} 9 & \multicolumn{10}{c|}{$U(4)\times U(2)_{L_1}\times U(2)_R\times U(2)_{L_2} $}\\
\hline \hline
\rm{stack} & $N$ & $(n^1,l^1)\times(n^2,l^2)\times(n^3,l^3)$ & $n_{\Ysymm}$ & $n_{\Yasymm}$ & $b$ & $b'$ & $c$ & $c'$ & $d$ & $d'$ \\
\hline
$a$ & 8 & $(1, -1) \times (0, -1) \times (-1, 1)$ & 0 & 0  & 2 & 1 & -3 & 0 & 0 & 0\\
$b$ & 4 & $(-1, -3) \times (-1, -1) \times (0, -1)$ & -2 & 2  & - & - & 4 & 8 & 1 & -2\\
$c$ & 4 & $(-1, -1) \times (-1, 0) \times (4, -1)$ & -3 & 3  & - & - & - & - & 0 & 3\\
$d$ & 4 & $(-1, -1) \times (0, -1) \times (-1, -1)$ & 0 & 0  & - & - & - & - & - & -\\
\hline
 & & & \multicolumn{8}{c|}{$x_A = \frac{4}{3}x_B = x_C = \frac{1}{3}x_D$}\\
 & & & \multicolumn{8}{c|}{$\chi_1=1$, $\chi_2=\frac{2}{3}$, $\chi_3=2$}\\
\hline
\end{tabular}
\end{center}
\end{table}

\begin{table}[h!]
\scriptsize
\caption{D6-brane configurations and intersection numbers of Class 10, and its MSSM gauge coupling relation is $g_a^2 = 2g_L^2 = \frac{1}{3}g_c^2 = \frac{3}{5} (\frac{5}{3}g_Y^2) = \frac{16}{5\sqrt{3}}\pi e^{\phi_4}$, with $61440\,(6144\times 10)$ dual models.}
\label{Lmodel4}
\begin{center}
\begin{tabular}{|c||c|c||c|c|c|c|c|c|c|c|}
\hline
\rm{Model} 10 & \multicolumn{10}{c|}{$U(4)\times U(2)_{L_1}\times U(2)_R\times U(2)_{L_2} $}\\
\hline \hline
\rm{stack} & $N$ & $(n^1,l^1)\times(n^2,l^2)\times(n^3,l^3)$ & $n_{\Ysymm}$ & $n_{\Yasymm}$ & $b$ & $b'$ & $c$ & $c'$ & $d$ & $d'$ \\
\hline
$a$ & 8 & $(1, -1) \times (0, -1) \times (-1, 1)$ & 0 & 0  & 3 & 0 & -3 & 0 & 0 & 0\\
$b$ & 4 & $(-1, -1) \times (-3, -1) \times (0, -1)$ & 2 & -2  & - & - & 0 & 4 & 0 & -3\\
$c$ & 4 & $(-1, -1) \times (-1, 0) \times (4, -1)$ & -3 & 3  & - & - & - & - & 0 & 3\\
$d$ & 4 & $(-1, -1) \times (0, -1) \times (-1, -1)$ & 0 & 0  & - & - & - & - & - & -\\
\hline
 & & & \multicolumn{8}{c|}{$x_A = 12x_B = x_C = 3x_D$}\\
 & & & \multicolumn{8}{c|}{$\chi_1=1$, $\chi_2=6$, $\chi_3=2$}\\
\hline
\end{tabular}
\end{center}
\end{table}

\begin{table}[h!]
\scriptsize
\caption{D6-brane configurations and intersection numbers of Class 11, and its MSSM gauge coupling relation is $g_a^2 = 2g_L^2 = g_c^2 = (\frac{5}{3}g_Y^2) = 4\sqrt{\frac{2}{3}}\pi e^{\phi_4}$, with $61440\,(6144\times 10)$ dual models.}
\label{Lmodel5}
\begin{center}
\begin{tabular}{|c||c|c||c|c|c|c|c|c|c|c|c|c|}
\hline
\rm{Model} 11 & \multicolumn{12}{c|}{$U(4)\times U(2)_{L_1}\times U(2)_R\times U(2)_{L_2}\times USp(2)^2 $}\\
\hline \hline
\rm{stack} & $N$ & $(n^1,l^1)\times(n^2,l^2)\times(n^3,l^3)$ & $n_{\Ysymm}$ & $n_{\Yasymm}$ & $b$ & $b'$ & $c$ & $c'$ & $d$ & $d'$ & 1 & 3\\
\hline
$a$ & 8 & $(-1, -1) \times (0, -1) \times (-1, -1)$ & 0 & 0  & 0 & 0 & 0 & -3 & 0 & 3 & 1 & -1\\
$b$ & 4 & $(-1, -1) \times (-1, 0) \times (1, -1)$ & 0 & 0  & - & - & 0 & 1 & 0 & -1 & 0 & 0\\
$c$ & 4 & $(-1, -1) \times (-3, -1) \times (0, -1)$ & 2 & -2  & - & - & - & - & 0 & -4 & 1 & 0\\
$d$ & 4 & $(-1, -1) \times (-1, 1) \times (2, 1)$ & 2 & 6  & - & - & - & - & - & - & 1 & -2\\
\hline
1 & 2 & $(2, 0) \times (1, 0) \times (1, 0)$ & \multicolumn{10}{c|}{$x_A = 3x_B = x_C = 3x_D$}\\
3 & 2 & $(0, -2) \times (1, 0) \times (0, 1)$ & \multicolumn{10}{c|}{$\beta^g_1=-2$, $\beta^g_3=-2$}\\
 & & & \multicolumn{10}{c|}{$\chi_1=2$, $\chi_2=3$, $\chi_3=1$}\\
\hline
\end{tabular}
\end{center}
\end{table}

\begin{table}[h!]
\scriptsize
\caption{D6-brane configurations and intersection numbers of Class 12, and its MSSM gauge coupling relation is $g_a^2 = \frac{4}{3}g_L^2 = \frac{5}{3}g_c^2 = \frac{7}{5} (\frac{5}{3}g_Y^2) = 4\sqrt{\frac{2}{3}}\pi e^{\phi_4}$, with $61440\,(6144\times 10)$ dual models.}
\label{Lmodel8}
\begin{center}
\begin{tabular}{|c||c|c||c|c|c|c|c|c|c|c|c|c|}
\hline
\rm{Model} 12 & \multicolumn{12}{c|}{$U(4)\times U(2)_{L_1}\times U(2)_R\times U(2)_{L_2}\times USp(2)^2 $}\\
\hline \hline
\rm{stack} & $N$ & $(n^1,l^1)\times(n^2,l^2)\times(n^3,l^3)$ & $n_{\Ysymm}$ & $n_{\Yasymm}$ & $b$ & $b'$ & $c$ & $c'$ & $d$ & $d'$ & 1 & 3\\
\hline
$a$ & 8 & $(1, -1) \times (0, -1) \times (-1, 1)$ & 0 & 0  & 3 & 0 & -3 & 0 & 0 & 0 & -1 & 1\\
$b$ & 4 & $(-1, -1) \times (-3, -1) \times (0, -1)$ & 2 & -2  & - & - & 0 & -4 & 0 & 1 & 1 & 0\\
$c$ & 4 & $(-1, -1) \times (-1, 1) \times (2, 1)$ & 2 & 6  & - & - & - & - & 0 & -1 & 1 & -2\\
$d$ & 4 & $(-1, -1) \times (-1, 0) \times (1, -1)$ & 0 & 0  & - & - & - & - & - & - & 0 & 0\\
\hline
1 & 2 & $(2, 0) \times (1, 0) \times (1, 0)$ & \multicolumn{10}{c|}{$x_A = 3x_B = x_C = 3x_D$}\\
3 & 2 & $(0, -2) \times (1, 0) \times (0, 1)$ & \multicolumn{10}{c|}{$\beta^g_1=-2$, $\beta^g_3=-2$}\\
 & & & \multicolumn{10}{c|}{$\chi_1=2$, $\chi_2=3$, $\chi_3=1$}\\
\hline
\end{tabular}
\end{center}
\end{table}

\begin{table}[h!]
\scriptsize
\caption{D6-brane configurations and intersection numbers of Class 13, and its MSSM gauge coupling relation is $g_a^2 = \frac{7}{3}g_L^2 =g_c^2  = (\frac{5}{3}g_Y^2) = 4\sqrt{\frac{2}{3}}\pi e^{\phi_4}$, with $147456\,(6144\times 24)$ dual models.}
\label{Lmodel11}
\begin{center}
\begin{tabular}{|c||c|c||c|c|c|c|c|c|c|c|}
\hline
\rm{Model} 13 & \multicolumn{10}{c|}{$U(4)\times U(2)_{L_1}\times U(2)_R\times U(2)_{L_2} $}\\
\hline \hline
\rm{stack} & $N$ & $(n^1,l^1)\times(n^2,l^2)\times(n^3,l^3)$ & $n_{\Ysymm}$ & $n_{\Yasymm}$ & $b$ & $b'$ & $c$ & $c'$ & $d$ & $d'$ \\
\hline
$a$ & 8 & $(-1, -1) \times (0, -1) \times (-1, -1)$ & 0 & 0  & 0 & 0 & 0 & -3 & 0 & 3\\
$b$ & 4 & $(-2, -2) \times (-1, 0) \times (1, -1)$ & 0 & 0  & - & - & 0 & 2 & 0 & -2\\
$c$ & 4 & $(-1, -1) \times (-3, -1) \times (0, -1)$ & 2 & -2  & - & - & - & - & 0 & -4\\
$d$ & 4 & $(-1, -1) \times (-1, 1) \times (2, 1)$ & 2 & 6  & - & - & - & - & - & -\\
\hline
 & & & \multicolumn{8}{c|}{$x_A = 3x_B = x_C = 3x_D$}\\
 & & & \multicolumn{8}{c|}{$\chi_1=2$, $\chi_2=3$, $\chi_3=1$}\\
\hline
\end{tabular}
\end{center}
\end{table}

\FloatBarrier

\begin{table}[h!]
\scriptsize
\caption{D6-brane configurations and intersection numbers of Class 14, and its MSSM gauge coupling relation is $g_a^2 = \frac{5}{3}g_L^2 = \frac{5}{3}g_c^2 = \frac{7}{5} (\frac{5}{3}g_Y^2) = 4\sqrt{\frac{2}{3}}\pi e^{\phi_4}$, with $184320\,(6144\times 30)$ dual models.}
\label{Lmodel1}
\begin{center}
\begin{tabular}{|c||c|c||c|c|c|c|c|c|c|c|}
\hline
\rm{Model} 14 & \multicolumn{10}{c|}{$U(4)\times U(2)_{L_1}\times U(2)_R\times U(2)_{L_2} $}\\
\hline \hline
\rm{stack} & $N$ & $(n^1,l^1)\times(n^2,l^2)\times(n^3,l^3)$ & $n_{\Ysymm}$ & $n_{\Yasymm}$ & $b$ & $b'$ & $c$ & $c'$ & $d$ & $d'$ \\
\hline
$a$ & 8 & $(1, -1) \times (0, -1) \times (-1, 1)$ & 0 & 0  & 3 & 0 & -3 & 0 & 0 & 0\\
$b$ & 4 & $(-1, -1) \times (-3, -1) \times (0, -1)$ & 2 & -2  & - & - & 0 & -4 & 0 & 2\\
$c$ & 4 & $(-1, -1) \times (-1, 1) \times (2, 1)$ & 2 & 6  & - & - & - & - & 0 & -2\\
$d$ & 4 & $(-2, -2) \times (-1, 0) \times (1, -1)$ & 0 & 0  & - & - & - & - & - & -\\
\hline
 & & & \multicolumn{8}{c|}{$x_A = 3x_B = x_C = 3x_D$}\\
 & & & \multicolumn{8}{c|}{$\chi_1=2$, $\chi_2=3$, $\chi_3=1$}\\
\hline
\end{tabular}
\end{center}
\end{table}
\FloatBarrier

The full list of 14 independent classes of models with extra $d$-stack in the $SU(2)_L$ gauge extension can be summarized into Table~\ref{L} as follows.
\begin{table}[h!] 
\centering
\caption{Multiplicity of equivalent models in each gauge coupling class of the left‑handed $SU(2)_L$ extension. The basic factor $6144$ originates from torus permutations, orientifold image, T‑duality and the D6‑brane sign equivalence principle.}
\label{L}
\label{tab:multiplicityR}
\begin{tabular}{|c||c||c||c|}
\hline
Class & Total models & Factor $/6144$ & Gauge coupling relation \\
\hline\hline
1 & 12288 & 2 & $g_c = g_d$ \\
2 & 12288 & 2 & $g_c = g_d$ \\
3 & 12288 & 2 & $g_b = g_c = g_d$ \\
4 & 12288 & 2 & $g_b = g_c$ \\
5 & 12288 & 2 & $g_c = g_d$ \\
6 & 12288 & 2 & $g_c = g_d$ \\
7 & 12288 & 2 & $g_c = g_d$ \\
8 & 30720 & 5 & - \\
9 & 61440 & 10 & - \\
10 & 61440 & 10 & $g_b = g_d$ \\
11 & 61440 & 10 & $g_b = g_c = g_d$ \\
12 & 61440 & 10 & - \\
13 & 147456 & 24 & $g_c = g_d$ \\
14 & 184320 & 30 & - \\
\hline
\end{tabular}
\end{table}

\end{document}